\documentclass[a4paper,10.5pt]{article}

\usepackage{blkarray}
\usepackage[hyphens]{url}
\usepackage{mathtools,amssymb}
\usepackage[usenames,dvipsnames,table]{xcolor}
\usepackage[titletoc]{appendix}
\usepackage{mathtools,amssymb}
\usepackage[makeroom]{cancel}
\usepackage{graphicx}
\usepackage{epstopdf}
\usepackage{mathrsfs}
\usepackage{paralist}
\numberwithin{equation}{section} 
\usepackage{enumitem}
\usepackage{stackrel}
\usepackage{float}
\usepackage{booktabs}
\usepackage[normalem]{ulem}  
\usepackage{cancel} 
\usepackage{hyperref}
\usepackage[final]{showkeys} 
\usepackage{longtable}
\usepackage{physics}
\usepackage{empheq}
\usepackage{cite} 
\usepackage{comment}
\usepackage{tikz}
\usetikzlibrary{decorations.pathmorphing}
\usetikzlibrary{decorations.pathreplacing}
\usetikzlibrary{decorations.text}
\usepackage{braket}
\usepackage{fullpage}
\usepackage{subcaption}
\usepackage{pgfplots}
\usepackage{amsmath}
\usepackage{ytableau}
\usepackage{dsfont}
\usetikzlibrary{patterns}

\DeclareMathAlphabet{\mathpzc}{OT1}{pzc}{m}{it}
\DeclareMathSymbol{\mlq}{\mathord}{operators}{``}
\DeclareMathSymbol{\mrq}{\mathord}{operators}{`'}

\allowdisplaybreaks
\allowbreak

\definecolor{refkey}{gray}{0.75}
\definecolor{labelkey}{RGB}{155,48,48}
\renewcommand*\showkeyslabelformat[1]{%
	\fbox{\parbox[t]{0.8\marginparwidth}{\raggedright\normalfont\scriptsize\url{#1}}}}

\usepackage{etoolbox}
\makeatletter
\patchcmd{\hyper@makecurrent}{%
	\ifx\Hy@param\Hy@chapterstring
	\let\Hy@param\Hy@chapapp
	\fi
}{%
	\iftoggle{inappendix}{
		\@checkappendixparam{chapter}%
		\@checkappendixparam{section}%
		\@checkappendixparam{subsection}%
		\@checkappendixparam{subsubsection}%
		\@checkappendixparam{paragraph}%
		\@checkappendixparam{subparagraph}%
	}{}%
}{}{ \errmessage{failed to patch}}

\newcommand*{\@checkappendixparam}[1]{%
	\def\@checkappendixparamtmp{#1}%
	\ifx\Hy@param\@checkappendixparamtmp
	\let\Hy@param\Hy@appendixstring
	\fi
}
\makeatletter

\newtoggle{inappendix}
\togglefalse{inappendix}

\apptocmd{\appendix}{\toggletrue{inappendix}}{}{\errmessage{failed to patch}}
\apptocmd{\subappendices}{\toggletrue{inappendix}}{}{\errmessage{failed to patch}}

\newcommand{\lsim}{\mathrel{\hbox{\rlap{\lower .55ex
				\hbox{$\sim$}} \kern-.3em \raise.4ex \hbox{$<$}}}}
\newcommand{\gsim}{\mathrel{\hbox{\rlap{\lower.55ex
				\hbox{$\sim$}} \kern-.3em \raise.4ex \hbox{$<$}}}}

\begin{document}
	
	
\newcommand{\partiald}[2]{\dfrac{\partial #1}{\partial #2}}
\newcommand{\be}{\begin{equation}}
\newcommand{\ee}{\end{equation}}
\newcommand{\f}{\frac}
\newcommand{\s}{\sqrt}
\newcommand{\lm}{\mathcal{L}}
\newcommand{\wm}{\mathcal{W}}
\newcommand{\om}{\omega}
\newcommand{\bea}{\begin{eqnarray}}
\newcommand{\eea}{\end{eqnarray}}
\newcommand{\ba}{\begin{align}}
\newcommand{\ea}{\end{align}}
\newcommand{\ep}{\epsilon}
\newcommand{\h}{\hat}
\def\ad{a^\dagger}
\def\psid{\psi^\dagger}
\def\ads{AdS$_{\text{2}}$~}
\def\gap#1{\vspace{#1 ex}}
\def\be{\begin{equation}}
\def\ee{\end{equation}}
\def\bal{\begin{array}{l}}
\def\ba#1{\begin{array}{#1}}  
\def\ea{\end{array}}
\def\bea{\begin{eqnarray}}
\def\eea{\end{eqnarray}}
\def\beas{\begin{eqnarray*}}
\def\eeas{\end{eqnarray*}}
\def\del{\partial}
\def\eq#1{(\ref{#1})}
\def\fig#1{Fig \ref{#1}} 
\def\re#1{{\bf #1}}
\def\bull{$\bullet$}
\def\nn{\nonumber}
\def\ub{\underbar}
\def\nl{\hfill\break}
\def\bibi{\bibitem}
\def\vev#1{\langle #1 \rangle} 
\def\mattwo#1#2#3#4{\left(\begin{array}{cc}#1&#2\\#3&#4\end{array}\right)} 
\def\tgen#1{T^{#1}}
\def\half{\frac12}
\def\floor#1{{\lfloor #1 \rfloor}}
\def\ceil#1{{\lceil #1 \rceil}}
	
\def\Tr{{\rm Tr}}
		
\def\mysec#1{\gap1\ni{\bf #1}\gap1}
\def\mycap#1{\begin{quote}{\footnotesize #1}\end{quote}}
		
\def\Red#1{{\color{red}#1}}
\def\blue#1{{\color{blue}#1}}
\def\Om{\Omega}
\def\a{\alpha}
\def\b{\beta}
\def\l{\lambda}
\def\g{\gamma}
\def\ep{\epsilon}
\def\Si{\Sigma}
\def\p{\phi}
\def\z{\zeta}

\def\lan{\langle}
\def\ran{\rangle}

\def\bit{\begin{item}}
\def\eit{\end{item}}
\def\benu{\begin{enumerate}}
\def\eenu{\end{enumerate}}
\def\fr#1#2{{\frac{#1}{#2}}}
\def\gsq{{{\tilde g}^2}}
	
\def\tr{{\rm tr}}
\def\intk#1{{\int\kern-#1pt}}


\newcommand{\com}{\textcolor{red}}
\newcommand{\new}[1]{{\color[rgb]{1.0,0.,0}#1}}
\newcommand{\old}[1]{{\color[rgb]{0.7,0,0.7}\sout{#1}}}
		
\renewcommand{\real}{\ensuremath{\mathbb{R}}}
		
\newcommand*{\Cdot}[1][1.25]{%
	\mathpalette{\CdotAux{#1}}\cdot%
		}
		\newdimen\CdotAxis
		\newcommand*{\CdotAux}[3]{%
			{%
				\settoheight\CdotAxis{$#2\vcenter{}$}%
				\sbox0{%
					\raisebox\CdotAxis{%
						\scalebox{#1}{%
							\raisebox{-\CdotAxis}{%
								$\mathsurround=0pt #2#3$%
							}%
						}%
					}%
				}%
				\dp0=0pt %
				\sbox2{$#2\bullet$}%
				\ifdim\ht2<\ht0 %
				\ht0=\ht2 %
				\fi
				\sbox2{$\mathsurround=0pt #2#3$}%
				\hbox to \wd2{\hss\usebox{0}\hss}%
			}%
		}
		
\newcommand\hcancel[2][black]{\setbox0=\hbox{$#2$}%
\rlap{\raisebox{.45\ht0}{\textcolor{#1}{\rule{\wd0}{1pt}}}}#2} 
		
\renewcommand{\arraystretch}{2.5}%
\renewcommand{\floatpagefraction}{.8}%
	
\def\newthing{\marginpar{{\color{red}****}}}
\def\doubt#1{{\color{red}{[*** #1]}}}
\def\vp{\varphi}
\def\vep{\varepsilon}
\reversemarginpar

\def\comments#1{({\color{red}{#1}}*****)}

		
\def\tu{\tau}
\def\ze{z}
\def\d{\partial}
\def\L{\varphi}  
		
\DeclareRobustCommand{\rchi}{{\mathpalette\irchi\relax}}
\newcommand{\irchi}[2]{\raisebox{\depth}{$#1\chi$}}


\hypersetup{pageanchor=false}

\vspace{.4cm}
\begin{center}
  \noindent{\Large \bf{2D or not 2D:\\
      a ``holographic dictionary'' for Lowest Landau Levels}}\\
\vspace{1cm} 
Gautam Mandal $^a$ \footnote{mandal@theory.tifr.res.in}, 
Ajay Mohan $^{a,b}$ \footnote{ajay.mohan@tifr.res.in} and
Rushikesh Suroshe $^b$ \footnote{rushikesh.suroshe@tifr.res.in}
\vspace{.3cm}
\begin{center}
$^a$ {\it International Centre for Theoretical Sciences}\\
{\it Shivakote, Bangalore 560089, India.}\\
{~}\\
$^b$ {\it Department of Theoretical Physics}\\
{\it Tata Institute of Fundamental Research, Mumbai 400005, India.}\\
\end{center}

\gap2
\end{center}

\begin{abstract}
We consider two-dimensional fermions on a plane with a perpendicular magnetic field, described by Landau levels. It is well-known that, semiclassically, restriction to the lowest Landau levels (LLL) amounts to imposing two constraints on a 4D phase space, which transforms the 2D coordinate space $(x,y)$ into a 2D phase space, thanks to the non-zero Dirac bracket between $x$ and $y$. A straightforward application of Dirac's prescription of quantizing LLL in terms of $L^2$ functions of $x$ (or of $y$) fails because the wavefunctions are clearly functions of $x$ {\it and} $y$. We find it possible, however, to construct a different 1D QM, sitting differently inside the 2D QM, which describes the LLL physics. The construction includes an exact 1D-2D correspondence between the fermion density $\rho(x,y)$ and the Wigner distribution of the 1D QM. In an appropriate large $N$ limit, (a) the Wigner distribution is upper bounded by 1, reflecting the semiclassical intuition that a phase space cell can have at most one fermion (Pauli exclusion principle) and (b) the 1D-2D correspondence becomes an identity transformation. (a) and (b) then imply an upper bound for the fermion density $\rho(x,y)$ which verifies known facts from LLL physics. We also explore the entanglement entropy (EE) of subregions of the 2D noncommutative space which displays behaviour distinct from conventional 2D systems as well as from conventional 1D systems, falling somewhere between the two. The main distinguishing feature of the EE, which is directly attributable to the noncommutative nature of space, is the absence of a logarithmic dependence on the size of the entangling region, even though there is a Fermi surface. In this paper, instead of working directly with the Landau problem, we consider a more general problem, namely 2D fermions in a rotating harmonic trap, which reduces to the Landau problem in a special limit. Among other consequences of the emergent 1D physics, we find that post-quench dynamics of the (generalized) LLL system is computed more simply in 1D terms, which is described by well-developed methods of 2D phase space hydrodynamics (see, e.g. \cite{Kulkarni:2018ahv} for a recent application). 
\end{abstract}

\pagenumbering{roman}

\newpage

\tableofcontents
\pagenumbering{arabic}
\setcounter{page}{1}



\setcounter{footnote}{0} 


\section{Introduction and Summary}\label{sec:intro}

Classically it is argued that an electron moving in a lowest Landau level (LLL)\footnote{Landau levels refer to energy levels of electrons on a plane with a constant, perpendicular magnetic field. This system and its generalizations are described in detail in Appenndix \ref{app:ll} and Section \ref{sec2}.} has a 2D phase space description \cite{Dunne:1989hv,Guralnik:2001ax}. The argument is that the LLL constraints (two in number) reduce the original 4D phase space to 2D. The 2D reduced phase space can be coordinatized by the original coordinates $(x,y)$ themselves, since they develop a nontrivial Dirac bracket (see, e.g. \cite{Dirac_1950} and the review in Appendix \ref{app:dirac}): namely, $\{x,y\}_{_{DB}}= \varepsilon \equiv \frac1{2mw}$.

\noindent Ordinarily, the appearance of such a symplectic structure in the ($x,y$) plane would lead one to expect the following:

\begin{enumerate}

\item[(a)] The LLL system should be describable in terms of a 1D QM in which \footnote{following Dirac's prescription of quantizing a constrained system \cite{Dirac_1950, DiracBook1964}.} the Dirac bracket goes over to a commutator bracket:
\begin{align}
  [x,y]= i\hbar \varepsilon \equiv i\hbar_{\rm eff}, \quad \hbar_{\rm eff}= \hbar \varepsilon= \fr{\hbar}{2 m \om}.
  \label{hbar-eff}
\end{align}
In other words, the LLL Hilbert space $H_{LLL}$ should be describable by square integrable ($L^2$) functions of $x$, on which $y$ behaves as $-i\hbar_{\rm eff} \fr{\del}{\del x}$.\footnote{or, by $L^2$ functions of $y$ on which $x$ behaves as $i\hbar_{\rm eff} \fr{\del}{\del y}$.}

{\it We will find that this does not work} (see Section (\ref{sec:LLL-classical}) for details and to see what version of the equation \eq{hbar-eff} does work out).\footnote{Note the contrast with gauge theories where the constraints define the full Hilbert space and the Dirac brackets do go over to commutator brackets.}

\item[(b)] Since $(x,y)$ is supposed to behave like a phase space, one would expect that the fermion density $\rho(x,y)$ for an $N$-fermion state should obey an upper bound following from the Pauli exclusion principle, namely that a phase space cell of size $\hbar_{\rm eff}$ cannot hold more than one fermion. Quantitatively, one would expect  
\begin{align}
  0 \le \lan \rho(x,y) \ran \le \rho_{\rm max}, \quad
  \rho_{\rm max} = \fr1{\hbar_{\rm eff}}= \fr{2m\om}{\hbar}
  \label{rho-max}
\end{align}

{\it We will find that, even though (a) does not hold, (b) does hold in a suitable semiclassical limit} (see Section (\ref{sec:N-particles}) for details).

\end{enumerate}

This is a puzzle: if (a) does not hold, that would appear to imply that quantum mechanically $x$ and $y$ are not conjugate variables; in that case, how does Pauli exclusion principle still hold for fermions in the $(x,y)$ plane? This is one of the main questions we explore in this paper.

The answer to this question is based on the fact that there is a {\it different} 1D QM sitting inside the full Hilbert space (in terms of a coordinate which is neither $x$ nor $y$, see \eq{arbit-lll}, \eq{iso}), which is isomorphic to the LLL Hilbert space. This allows us to derive a 1D-2D correspondence which leads to expressions for LLL observables in terms of this specific 1D QM. In particular, it leads to an exact expression for the fermion density $\lan \rho(x,y) \ran$ in LLL states in terms of (an integral transform of) the corresponding Wigner function of the 1D problem (see \eq{1d-2d-single}, \eq{1d-2d-N}).

We find that in the semiclassical limit implemented by large $N$ (see \eq{large-N})\footnote{In this paper we will follow \cite{Das:1991uta, Das:1991ba, Dhar:1992rs, Dhar:1992hr} to describe the semiclassical limit of fermions in terms of the Wigner distribution.}, the above integral transform between the fermion density and the 1D Wigner function becomes an identity transformation (see, e.g. \eq{rho-vs-u-classical2}). In this limit, the phase space corresponding to the 1D QM gets identified with $(x,y)$ and the corresponding 1D Wigner function obeys an upper bound of 1 (where the bound is saturated for single Slater states \footnote{By a Slater state, we refer to an $N$-particle state whose wavefunction $\Psi$ can be written as a Slater determinant of $N$ single-particle wavefunctions $\psi_i$: $ \Psi(\vec x_1, ...,\vec x_N)= {\rm Det}_{i,k}[\psi_{i}(\vec x_k)]$. In the second quantized notation this is a Fock space state of the form $|F\rangle = c^\dagger_{\psi_N}... c^\dagger_{\psi_1}|0\rangle.$ }). Consequently, $\rho(x,y)$ satisfies \eq{rho-max} above, since it gets identified with the Wigner function $\left( \text{times} \fr1{\hbar_{\rm eff}}\right)$. 

Besides the above question of the bound, the 1D-2D correspondence allows us to address dynamical questions of the LLL system in 1D terms, which is described by well-developed methods of 2D phase space hydrodynamics (for a recent application see, e.g. \cite{Kulkarni:2018ahv}).

In this paper, we also discuss if the effective 1D property of the LLL system is reflected in the ground state entanglement entropy (EE) of a subregion\footnote{Taken to be disc shaped for simplicity.} of size $R$ of the $(x,y)$ plane, where the Fermi energy is well within the LLL band. Unlike in a conventional free fermion system with a Fermi surface (such as free fermions in a 2D harmonic trap), where the EE is supposed to go as $R \log R$, in the LLL context the EE goes as $R$. So, the EE does not behave like in a standard 2D system; in fact, nor does it behave in a standard 1D system where the EE of a subinterval of size $R$ of the real line would go as $\log R$. So as far as the EE is concerned, the noncommutative 2D is somewhere between 2D and 1D! The main distinguishing feature of the EE, which is directly attributable to the noncommutative nature of space, is the absence of a logarithmic dependence on the size of the entangling region, even though there is a Fermi surface. The reason for this is the appearance of short range correlators with a range set by the scale of noncommutativity.

We note that the idea of a 1D-2D correspondence has appeared recently in \cite{Das:2022mtb}, although in a slightly different way; the difference with our work is that while \cite{Das:2022mtb} invokes a formal auxiliary 1D fermion system, our 1D QM is embedded as a one dimensional subspace of the existing 2D QM.

\paragraph{The plan for the rest of the paper}

We describe in Section [\ref{sec2}] and Appendix [\ref{app:ll}] some basics of Landau level physics\cite{1930ZPhy...64..629L} and its generalization in terms of 2D fermions in a rotating harmonic trap \cite{Lacroix-A-Chez-Toine:2018rwb,PhysRevA.107.023302}. We describe the classical LLL constraints and Dirac brackets\cite{Dirac_1950} in Section [\ref{sec:LLL-classical}] and Appendix [\ref{app:dirac}], where we also point out how low energy constraints are different from gauge constraints and why na{\i}ve Dirac quantization does not work. In Section [\ref{sec:single-1d-2d}], we discuss a quantum isomorphism between the LLL subspace of the 2D QM and a 1D QM (see \eq{arbit-lll},\eq{iso}), which leads to an exact integral relation between the 2D fermion density $\rho(x,y)$ and the Wigner distribution \cite{PhysRev.40.749} following from the equivalent 1D QM. Derivation of this relation requires some preliminaries about Weyl correspondence between Hilbert space operators and phase space functions, and a discussion about Wigner distributions; these are provided in Appendix [\ref{app:weyl-wigner}]. In Section \ref{sec:N-particles}, we extend the discussion to $N$-particle states. These include (i) Slater states (with wavefunctions given by Slater determinants), including the Ground state, (ii) linear combination of Slater states, (iii) $W_\infty$ coherent states and (iv) mixed states, like the thermal state. We show how for (i), (iii) and (iv), the exact 1D-2D correspondence continues to hold. The result (b) above, about the upper bound for $\lan\rho(x,y)\ran$ continues to hold in all cases, as shown in Section  (\ref{sec:N-particles}) , under the large $N$ limit \eq{large-N}. In section [\ref{sec:dynamics}], the 1D-2D correspondence is used to show how the simple dynamics of the Wigner distribution at the large $N$ limit (in terms of phase space hydrodynamics) is inherited by $\rho(x,y,t)$ to allow computation of the time-evolution of the fermion density by classical fluid flow. In Section [\ref{sec:EE}] we discuss the ground state entanglement entropy (EE) of a disk shaped region of size $R$ in the $(x,y)$ plane; the result ($\propto R$) is different from conventional 2D systems with a Fermi surface (where EE is $\propto R\log R$ \cite{Klich_2006,PhysRevLett.105.050502}) and also from conventional 1D system (where an interval of size $R$ will have EE $\propto \log R$). We conclude in Section (\ref{sec:conclusion}) with some remarks on the relation with the integer Quantum Hall system\cite{tong}. We collect some additional mathematical results in the remaining appendices.

\section{Generalized Landau system}\label{sec2}

The system of Landau fermions refer to 2D fermions subject to a uniform transverse magnetic field $B$ (see Appendix \ref{app:ll} for more details). In the symmetric gauge \cite{PhysRevB.27.3383}, the Hamiltonian becomes \eq{h-ll} which we reproduce here
\begin{align}
  H_0  =\frac{1}{2m}\left( \left({p_{x}}-\frac{eB {y}}{2}\right)^{2}+\left({p_{y}}+\frac{eB{x}}{2}\right)^{2}\right)
  \label{h-ll-text}
\end{align}

\subsection{Generalization: fermions in a rotating harmonic trap}

It is an interesting fact that the Landau system \eq{h-ll-text} is related to free fermions rotating in a 2D harmonic trap \cite{Lacroix-A-Chez-Toine:2018rwb,PhysRevA.103.033321}, for which the Hamiltonian is
\begin{align}
    H = \frac{p_{x}^{2}+p_{y}^{2}}{2m} + \frac{m \omega^{2}(x^{2}+y^{2})}{2} +\Omega L_z, \quad L_z= (xp_{y}-yp_{x}) \label{trap-ham}
\end{align}
The two Hamiltonians in fact coincide if $\omega= \Omega= \fr{eB}{2m}$. This suggests a natural generalization of the Landau system in which we take $\Omega \le \omega$. \footnote{For $\Omega > \omega$ the centrifugal force causes the fermions to fly off and the system is unstable.}

Let us parameterize $\Omega = \nu \omega$, with $\nu \le 1$; in the following we will mostly take $\nu$ to be close to 1. The Hamiltonian \eq{trap-ham} can be regarded as a generalization of the Landau Hamiltonian \eq{h-ll-text} in two different ways. If we regard the coefficient $\omega^2$ of the trap potential as $\Omega^2 + (\omega^2 - \Omega^2)$ and choose $\Omega=eB/(2m)$, then $H$ describes Landau fermions in a harmonic potential:
\begin{align}
  H= H_0 + \Delta H_{trap}, \hspace{7ex} \Delta H_{trap}= \frac{m (\omega^{2} - \Omega^2)(x^{2}+y^{2})}{2} = \frac{e^2 B^2}{8m}(1-\nu^2) (x^{2}+y^{2})
  \label{delta-h-trap}
\end{align}
Alternatively, if we regard the rotation term as $\left(-\omega +(\omega - \Omega)\right)L_z$ and choose $\omega=eB/(2m)$, then $H$ describes Landau fermions in a rotating frame:
\begin{align}
  H= H_0 - \Delta H_{rot}, \hspace{7ex} \Delta H_{rot}= (\omega - \Omega))L_z = \frac{eB}{2m}(1- 1/\nu)L_z
  \label{delta-h-rot}
\end{align}

With the above understanding, in the rest of the paper, we will work with the Hamiltonian \eq{trap-ham}, which describes a generalized Landau problem. It is not difficult to see, by using the ladder operators \eq{b-ll} and \eq{a-ll}, that the Hamiltonian is now given by (see Appendix A of \cite{PhysRevA.103.033321} for more details)
\begin{align}
    H =  \hbar \omega(1 + (1+\nu) a^\dagger a + (1-\nu) b^\dagger b ) \label{trap-spec}
\end{align}
It is clear that the eigenvectors of $H$ are given by the states \eq{n1-n2}
\begin{align}
  |n_1, n_2\ran =\frac{1}{\sqrt{n_{1}!}\sqrt{n_{2}!}} (a^\dagger)^{n_1} (b^\dagger)^{n_2} |0\ran, \quad
  a|0\ran = b|0\ran=0
  \label{n1-n2-trap}
\end{align}
with the energy spectrum
\begin{align}
    E_{n_1, n_2} =  \hbar \omega(1 + (1+\nu)n_1 + (1-\nu)n_2 ) \label{trap-eigen}
\end{align}
Note that for $\nu=1$, the Hamiltonian and the eigenvalues reduce to those of the Landau Hamiltonian, viz. \eq{spectrum} and \eq{spect}. 

As in the original Landau problem (see Appendix \ref{app:x1-x2}), the two sets of ladder operators can be expressed in terms of two sets of phase space variables $(x_1,p_1, x_2,p_2)$ \eq{a-b-x1-x2}, which are related to the original phase space variables $(x,y,p_x,p_y)$ by \eq{x1-x2-x-y}. In terms of these new phase space variables, the Hamiltonian $H$ \eq{trap-ham} or \eq{trap-spec} is written as
\begin{align}
  H = (1+\nu)\frac{1}{2m}(m^2\omega^2x_1^2 + p_1^2) + (1-\nu)\frac{1}{2m}(m^2\omega^2x_2^2 + p_2^2)
  \label{trap-ham-x1-x2}
\end{align}
It is clear that the wavefunctions corresponding to \eq{n1-n2-trap} are the same as in the original Landau problem, viz. \eq{n1-n2-xy} and \eq{n1-n2-x1-x2} in the $(x,y)$ and $(x_1,x_2)$ representations respectively.

\subsection{Lowst Landau level (LLL) of the generalized problem}\label{sec:ll-gen-band}

The band structure of the original Landau problem (see Fig \ref{fig:ll-band} in Appendix \ref{app:lll}) is now generalized to Fig \ref{fig:lll-gen-band}, in accordance with \eq{trap-eigen}.
\begin{figure}[H]
  \begin{center}
    \begin{tikzpicture}[scale=0.65]

\draw[->] (0,0) -- (0,10) node[left] {$E$};
\draw[->] (0,0) -- (9,0) node[right] {$n_2$};

\draw[blue] (0,3) -- (9,3);
\draw[blue] (0,6) -- (9,6);
\node[left,blue] at (5,0.2) {$n_1=0$};
\node[left,blue] at (0,0) {$E_0=\hbar \omega$};
\node[left,blue] at (5,3.2) {$n_1=1$};
\node[left,blue] at (0,3) {$E_1=(2+\nu)\hbar \omega$};
\node[left,blue] at (5,6.2) {$n_1=2$};
\node[left,blue] at (0,6) {$E_2=(3+2\nu)\hbar \omega$};
\draw[red] (0,0) -- (8,1);
\foreach \i in {0,0.2,...,6.2} {
    \pgfmathsetmacro{\y}{1/8*\i} 
    \fill[black] (\i,\y) circle (0.08);
}

\draw[red] (0,3) -- (8,4);
\draw[red] (0,6) -- (8,7);

\end{tikzpicture}
  \end{center}
  \caption{\footnotesize{Landau levels in the generalized Landau problem. The energy spectra, for various values of $n_1=0,1,2,...$, are given by the red lines, which define the Landau levels in the generalized problem. The lowest Landau level corresponds to $n_1=0$ and the black dots represent the allowed $n_2$ values. To be contrasted with the energy levels of the original Landau problem depicted in Figure \ref{fig:ll-band}.}}
  \label{fig:lll-gen-band}
\end{figure}
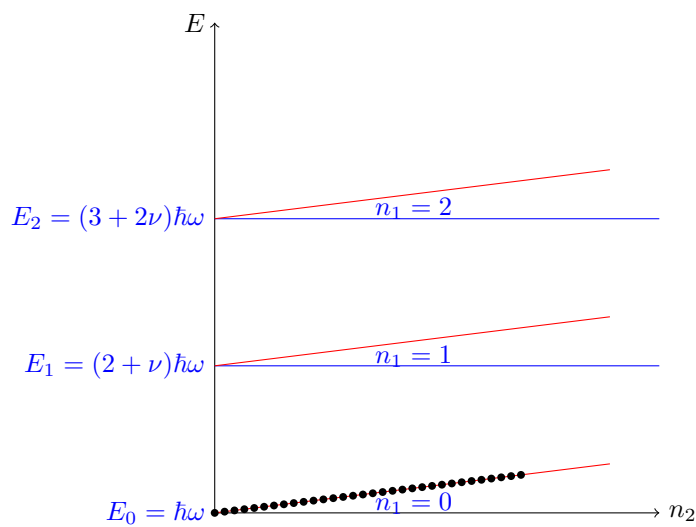
The lowest energy band corresponds to $n_1=0$, which we will call the lowest Landau level (LLL). States in the LLL are not degenerate; they have a spectrum
\begin{align}
    E_{0, n_2} =  \hbar \omega(1 + (1-\nu)n_2 ). \label{trap-lll}
\end{align}  
We will consider $\nu$ to be close to 1, with $1-\nu \ll 1$, so that the degeneracy breaking term in \eq{trap-lll} is very small. We will work with values $n_2$ which are such that $E_{0,n_2} \ll E_{1,0}$ (i.e. $n_2 \ll (1-\nu)/(1+\nu)$) so that the LLL state of interest is always less energetic that any state in the first Landau level corresponding to $n_1=1$. 

The LLL wave functions are given by the same expressions \eq{lll-wavefunctions} as in the original Landau problem. We reproduce the expressions here for convenince:
\begin{align}
& \psi_{n}(x,y)= \langle x,y|0,n \rangle=  \frac{e^{\frac{-1}{2l_{0}^
        {2}} \left(x^2+y^2\right)} \left((x-i y)/l_{0}\right)^{{n}}}{l_{0}\sqrt{\pi {n}!}} \label{lll-gen-psi}\\
  & \tilde\chi_{n}(x_{1},x_{2}):= \langle x_1, x_2 |0,n \rangle =\bar\chi_{0}(x_1) \chi_n(x_2), \nonumber\\
  &\bar\chi_{0}(x_1) := \frac {1}{\sqrt{l_0\sqrt{\pi}} } e^{-  x_{1}^{  2 } /(2 l_0^2)},\; \chi_n(x_2):=  \frac {1}{\sqrt{  2^n n!l_0\sqrt{\pi}} }
   e^{-   x_{ 2}^ 2 /  (2 l_0) } H_{n}(   x_{  2 } /l_{0})
  \label{lll-gen-chi}
\end{align}
In the above
\begin{align}
  l_{0}=\sqrt{\frac{\hbar}{m\om}}
  \label{def-l0}
\end{align}
defines a characteristic length scale in the Landau problem.

\subsection{LLL condition and constrained 2D phase space}\label{sec:LLL-classical}

Let us define the projection operator $\hat P$ to the LLL states:
\begin{align}
  \hat P \equiv \sum_{n=0,1,...,\infty} |0,n\rangle \langle 0,n|
  \label{p-lll}
\end{align}
where we wrote $n$ for $n_2$. The projection of the full Hilbert space ${\cal H}$ onto the LLL sector will be denoted as ${\cal H}_{LLL}= \hat P {\cal H}$. We will define the projection of an arbitrary state $|\psi\rangle$ onto ${\cal H}_{LLL}$ as
\[
|\psi \rangle_P = \hat P |\psi\rangle =\sum_{n=0,1,...,\infty} \langle 0,n|\psi\rangle |0,n\rangle
\]
Similarly operators projected to the LLL sector of the Hilbert space are given by
\begin{align}
  \hat O_P \equiv \hat P \hat O \hat P = \sum_{n,m=0,1,...,\infty} \langle 0,n| \hat O |0,m \rangle |0,n \rangle \langle 0,m|
  \label{o-p}
\end{align}
It is easy to prove that the operators $\hat x_1, \hat p_1$, projected to the LLL, vanish:\footnote{We use below the notation $\hat O_P \equiv \hat P \hat O \hat P$ \label{ftnt:projection}}
\begin{align}
  (\hat x_1)_P=0= (\hat p_1)_P.
  \label{x1-p1=0-qm}
\end{align}
This follows simply by noting that on every LLL state $a|0,n\rangle=0$ (by definition). This, of course, implies $\langle 0,n| a |0,m \rangle=0$ which further implies $\langle 0,n| a^\dagger |0,m \rangle=0$. By taking sums and difference of these equations, and using \eq{a-b-x1-x2}, we find that the LLL matrix elements of $\hat x_1$ and $\hat p_1$ vanish. Therefore, upon using \eq{o-p}, \eq{x1-p1=0} follows.

\subsubsection{Classical constraints}\label{sec:constranits}

Classically, the LLL constraints \eq{x1-p1=0-qm} correspond to the phase space constraints
\begin{align}
  x_1=0=p_1.
  \label{x1-p1=0}
\end{align}
By using \eq{x1-x2-x-y}, these can be rewritten in terms of the original phase space coordinates, as
\begin{align} C_1 =\frac{1}{\sqrt{2}} ( x+ \frac{p_y}{m\omega}) = 0, \quad
  C_2 = \frac{1}{\sqrt{2}}(p_x - m\omega y)=0
  \label{c1-c2}
\end{align}
Thus the LLL physics is described by a reduced two-dimensional phase space ${\cal M}$ \cite{Dunne:1989hv} (instead of the original 4D phase space) defined by the two constraints \eq{c1-c2}.

In Appendix [\ref{app:dirac}] we have shown that the reduced phase space can be parameterized by the coordinates $(x,p)$ which become non-commutative in the sense that they pick up a non-trivial Dirac bracket \eq{x-y-db}:
\begin{align}
  \{x,y\}_{DB}= \fr1{2m\om}
  \label{x-y-db-text}
\end{align}

\subsubsection{Dirac's quantization}\label{sec:dirac-qn}

We note that the LLL constraints are basically some ``effective constraints'' valid at an appropriate range of low energies ($E \ll E_{1,0}$). This is to be contrasted with ``genuine constraints'' such as in gauge theories, where the part of the Hilbert space satisfying the constraints, ${\cal H}_{phys}$, is the entire allowed part of the Hilbert space (see Figure \ref{fig:gauge}).

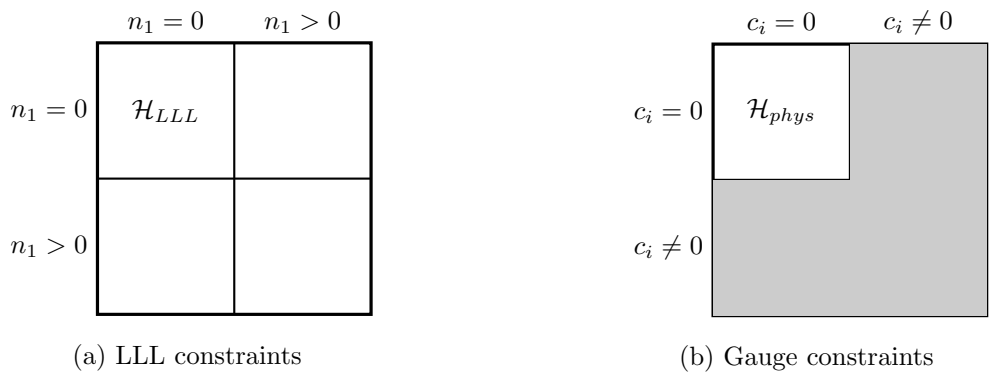
\begin{figure}[H]
    \centering
    \begin{minipage}{0.5\textwidth}
        \centering
        \begin{tikzpicture}[scale=0.9, thick]
            \draw[very thick] (0,0) rectangle (4,4);
            \draw (2,0) -- (2,4);
            \draw (0,2) -- (4,2);

            \node at (1,3) {${\cal H}_{LLL}$};
            \node[above] at (1,4) {$n_1=0$};
            \node[above] at (3,4) {$n_1>0$};
            \node[left] at (0,3) {$n_1=0$};
            \node[left] at (0,1) {$n_1>0$};
        \end{tikzpicture}\\
        (a) LLL constraints
    \end{minipage}%
    \begin{minipage}{.5\textwidth}
        \centering
        \begin{tikzpicture}[scale=0.9, thick]
            \draw[very thick] (0,0) rectangle (4,4);
            \draw (2,0) -- (2,4);
            \draw (0,2) -- (4,2);

            \fill[gray!40] (2,0) rectangle (4,2);
            \fill[gray!40] (0,0) rectangle (2,2);
            \fill[gray!40] (2,2) rectangle (4,4);

            \node at (1,3) {${\cal H}_{phys}$};
            \node[above] at (1,4) {$c_i=0$};
            \node[above] at (3,4) {$c_i\neq 0$};
            \node[left] at (0,3) {$c_i=0$};
            \node[left] at (0,1) {$c_i\neq 0$};
        \end{tikzpicture}\\
        (b) Gauge constraints
    \end{minipage}    
    \caption{\footnotesize{Symbolic representation of constrained Hilbert spaces in terms of projection operators (which are the top left diagonal blocks). Panel (a) represents the LLL constraints which are effective constraints arising from a low energy approximation. Panel (b) represents gauge constraints which are genuine constraints: here the entire Hilbert space is in the top left diagonal block.}}
    \label{fig:gauge}
\end{figure}

Dirac's prescription for quantizing constrained systems \cite{Dirac_1950, DiracBook1964} would be to promote a Dirac bracket to a  commutator:
\begin{align}
  [\hat x,\hat y] \stackrel{\rm\tiny prescription}{=}  i\hbar \{x,y\}_{DB}=i\hbar_{\rm eff}, \quad \hbox{where}~ \hbar_{\rm eff} := \fr{\hbar}{2m\om} = \fr{l_0^2}{2}
  \label{x-y-comm}
\end{align}
where $l_0= \sqrt{\hbar/(m\om)}$ is the characteristic length scale of the generalized Landau problem (see \eq{def-l0}) which defines the scale of non-commutativity. This, of course, cannot be true in the full Hilbert space since $[\hat x,\hat y]=0$. It turns out that the correct version of \eq{x-y-comm} is
\begin{align}
  [\hat x_P,\hat  y_P]= \fr{l_0^2}{2} \hat P
  \label{x-y-comm-true}
\end{align}
which follows from Section \ref{sec:LLL-classical} (see especially \eq{x1-p1=0-qm}).  

\subsubsection{An alternative route to the constraints from the Weyl correspondence}\label{sec:weyl-constraints}

Consider a phase space function $f(x_1,x_2,p_1,p_2)$. By using the Weyl correspondence \cite{Weyl1931Theory} described in Appendix [\ref{app:weyl-wigner}], we can assign to this an operator $\hat f$. As shown in that Appendix, the expectation value of $\hat f$ in any LLL state is given by
\begin{align}
\langle 0,n_2| \hat f |0,n_2\rangle = \int \fr{d\vec x d\vec p}{(2\pi \hbar)^2} u_{n_2}(\vec x, \vec p) f(x_1,x_2,p_1,p_2)
\label{n2-ev}
\end{align}
where $u_{n_2}(\vec x, \vec p)$ is the Wigner distribution \cite{PhysRev.40.749}  for the state $|0,n_2\rangle$. We will find shortly that in the $\hbar \to 0$ limit (see \eq{u-4d},\eq{val-u0}, \eq{u0-delta}),
\begin{align}
  u_{n_2}(\vec x, \vec p) \propto \delta(x_1)\delta(p_1)
  \label{u-n2-delta}
\end{align}
which amounts to putting $x_1=0=p_1$ in the phase space observable $f(x_1,x_2,p_1,p_2)$.

\section{1D-2D correspondence for single-particle states}\label{sec:single-1d-2d}

In the previous section we argued that in the semiclassical limit the LLL system should have a description in terms of a 2D phase space. In this section, we will pursue this, first in the quantum theory and then in the $\hbar \to 0$ limit for single-particle states ($N=1$).

\subsection{Quantum correspondence}

The main result of this section will be the quantum correspondence \eq{1d-2d-single}.

Let us consider a single-particle state $\tilde \chi(x_1,x_2)$ belonging to the LLL Hilbert space. Such a state will be an arbitrary linear combination of the states \eq{lll-gen-chi}:(we will use the notation $\vec x=(x_1,x_2), \vec p=(p_1,p_2),$ etc. below)
\begin{align}
  & \tilde \chi(\vec x) =\sum_n A_n \tilde\chi_n(x_1,x_2) =: \bar\chi_0(x_1) \chi(x_2), \quad \bar\chi_{0}(x_1) := \left({\frac { m  \omega } {  \pi  \hbar }}\right)^{1/4} e^{- m  \omega  x_{1}^{  2 } /(2\hbar)},\ \,
   \chi(x_2)= \sum_n A_n \chi_n(x_2)
  \label{arbit-lll}
\end{align}
where $\tilde\chi_n(x_1,x_2)$ and $\chi_n(x_2)$ are defined as in \eq{lll-gen-chi}.

We can rephrase the above equation more abstractly as an isomorphism between (the LLL sector of) the 2D QM and a 1D QM
\begin{align}
 \boxed{ L^2({\bf R}^2) \ni |\tilde\chi_n \ran = |\bar\chi_0 \ran |\chi_n \ran, \quad |\chi_n \ran \in L^2({\bf R})}.
  \label{iso}
\end{align}
The basis states on the left side are in 1-1 correspondence with the basis states on the right side, thus establishing the isomorphism. 

The wavefunction $\tilde\psi(x,y)$ $= \lan x,y | \tilde \chi \ran$, in the $(x,y)$ representation, is given by (see Appendix [\ref{app:transform}])
\begin{align}
  \tilde\psi(x,y)=\int dx_1 dx_2 \langle x,y | x_1, x_2 \rangle \tilde \chi(x_1,x_2)
  \label{arbit-psi}
\end{align}
We will write explicit expressions only in specific cases; we will not need them for our general results below.

Note that an isomorphism like \eq{iso} exists in the $(x_1,x_2)$ representation, but not in the $(x,y)$ representation. 

The Wigner distribution $\tilde u(\vec x, \vec p)$ (explained in detail in Appendix [\ref{app:weyl-wigner}]) corresponding to an arbitrary state $\ket{\psi}$ is defined as
\begin{align}
  \tilde u(\vec x, \vec p) &= \bra{\psi} \hat g(\vec x, \vec p)| \ket{\psi} = \int d\vec \eta \, {\psi}^*(\vec x+ \vec \eta/2)\, \psi(\vec x- \vec \eta/2)\, \exp[i\vec \eta.\vec \p/\hbar],\\
  \hbox{where  } \hat g(\vec x, \vec p)& = \int \fr{d\vec \a\,d\vec \b}{(2\pi\hbar)^2} \exp\left[i/\hbar\left( \vec a.(\hat{\vec x}-\vec x)+ \vec \b.(\hat{\vec p}-\vec p)\right)\right] \nonumber
\end{align}
Because of the factorized form of the state $\tilde \chi$ in \eq{arbit-lll}, the Wigner distribution corresponding to state $\tilde \chi$ also has a factorized form:
\begin{align}
  \tilde u(\vec x, \vec p) &= \int d\vec \eta \, {\tilde \chi}^*(\vec x+ \vec \eta/2)\, \tilde \chi(\vec x- \vec \eta/2)\, \exp[i\vec \eta.\vec \p/\hbar]   \label{u-4d-a}\\
  &= \bar u_0(x_1, p_1) u(x_2,p_2), \label{u-4d}\\
  \bar u_0(x_1, p_1)&=\int d\eta_1 {\bar\chi_0}^*(x_1+\eta_1/2)\,\bar\chi_0(x_1-\eta_1/2)\exp[i \eta_1 p_1/\hbar]=  \left(2 \exp \left(-\frac{p_1^2}{ (m \omega \hbar)}- \frac{x_1^2 m \omega}{\hbar }\right) \right)\label{val-u0}\\
  u(x_2,p_2)&= \int d\eta_1 {\chi}^*(x_2+\eta_2/2)\,\chi(x_2-\eta_2/2)\exp[i \eta_2 p_2/\hbar]\label{u-x2-p2}
\end{align}
The computation in the second step of \eq{val-u0} involves Gaussian integrals and is straightforward.


We will now prove the following quantum correspondence between the 2D real space fermion density $\rho(x,y)$ and the 2D Wigner distribution $u(x_2, p_2)$:
\begin{align}
  &\boxed{\rho(x,y)\equiv |\tilde \psi(x,y)|^2 = \int \fr{d x_2\,d p_2}{\pi\hbar} K(x,y,x_2,p_2) u(x_2,p_2)} \label{1d-2d-single}\\
  &\hbox{where   } K(x,y,x_2,p_2)= \frac{m \omega}{( \pi \hbar)} \exp\left( -\frac{( \sqrt{2} m\omega y + p_2)^2}{\hbar m \omega} \right) \exp\left( - \frac{(\sqrt{2}x - x_2)^2 m \omega}{\hbar} \right) \label{kernel}
\end{align}
The Wigner distribution $u(x,y,p_x,p_y)$ corresponding to the state \eq{arbit-psi} is given by
\begin{align}
u(x,y,p_x,p_y)= \int d\eta_x d\eta_y \tilde\psi^*(x+\eta_x/2,y+\eta_y/2)\, \tilde\psi(x-\eta_x/2,y-\eta_y/2)\,\exp[i/\hbar(\eta_x p_x + \eta_yp_y)]\nonumber
\end{align}
and the Wigner distribution for a definite LLL state $\ket{0,n}$ is already known to be of the form (with the detailed proof given in Appendix[\ref{wign-xy}])
\begin{align}
u_n(\vec{x},\vec{p})=4\exp \left \{-\left(\frac{x^2+y^2}{l_0^2}+\frac{l_0^2}{\hbar^2}(p_x^2+p_y^2)\right)\right\}L_n\left(\left(\frac{l_0 p_x}{\hbar}-\frac{y}{l_0}\right)^2+\left(\frac{l_0 p_y}{\hbar}+\frac{x}{l_0}\right)^2\right)
\label{eq:wign-xy}\end{align}
According to the analysis in Appendix [\ref{app:weyl-wigner}] leading to \eq{u-inv-d}, we have
\[
u(x,y,p_x,p_y)= \tilde u(x_1,x_2,p_1,p_2)
\]
provided that $x_1,x_2,p_1,p_2$ are related to $x,y,p_x,p_y$ by \eq{x1-x2-x-y}. We can write the above equation in the following way
\begin{align}
u(x,y,p_x,p_y)=\int {d\vec x\,d\vec p} \delta_1 \delta_2 \delta_3 \delta_4 \tilde u(x_1,x_2,p_1,p_2), 
\label{uu1}
\end{align}
where, $\delta_1, \delta_2, \delta_3, \delta_4 $ represent delta functions, defined as
\begin{align}
  &\delta_1=\delta(x_1 -\frac{1}{\sqrt{2} m \omega}(m \omega x + p_y) ),\quad
  \delta_2= \delta(p_1 - \frac{1}{\sqrt{2}}(p_x - m \omega y)),\nonumber \\ 
  & \delta_3 =\delta(x_2 - \frac{1}{\sqrt{2} m \omega}(m \omega x - p_y)) \equiv \sqrt{2}m \omega \delta(p_y-m \omega (x-\sqrt{2}x_2)) , \nonumber \\ 
  &\delta_4=\delta(p_2 - \frac{1}{\sqrt{2}}(p_x + m \omega y)) \equiv \sqrt{2}\delta(p_x-(\sqrt{2}p_2-m \omega y)).  
\end{align}
which implement the transformation of phase space coordinates \eq{x1-x2-x-y}.
The strategy for the proof of \eq{1d-2d-single} is as follows (see Figure \ref{fig:flowchart})
\begin{figure}[H]
  \centering
 \begin{tikzpicture}[scale=0.45]

\draw[blue] (0.5,0) -- (9.5,0);
\draw[blue] (0,0) -- (0,9.5);
\draw[blue] (10,0.5) -- (10,9.5);
\draw[blue] (0,10) -- (10,10);
\node[left,black] at (0,0)  {$\rho(x,y)$};
\node[left,black] at (0,10) {$u(x,y,p_x,p_y)$};
\node[left,black] at (15.5,10) {$\tilde u(x_1,x_2,p_1,p_2)$};
\node[left,black] at (13,0) {$\tilde u(x_2,p_2)$};
\node[left,black] at (0,5) {$\int dp_x dp_y$};
\node[left,black] at (14,5) {$\int dx_1 dp_1$};

\end{tikzpicture}

  \caption{\footnotesize{The flow chart to illustrate the quantum 1D-2D correspondence for single particle LLL wave-function.}}
  \label{fig:flowchart}
\end{figure}
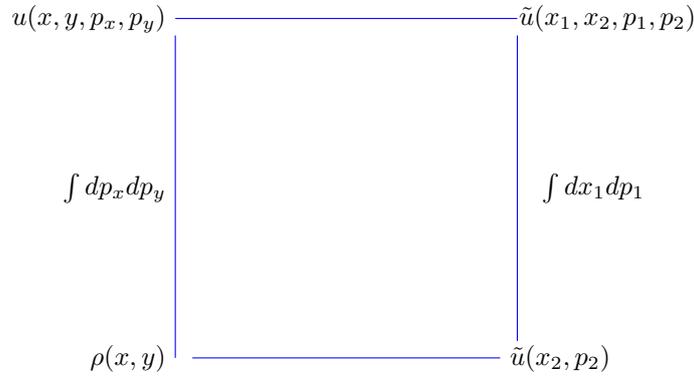
In equations,
\begin{align}
\rho(x,y)&=\int \fr{dp_x\, dp_y}{(2 \pi \hbar)^2} u(x,y,p_x,p_y) \nonumber\\ &=\int \fr{dp_x\, dp_y}{(2 \pi \hbar)^2} {d\vec x\,d\vec p}\delta_1 \delta_2 \delta_3 \delta_4 \tilde u(x_1,x_2,p_1,p_2)\nonumber\\ 
&=\int \fr{dp_x\, dp_y}{(2 \pi \hbar)^2} {d\vec x\,d\vec p}\delta_1 \delta_2 \delta_3 \delta_4 \bar u_0(x_1,p_1)u(x_2,p_2)\nonumber\\ 
&  =  \frac{m \omega}{\pi \hbar}\int \fr{dx_2dp_2}{\pi\hbar}  \exp\left( -\frac{(   p_2-\sqrt{2} m\omega y)^2}{\hbar m \omega} \right) \exp\left( - \frac{(\sqrt{2}x - x_2)^2 m \omega}{\hbar} \right)  u(x_2,p_2)
\label{proof}
\end{align}
which is the same as the equation \eq{1d-2d-single}. In the final step we have used the four delta-functions to integrate out $x_1, p_1, p_x, p_y$.

\subsubsection{Definite LLL eigenstates}\label{sec:examples-single}

Suppose, we specialize to the case of a single-particle occupying a definite LLL state (say $\ket{0,n}$) instead of the more general linear superposition of many LLLs as shown in \eq{arbit-lll}.
Then, the $x_2$ dependent factor in the wavefunction is
\[
\chi(x_2)= \chi_n(x_2):= \left({\frac { m  \omega } {  \pi  \hbar }}\right)^{1/4}
\frac { e^{- m  \omega  x_{  2 }^{  2 } /  (2   \hbar) } } {   \sqrt {  2^n n! } } H_{n}(   x_{  2 } \sqrt{m\om/\hbar})
\]
Using \eq{u-x2-p2} we now get
\begin{align}
  u(x_2,p_2)= u_n(x_2,p_2):=2 (-1)^{n} \exp\left(-\frac{p_2^2 l_0^2}{\hbar^2}- \frac{x_2^2}{l_0^2}\right) L_{n}\left(2\left(\frac{p_2^2 l_0^2}{\hbar^2}+ \frac{x_2^2}{l_0^2}\right) \right), 
  \label{un-x2-p2}
\end{align}
Is this related to $\rho(x,y)$ as in \eq{1d-2d-single}? In the present case, using \eq{lll-gen-psi} we get
\begin{align}
\rho(x,y)= \rho_n(x,y) = |\psi_{n}(x,y)|^2 =  \frac{e^{\frac{-1}{l_{0}^
      {2}} \left(x^2+y^2\right)} \left((x^2+ y^2)/l_{0}^2\right)^{{n}}}{l_{0}^2\pi {n}!}
\label{rhon-x-y}
\end{align}
Here $l_0= \sqrt{\frac{\hbar}{m\om}}$ (see \eq{def-l0}).
It is not difficult to show that \eq{rhon-x-y} is related to \eq{un-x2-p2} by the relation \eq{1d-2d-single}:
\begin{align}
  \rho_n(x,y)\equiv |\tilde \psi(x,y)|^2 = \int \fr{d x_2\,d p_2}{(\pi\hbar)} K(x,y,x_2,p_2) u_n(x_2,p_2)
  \label{un-rhon}
\end{align}
A representative plot of both $\rho_n(x,y)$ and $u_n(x_2,p_2)$ are reproduced below (Figure \ref{fig:un-rhon}).
\begin{figure}[H]
    \centering
    \begin{minipage}{.45\textwidth}
        \centering
        \includegraphics[width=0.45\textwidth, height=0.15\textheight]{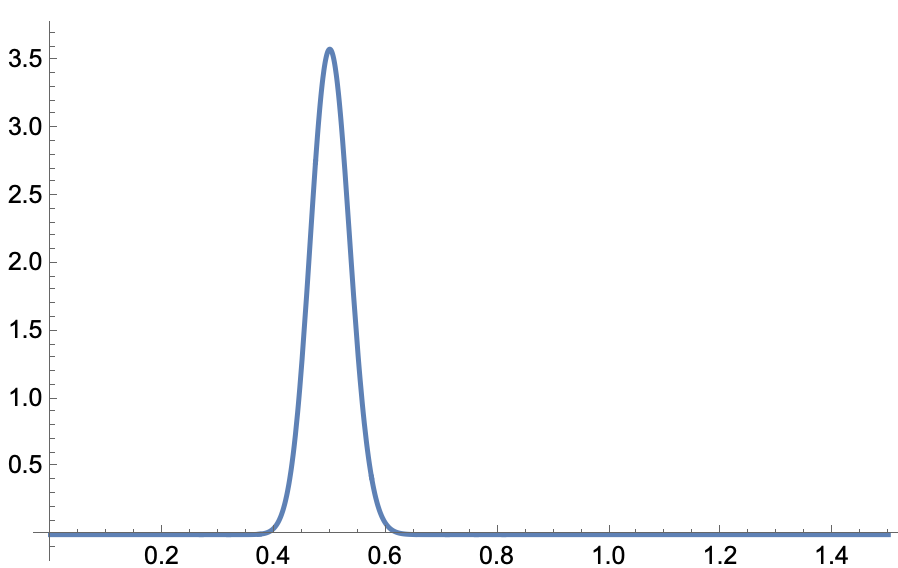}\\
        (a) $\rho_n(r) := \rho_n(x,y)$ for $n=50, \hbar=1/200$, $l_0=1$. Here $r= \sqrt{x^2 + y^2}$. The peak is at $r_{peak} \approx \sqrt{n\hbar}l_0= 0.5$.
    \end{minipage}%
    \begin{minipage}{0.5\textwidth}
        \centering
        \includegraphics[width=0.45\textwidth, height=0.15\textheight]{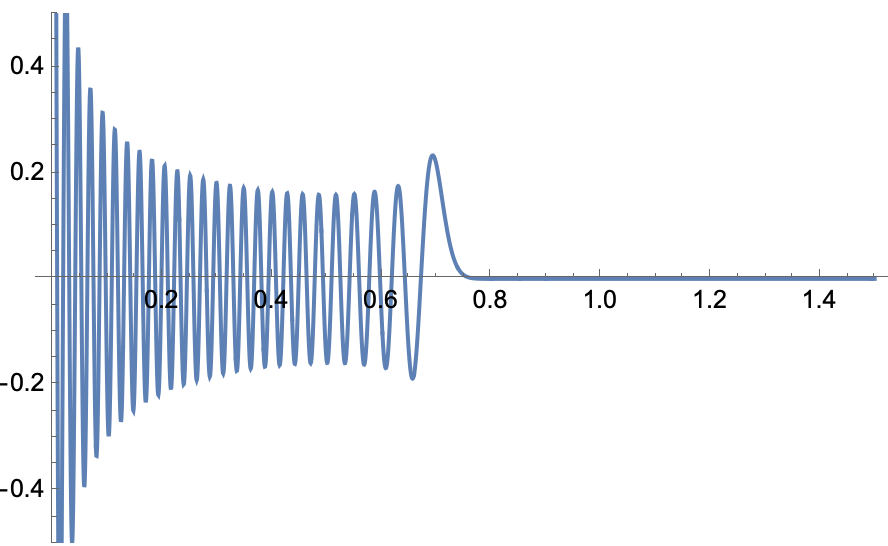}\\
        (b) $u_n(\tilde r) := u_n(x_2,p_2)$ for $n=50, \hbar=1/200$, $l_0=1$. Here $\tilde r= \sqrt{x_2^2/l_0^2 + p_2^2 l_0^2/\hbar^2}$. The dip is at $\tilde r_{dip} \approx \sqrt{2n\hbar}\approx 0.7.$
    \end{minipage}
    \caption{\footnotesize{The wiggles in (b) are suppressed by the integral transform in \eq{un-rhon} to yield (a).}}
    \label{fig:un-rhon}
\end{figure}

\subsubsection{ For a time-dependent state}\label{sec:time-dep}

Note that the general equation \eq{1d-2d-single} can be obtained by taking a linear combination $\sum_n A_n (...)$ on both sides of the eigenfunction-relation \eq{un-rhon} since the kernel $K(x,y,x_2,p_2)$ is independent of the quantum number $n$. A time-dependent LLL wave-function corresponds to making
\begin{align}
  A_n \to A_n(t)= A_n \exp[-i/\hbar E_{0,n}t],\; E_{0,n}= \hbar\om(1+ (1-\nu)n)
  \label{time-dep}
\end{align}
where we have used \eq{trap-lll}. Thus, we find that the relation \eq{1d-2d-single} continues to hold for time-dependent states:
\begin{align}
  \rho(x,y,t)\equiv |\tilde \psi(x,y)|^2 = \int \fr{d x_2\,d p_2}{\pi\hbar} K(x,y,x_2,p_2) u(x_2,p_2,t)
  \label{ut-rhot}
\end{align}

\subsection{The Kernel approximation in  $\hbar\to 0$ limit}

Note that in the limit $\hbar \to 0$, the first factor of \eq{u-4d}, namely \eq{val-u0}, becomes a delta-function
\begin{align}
  \bar u_0(x_1,p_1) \stackrel{\hbar\to 0}{\to} 2\pi\hbar\, \delta(x_1)\delta(p_1)
  \label{u0-delta}
\end{align}
due to the fact that the Gaussian expressions are precisely a representation of delta-functions in the $\hbar \to 0$ limit (see \eq{delta-rep}). As explained in Section \ref{sec:weyl-constraints}, this is one of the ways to understand the LLL constraints $x_1=0=p_1$.

Note that, in the same manner as above, the kernel in \eq{kernel} would represent the following delta-function in the $\hbar \to 0$ limit:
\begin{align}
  K(x,y,x_2,p_2)) \stackrel{\hbar\to 0}{\to} m \omega \delta(x_2- \sqrt 2 x) \delta(p_2 - \sqrt 2 m\om y)
  \label{kernel-delta}
\end{align}
Unfortunately, however, we cannot use the above result in, {\it e.g.} \eq{un-rhon}, since the function $u_n(x_2, p_2)$ depends on $\hbar$ (see \eq{un-x2-p2}) and therefore does not satisfy the criterion of a test function in \eq{delta-meaning}. This is the reason the two functions in Figure \ref{fig:un-rhon} are so different.

We will find in the next section that in the $N$-particle classical limit, the Wigner distribution in the $(x_2,p_2)$ plane becomes {\it independent of $\hbar$} \footnote{More precisely, it has a good $\hbar\to 0$ limit.}  and that enables us to use \eq{kernel-delta} to equate the functional forms of $\rho(x,y)$ and $u(x_2,p_2)$.

\section{1D-2D correspondence for the $N$-particle states}\label{sec:N-particles}

We will now consider putting $N$ fermions, all in the LLL states. In fact, we will consider situations where the energy of the highest occupied state $|0,n_N\ran$ is well short of the first excited level (see comments below \eq{trap-lll}):
\begin{align}
  E_{0,n_N} \ll E_{1,0} \; \Rightarrow n_N \ll \fr{1+\nu}{1-\nu}.
  \label{n-max}
\end{align}
To get going, it is useful to introduce the second quantized fermion fields:
\begin{align}
\Upsilon(x_1, x_2) &= \sum_{n_1,n_2} \chi_{n_1,n_2}(x_1,x_2) c_{n_1,n_2}
\quad
    {\Upsilon}^\dagger (x_1, x_2) = \sum_{n_1, n_2} \chi_{n_1,n_2}(x_1,x_2)
    c^\dagger_{n_1,n_2}
    \label{2nd-x1-x2}\\
\Psi(x, y) &= \sum_{n_1,n_2} \psi_{n_1,n_2}(x,y) c_{n_1,n_2}
\quad
    {\Psi}^\dagger (x, y) = \sum_{n_1, n_2} \psi_{n_1,n_2}(x,y)
    c^\dagger_{n_1,n_2}
    \label{2nd-x-y}    
\end{align}
where we have used the wave-functions \eq{n1-n2-x1-x2} and \eq{n1-n2-xy}. Here $c_{n_1,n_2}, c^\dagger_{n_1,n_2}$ are the annihilation and the creation operators for a fermion at the state $|n_1, n_2\rangle$, i.e., $|n_1, n_2\rangle = c^\dagger_{n_1,n_2}|0\rangle$.

One can also define the corresponding second quantized Wigner distribution operators:
\begin{align}
  {\hat{\tilde U}}(x_1,x_2,p_1,p_2)&=\int d\eta_1 d\eta_2 {\Upsilon}^\dagger (\vec x + \vec \eta/2){\Upsilon} (\vec x - \vec \eta/2)\exp[i/\hbar \vec\eta.\vec p]
  \label{2nd-u-x1x2}\\
  \hat U(x,y,p_x,p_y)&=\int d\eta_x d\eta_y\, {\Psi}^\dagger (x + \eta_x/2,y + \eta_y/2){\Psi} (x -\eta_x/2, y-\eta_y/2)\exp[i/\hbar(\eta_x p_x + \eta_y p_y)]
  \label{2nd-u-xy}
\end{align}
Note that the second quantized fields appropriate to the LLL Hilbert space are
\begin{align}
  \Upsilon(x_1,x_2)&:=  \bar\chi_0(x_1) \Upsilon(x_2),\; \Upsilon(x_2)= \sum_n \chi_n(x_2) c_n \nonumber\\
 {\Upsilon}^\dagger(x_1,x_2)&:=  {\bar\chi_0}^*(x_1) {\Upsilon}^\dagger(x_2),\; \Upsilon^\dagger(x_2)= \sum_n \chi_n^*(x_2) c^\dagger_n
 \label{lll-2nd-x1x2}\\
  \Psi(x,y)&:= \sum_n \psi_n(x,y) c_n, \quad
 {\Psi}^\dagger(x,y):=  \sum_n \psi_n^*(x,y) c^\dagger_n
 \label{lll-2nd-xy}
\end{align}
where $c_n := c_{0,n}, c^\dagger_n := c^\dagger_{0,n}$. The corresponding Wigner distribution operators are
\begin{align}
  {\hat{\tilde U}}(x_1,x_2,p_1,p_2)&= \bar u_0(x_1,p_1) \hat U(x_2,p_2)                 , \;  \hat U(x_2,p_2) 
  =\int d\eta_2 {\Upsilon}^\dagger (x_2 +\eta_2/2){\Upsilon} (x_2 - \eta_2/2)\exp[i/\hbar \vec\eta.\vec p]
  \label{lll-u-x1x2}\\
  \hat U(x,y,p_x,p_y)&=\int d\eta_x d\eta_y\, {\Psi}^\dagger (x + \eta_x/2,y + \eta_y/2){\Psi} (x -\eta_x/2, y-\eta_y/2)\exp[i/\hbar(\eta_x p_x + \eta_y p_y)]
  \label{lll-u-xy}
\end{align}
The Wigner distribution in a particular $N$-fermion LLL state $|F\rangle$ will be
\begin{align}
\tilde U(\vec x, \vec p)=  \langle F| \hat{\tilde U}(\vec x, \vec p)|F\rangle= \bar u_0(x_1,p_1) u(x_2,p_2),\;
  u(x_2,p_2)= \langle F| \hat U(x_2, p_2)|F\rangle
  \label{lll-u-vev}
\end{align}
The second quantized fermion density is, e.g,
\begin{align}
  \hat\rho(x,y)=  \Psi^\dagger(x,y)  \Psi(x,y)
  \label{2nd-fermi-density}
\end{align}
In the following subsections, we will consider the $N$ fermions to be in the following states:
\begin{itemize}
\item A single Slater state: e.g., the $N$ particle ground state, states where the electrons occupy an arbitrary filling (such as band states), and $W_{\infty}$ coherent states. We call them single Slater states (or Slater-determinant states) because the $N$-particle wavefunction $\langle z_1,z_2,..., z_N| F\rangle$ is a determinant known as the Slater determinant.
\item Linear combination of single Slater states ($|F\rangle=\sum_{m} \alpha_m |F_m\rangle, \quad |F_m\rangle $ are general  single Slater states with arbitrary filling)
\item Thermal state
\end{itemize}
In each of these cases, we will show first the exact quantum correspondence and then the semiclassical correspondence defined by the following large $N$ limit 
\begin{align}
  N \to \infty, \; \hbar \to 0, \; \hbox{such that}\; N\hbar =1
  \label{large-N}
\end{align}
It is also assumed that, in this limit $\nu \equiv \Omega/\om$ is kept close enough to 1 so that the LLL condition \eq{n-max} remains satisfied.\\
It is convenient to express the fermion density and the Wigner distribution in radial coordinates as 
\[
x = r \cos \theta, \quad y = r \sin \theta, \quad x_2 \sqrt{m \omega} = \tilde{r} \cos \tilde{\theta}, \quad \frac{p_2}{\sqrt{m \omega}} = \tilde{r} \sin \tilde{\theta}.
\]
We will use this parametrization in the following sections, particularly in the large-$N$ limit.

\subsection{Single Slater state}
\subsubsection{Ground state}\label{sec:ground}

The ground state is defined by
\begin{align}
  |F \rangle= c^\dagger_{N-1}...c^\dagger_1 c^\dagger_0 |0\rangle
  \label{ground}
\end{align}
where we have chosen $N$ very large, and at the same time $(1-\nu)$  so small  that the $N$-th occupied level is still in the LLL (see discussion below \eq{trap-lll}).

It is easy to see that \eq{lll-u-vev} translates to
\begin{align}
  \tilde U(\vec x, \vec p)= \bar u_0(x_1,p_1) \underbar u(x_2,p_2), \quad
  \underbar u(x_2,p_2) = \sum_{n=0}^{N-1} u_n(x_2,p_2)
  \label{ground-u}
\end{align}
where $u_n(x_2,p_2)$ is given by \eq{un-x2-p2}.

In a similar way, one can show that the fermion density \eq{2nd-fermi-density}, evaluated in the ground state also becomes a sum over the occupied states:
\begin{align}
  \rho(x, y)=  \sum_{n=0}^{N-1} \rho_n(x,p)
  \label{ground-rho}
\end{align}
By using the 1d-2d correspondence for single-particle states \eq{1d-2d-single}, and by using the linear sum structure of \eq{ground-u} and \eq{ground-rho}, we get the following 1d-2d correspondence for $N$-particles:
\begin{align}
  \rho(x,y)= \int \fr{dx_2\, dp_2}{2\pi \hbar} K(x,y;x_2,p_2) \underbar u(x_2,p_2)
    \label{1d-2d-N}
\end{align}
In the large $N$ limit, the $N$-particle Wigner distribution \eq{ground-u} involves a sum over a large number of single-particle Wigner distributions \eq{un-x2-p2}. The effect of this sum is that (because of the alternating signs due to the factor $(-1)^n$ in $u_n(x_2,p_2)$) the rapid oscillations of the single-particle distributions destructively interfere, giving an approximate shape of a step function (see Fig \ref{fig:ground-classical} (a)). In fact, as we discuss in Appendix [\ref{rhoproof}],  the limiting value of the ground state Wigner distribution \eq{ground-u} is (see \eq{u-droplet} and \eq{uxp-V})
\begin{align}
  \underbar u(x_2,p_2)= \theta\left(\tilde r_0^2 -\tilde r^2\right), \quad \tilde r^2= \tilde x_2^2  + \tilde p_2^2, \quad \tilde x_2= x_2/l_0,\; \tilde p_2= p_2 l_0, \tilde r_0 \equiv \sqrt{2 N \hbar}=\sqrt 2
  \label{ground-u-classical}
\end{align}
Since in this limit, the $\hbar$-dependence of $\underbar u$ vanishes, we can use \eq{kernel-delta} in \eq{1d-2d-N} (see the discussion around \eq{kernel-delta}), and get
\begin{align}
  \rho(x,y) &= \fr{m\om}{\pi\hbar} \underbar u(x_2,p_2)|_{x_2= \sqrt 2 x, p_2= - \sqrt 2 y/l_0^2}
\label{rho-vs-u-classical}\\
  &= \fr{m\om}{\pi\hbar} \theta(r_0^2- r^2), \quad r_0^2= \fr{\tilde r_0^2}{2m\om}= \fr{N\hbar}{m\om}= \fr1{m\om}.
  \label{ground-rho-classical}
\end{align}
which can be independently verified (see Appendix \ref{rhoproof}). Note that the explicit computation of $\rho(x,y)$ bears this out (see Fig \ref{fig:ground-classical} (b)). 

\begin{figure}[H]
    \centering
    \begin{minipage}{.5\textwidth}
        \centering
        \includegraphics[width=0.48\textwidth, height=0.15\textheight]{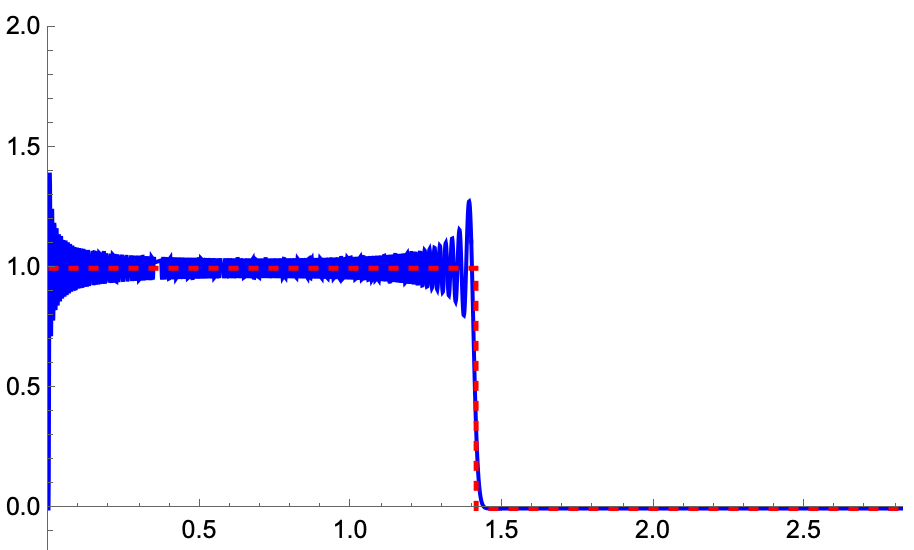}\\
        \vspace{1ex}\vbox{\begin{quote}\baselineskip=9pt{\footnotesize (a) The solid blue curve shows the large $N$ limit of the Wigner distribution.  The red dashed curve shows the theoretical limit \eq{ground-u-classical}.  On the $x$-axis is plotted $\tilde r$. Note the dip of the blue curve at $\tilde r= \sqrt 2$.}\end{quote}}
    \end{minipage}%
    \begin{minipage}{0.5\textwidth}
        \centering
        \includegraphics[width=0.48\textwidth, height=0.15\textheight]{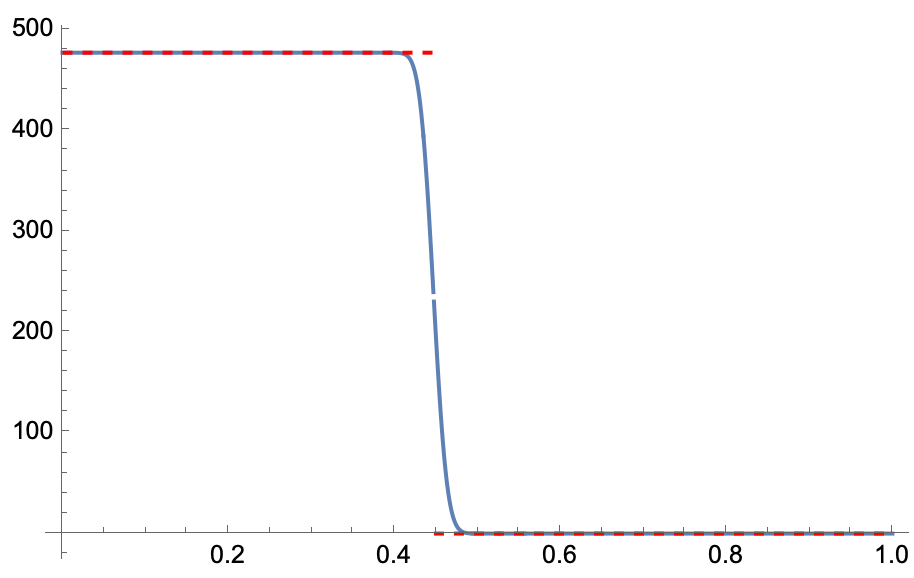}\\
        \kern5pt\vbox{\begin{quote}\baselineskip=9pt {\footnotesize (b) The solid blue curve shows the large $N$ limit of the Fermion density. The red dashed curve shows the theoretical limit \eq{ground-rho-classical}. On the $x$-axis is plotted $r$; note the dip of the blue curve at $r=1/\sqrt 5 \approx .45$, and the height $\approx 480$ which matches the value $\fr{m\om}{\pi\hbar}$.}\end{quote}}
    \end{minipage}
    \caption{\footnotesize{Classical limit of ground state properties. $N=100, \hbar=1/N, m\om=5$.}}
    \label{fig:ground-classical}
\end{figure}

We give below the 3D plots corresponding to the above.

\begin{figure}[H]
    \centering
    \begin{minipage}{.5\textwidth}
        \centering
        \includegraphics[width=0.48\textwidth, height=0.15\textheight]{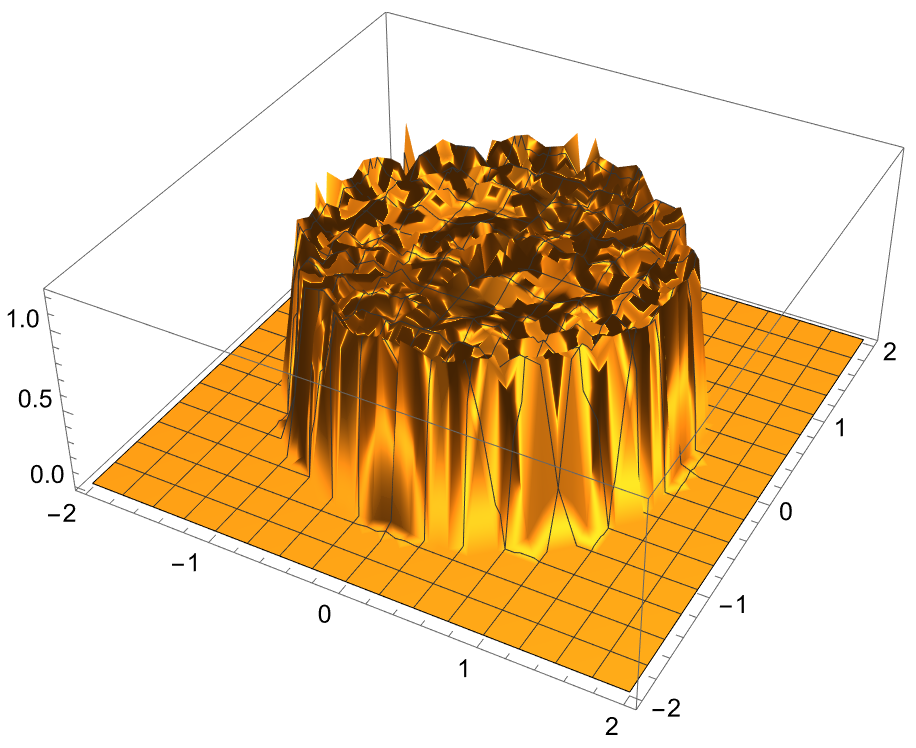}\\
        \vspace{1ex}\vbox{\begin{quote}\baselineskip=9pt{\footnotesize (a) 3D plot of the Wigner distribution. The horizontal axes represent $\tilde x_2, \tilde p_2$.}\end{quote}}
    \end{minipage}%
    \begin{minipage}{0.5\textwidth}
        \centering
        \includegraphics[width=0.48\textwidth, height=0.15\textheight]{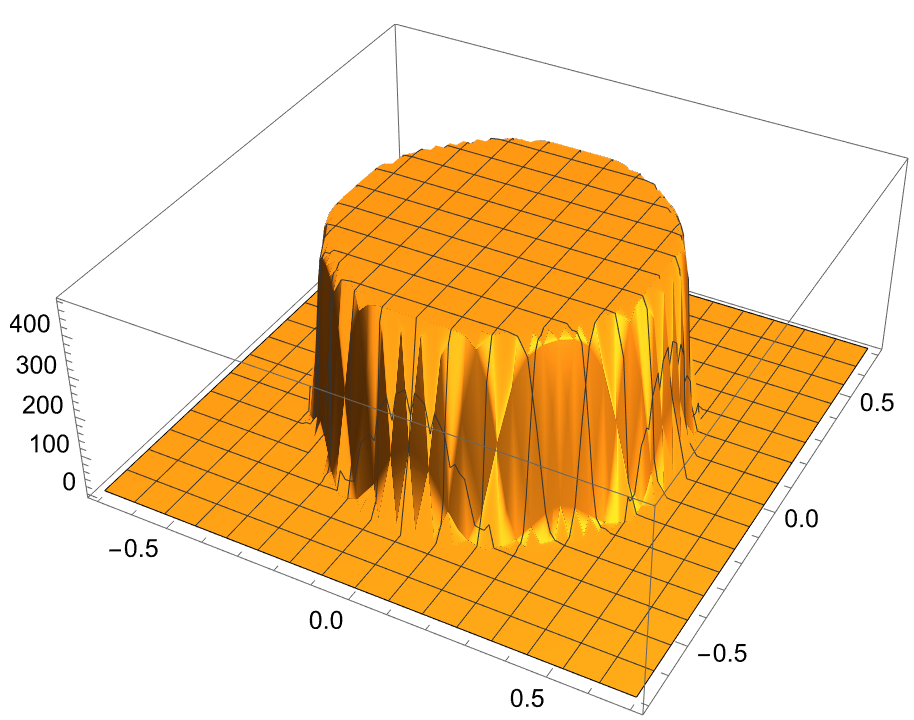}\\
        \kern5pt\vbox{\begin{quote}\baselineskip=9pt {\footnotesize (b) 3D plot of the fermion density. The horizontal axes represent $x,y$.}\end{quote}}
    \end{minipage}
    \caption{\footnotesize{3D plots corresponding to Figure \ref{fig:ground-classical}.}}
    \label{fig:ground-classical-3D}
\end{figure} Note that the fermion density \eq{ground-rho-classical} has the property
\begin{align}
  \rho(x,y) &= \fr1{2\pi\hbar_{\rm eff}},\, \hbar_{\rm eff}= \hbar/(2m\om) \, {\rm if} \, r<r_0 \nonumber\\
  &= 0,\, {\rm if} \, r<r_0
  \label{Pauli}
\end{align}
which reflects the semiclassical version of the Pauli principle that in an area element $\Delta x\,\Delta y= 2\pi \hbar_{\rm eff}$, $\hbar_{\rm eff}= \hbar/(2m\om)$ which is suggested by $\{x,y\}_{DB}$.

\subsubsection{General single Slater states}\label{sec:slater}
We will now consider $N$-particle states obtained by filling a certain orthonormal set of single-particle LLL states $|0,f_1\rangle, |0,f_2\rangle, ..., |0,f_N\rangle$. Here the single particle state $|0,f_i\rangle$ are not necessarily identical to the energy eigenstates $|0,n\rangle$ and more generally of the form $|0,f_i\rangle=\sum_{n=0}^\infty f_{i,n} |0,n\rangle$. By defining annihilation operators $c_{f_i}=\sum_{n=0}^\infty f_{i,n}^* c_{n}$, such a multi-particle state can be represented as
\begin{align}
  |F \rangle= |\{c_{f_{i}}^{\dagger}\}\rangle=c^{\dagger}_{f_N}...c^{\dagger}_{f_2} c^{\dagger}_{f_1} |0\rangle
  \label{slater-state}
\end{align}
where $|0\rangle$ is the zero particle state. 
The single-particle Wigner distribution corresponding to the state $|f_i\rangle$ (omitting the $\bar u_0(x_1,p_1)$ factor) is given by
\begin{align}
  u_{f_i}(x_2,p_2)= \int d\eta_2 f_i^{*}(x_2+\eta_2/2) f_i(x_2-\eta_2/2)\exp[i/\hbar \eta_2 p_2]
  \label{u-f}
\end{align}
 and the $N$-particle Wigner distribution \eq{lll-u-vev} corresponding to $\ket{F}$ looks as
 \begin{align}
\tilde U(\vec x, \vec p)=  \bra{F} \hat{ \tilde U}(\vec x, \vec p)\ket{F}= \bar u_0(x_1,p_1) \underbar u(x_2,p_2), \quad
  \underbar u(x_2,p_2) = \sum_{i=1}^{N} u^{}_{f_i}(x_2,p_2)
  \label{lll-u-slater}
\end{align}
The fermion density is again given by a linear sum
\begin{align}
  \rho(x, y)=  \sum_{i=1}^{N} \rho^{}_{f_i}(x,y),\quad \rho^{}_{f_i}(x,y)= |f_i(x,y)|^2,\quad
  f^{}_i(x,y)= \langle x,y | f^{}_i \rangle
  \label{slater-rho}
\end{align}
Because of the linear structure of \eq{lll-u-slater} and \eq{slater-rho}, the quantum equivalence between the many-particle density and Wigner, \eq{1d-2d-N} still holds and we can write
\begin{align}
  \rho(x,y) &= \fr{m\om}{\pi\hbar} \underbar u(x_2,p_2)|_{x_2= \sqrt 2 x, p_2= - \sqrt 2 y/l_0^2}
\label{rho-vs-u-classical2}\end{align}
in the large $N$ limit.

\subsubsection*{Band states}
The most common example of states with arbitrary filling are the so-called ``band'' states, where the fermions in the ground state leave out and occupy bands of states, (see, e.g. Figure (\ref{fig:band}))
\begin{align}
n_1=N_1, n_2= N_1+2, ... , n_N = N_2 \equiv N_1+N-1
\label{band}
\end{align}
\begin{figure}[H]
  \begin{center}
    \begin{tikzpicture}[scale=0.65]

\draw[->] (0,0) -- (0,5) node[left] {$E$};
\draw[->] (0,0) -- (12,0) node[right] {$n_2$};

\node[left,blue] at (0,0) {$E_0=\hbar \omega$};

\draw[red] (0,0) -- (12,4);
\foreach \i in {4.6,4.8,...,10.2} {
    \pgfmathsetmacro{\y}{1/3*\i} 
    \fill[black] (\i,\y) circle (0.08);
}
\draw[dashed] (4.6,0) -- (4.6,5);
\node[below] at (4.6,0) {$N_{1}$};
\draw[dashed] (10.2,0) -- (10.2,5);
\node[below] at (10.2,0) {$N_{2}$};
\end{tikzpicture}
  \end{center}
  \caption{\footnotesize{A ``band state'' with fermions occupying $N$ levels from $N_1$ to $N_2$.}}
  \label{fig:band}
\end{figure}
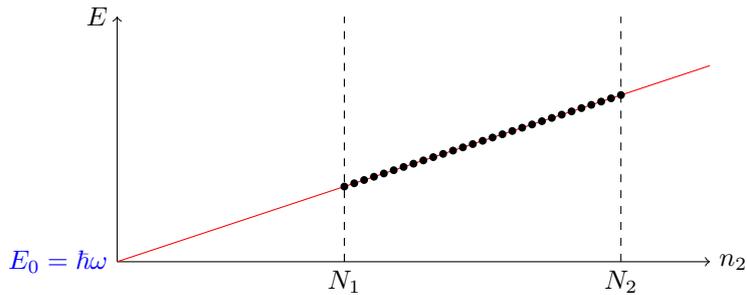

For the band state \eq{band}, where $N_1/N, N_2/N$ are held fixed in the large $N$ limit, both the Wigner distribution and the fermion density become step functions, related  by \eq{rho-vs-u-classical2}. For $N=300, N_1=50, N_2=350$, we obtain the following numerical plots:
\begin{figure}[H]
    \centering
    \begin{minipage}{.5\textwidth}
        \centering
        \includegraphics[width=0.48\textwidth, height=0.15\textheight]{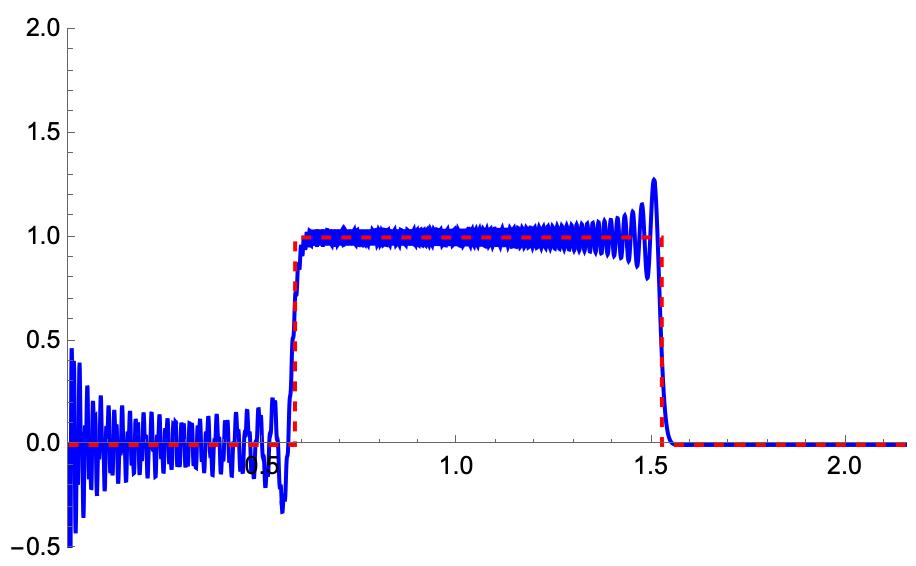}\\
        \vspace{1ex}\vbox{\begin{quote}\baselineskip=9pt{\footnotesize (a) The solid blue curve shows the large $N$ limit of the Wigner distribution.  The red dashed curve shows the theoretical limit \eq{band-u-classical}.  On the $x$-axis is plotted $\tilde r$.}\end{quote}}
    \end{minipage}%
    \begin{minipage}{0.5\textwidth}
        \centering
        \includegraphics[width=0.48\textwidth, height=0.15\textheight]{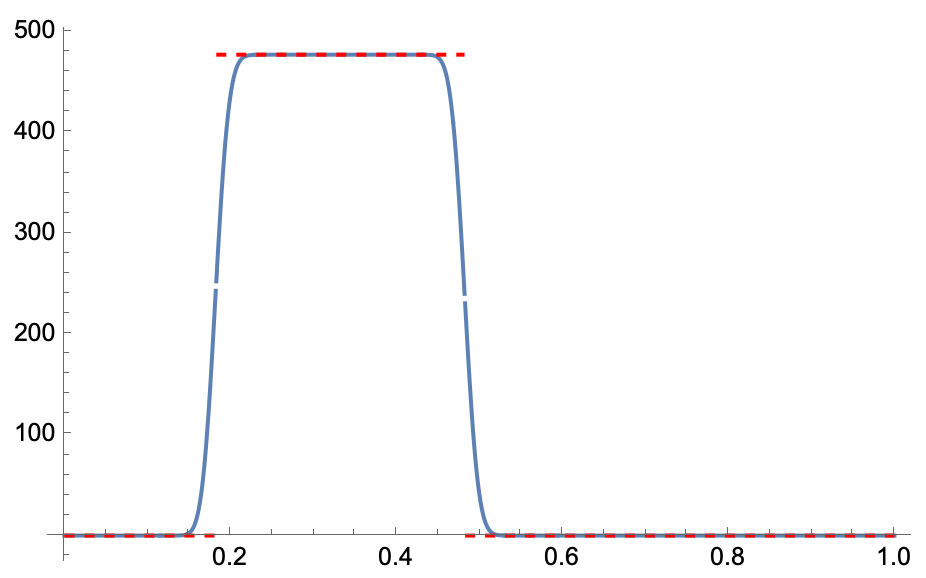}\\
        \kern5pt\vbox{\begin{quote}\baselineskip=9pt {\footnotesize (b) The solid blue curve shows the large $N$ limit of the Fermion density. The red dashed curve shows the theoretical limit \eq{band-rho-classical}. On the $x$-axis is plotted $r$.}\end{quote}}
    \end{minipage}
    \caption{\footnotesize{Classical limit of band states. $N=300, N_1=50, N_2=350, \hbar=1/N, m\om=5$.}}
    \label{fig:band-classical}
\end{figure}
The large $N$ limit is given by the following theoretical curves:
\begin{align}
  u(x_2,p_2) &= \theta(\tilde r_{high} - \tilde r)  \theta( \tilde r-\tilde r_{low} ),\; \tilde r_{high}= \sqrt{2N_2\hbar},\, \tilde r_{low}= \sqrt{2N_1\hbar}
  \label{band-u-classical}\\
  \rho(x,y) &= \fr{m\om}{\pi\hbar}(\theta(r_{high} - r)  \theta(r-r_{low})),\;
  r_{high}= \sqrt{\fr{N_2\hbar}{m\om}},\, r_{low}=\sqrt{ \fr{N_1\hbar}{m\om}}.
  \label{band-rho-classical}
\end{align}
Note that these two functions are related by \eq{rho-vs-u-classical2}.
\subsubsection*{$W_\infty$ coherent states }
For a time-evolving state
\[
|\psi,t\rangle := U(t)|\psi \rangle,\; U(t)= \exp[i/\hbar \hat H t]
\]
the Wigner distribution satisfies \eq{u-eom}:
\begin{align}
\del_t u(x,p,t)= \{u,H\}_{MB}
\label{dynamics}
\end{align}
The finite time $u(x,p,t)$ can be obtained by solving this differential equation.

If we apply the time-evolution operator to an energy eigenstate $|n\rangle$, however, then $u(x,p)$ does not evolve. But, if we instead consider a more general unitary transformation
\begin{align}
|n,s\rangle =U(s)|n\rangle,\; U(s)= \exp[i/\hbar \hat K s]
\label{K-transform}
\end{align}
then the Wigner distribution of $|n,s\rangle$  satisfies
\begin{align}
  \del_s u(x,p,s)= \{u,K\}_{MB}
  \label{K-evolution}
\end{align}
A general unitary transform of the form \eq{K-transform} (equivalently \eq{K-evolution}) is called a $W_\infty$ transformation \cite{Pope:1992ts}.
Here $K(x,p)$ and $\hat K$ are related by the Weyl correspondence (see Appendix \ref{app:weyl-wigner}).  We solve \eq{K-evolution} for some suitable initial condition ($u(x,p,t=0)=u_0(x,p)$) to obtain a finite $W_\infty$ transformation of the initial state. Such $W_\infty$ states can then be made to evolve with our original LLL Hamiltonian and since they are no longer eigenstates, they are now guaranteed to exhibit non-trivial time evolution.

\subsection{Linear combination of single Slater states}
A general multi-particle  state, $|F\rangle_{gen}$, is defined as a linear combination of Slater states:
\begin{align}
    |F\rangle_{gen}=\sum_{m}\alpha_m|F_m\rangle
\label{gen-slate}
\end{align}
where $|F_m\rangle$ is an N-particle Slater state with an arbitrary filling (see (\ref{slater-state})), and the coefficients satisfies the normalization condition $\sum_{m}|\alpha_m|^2=1$. Before proceeding further, it is convenient to evaluate the term
\begin{align}
\langle F_m |\Psi^\dagger(x,y)\Psi(x',y')|F_n\rangle=\sum_{i,j} \langle F_m|c_i^\dagger c_j|F_n\rangle \psi^*_{i}(x,y)\psi_{j}(x',y')
\label{eq:bilinear-gen}\end{align}  
where the matrix elements $\langle F_m|c_i^\dagger c_j|F_n\rangle$ are:
\begin{align}
    \langle F_m|c_i^\dagger c_j|F_n\rangle=\langle 0|c^{}_{f^m_1}c^{}_{f^m_{2}}...c^{}_{f^m_N}(c_i^\dagger c_j)c_{f^n_N}^{\dagger}...c_{f^n_2}^{\dagger} c_{f^n_1}^{\dagger} |0\rangle
\end{align}
The matrix elements are non-zero if and only if \begin{enumerate}[label=(\roman*)]
    \item $\{c^{}_{f^m_i}\}= \{c^{}_{f^n_i}\}$ or
    \item There is only one element different between the two sets: $\{c^{}_{f^m_i}\}$ and $\{c^{}_{f^n_i}\}$
\end{enumerate}
The fermion density of this state $|F\rangle_{gen}$, is given by:
\begin{align}
&\rho_{gen}(x,y)=
 \sum_{m,n}\alpha_m^*\alpha_n \langle F_m |\Psi^\dagger(x,y)\Psi(x,y)|F_n\rangle \nonumber \\
 &= \sum_{m}|\alpha_m|^2\langle F_m |\Psi^\dagger(x,y)\Psi(x,y)|F_m\rangle+   \sum_{m \neq n}\alpha_m^*\alpha_n \langle F_m |\Psi^\dagger(x,y)\Psi(x,y)|F_n\rangle \label{gen-rho}
\end{align}
The N-particle Wigner distribution (from Eq. (\ref{lll-u-vev})) is :
\begin{align}
    \tilde U_{gen}= {}_{gen}\langle F| \hat{\tilde U}(\vec{x},\vec{p})|F\rangle_{gen}=\sum_{m,n}\alpha^*_m \alpha _nU_{mn}=\sum_{m}|\alpha_m|^2 U_{mm}+\sum_{m \neq n}\alpha_{m}^*\alpha_{n}U_{mn}
\label{gen-u}
\end{align}
where
\begin{align}
    U_{mn}=\langle F_{m}| \hat{\tilde U}(\vec x, \vec p)|F_{n}\rangle=\sum_{i,j}\int d \vec{\eta}\exp(-i \vec{\eta}\cdot\vec{p}/\hbar)\psi_i^*\left(\vec{x}-\vec{\eta}/2\right)\psi_j\left(\vec{x}+\vec{\eta}/2\right) \langle f_{m}| c_i^{\dagger}c_j|f_{n}\rangle
\label{def}
\end{align}
Because of the presence of off-diagonal terms, the $1\text{D}$–$2\text{D}$ correspondence between Eqs.~(\ref{gen-rho}) and (\ref{gen-u}) is not straightforward. However, provided the Slater states we consider are predominantly different, which is almost always the case when we consider two different band states, then these cross-terms become zero and don't matter. The more nuanced situation in which the Slater states in question are almost identical (except for the filling of one or two electrons) is discussed in Appendix (\ref{app:lin-comb-slater}).

\subsubsection*{Two Slater states}
Consider the linear combination of the ground state and the band state discussed in \eq{band}:
\begin{align}
\ket{F} = \alpha_1 \ket{F_{\text{ground}}} + \alpha_2 \ket{F_{\text{band}}}
\end{align}
Since the fillings of both these Slaters differ in more than one place, the cross-terms mentioned earlier do not contribute and we are left with 
\begin{align}
\underbar u(x_2,p_2) = |\alpha_1|^2 \underbar u^{\text{ground}}(x_2,p_2) + |\alpha_2|^2 \underbar u^{\text{band}}(x_2,p_2) 
\end{align}
\begin{figure}[H]
  \centering
  \includegraphics[scale=0.34]{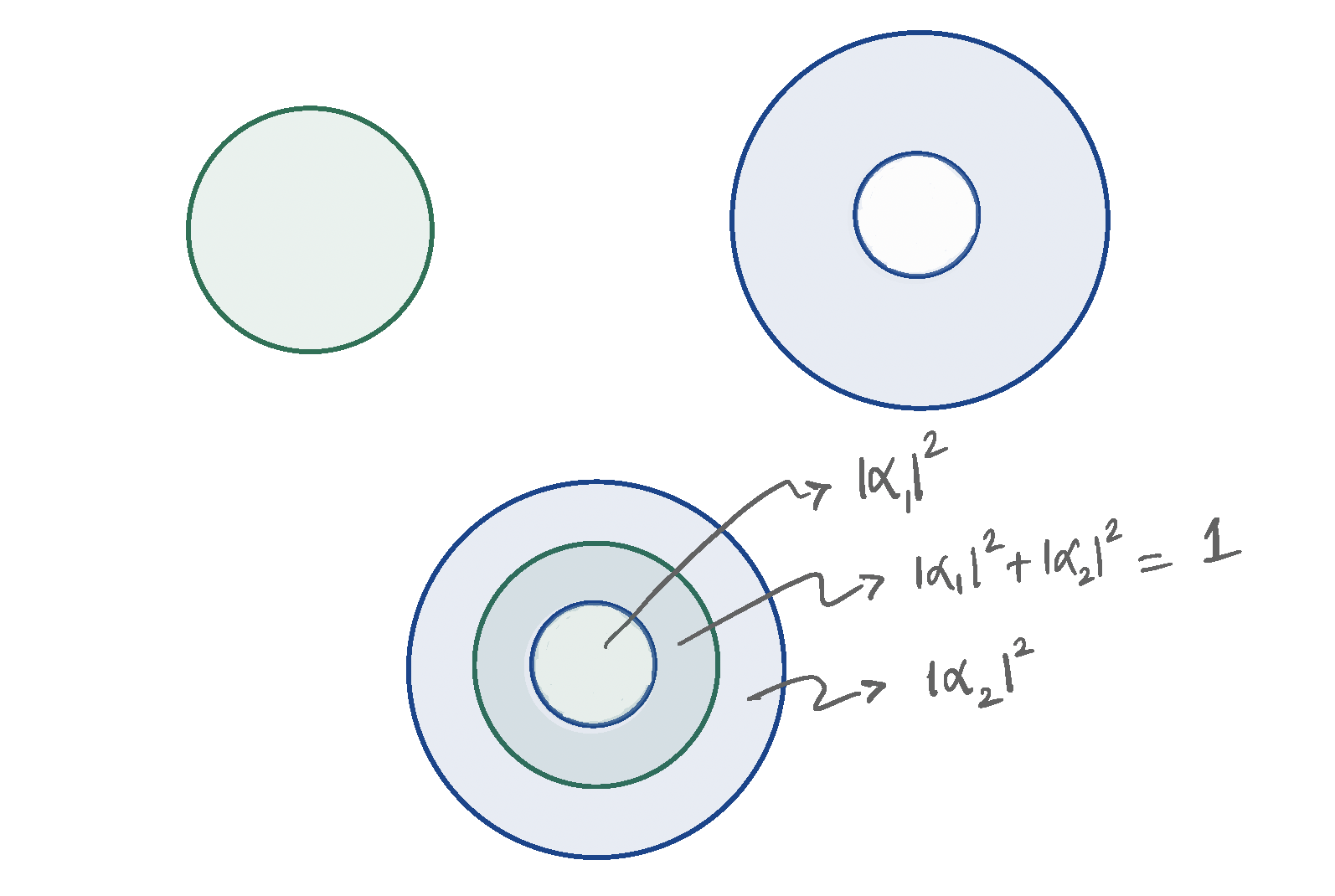}
  \caption{\footnotesize{The green disc and the blue annulus represent the Wigner distributions of the ground state and a band state respectively. In the region of overlap, the net Wigner distribution is $1$, and elsewhere, it is less than $1$.}}
  \label{fig:2-slater}
\end{figure}

\subsubsection*{Three Slater states}
Here we consider the linear combination of three distinct Slater states with insignificant overlap in their fillings.
\begin{align}
\ket{F} = \alpha_1 \ket{F_1} + \alpha_2 \ket{F_2} + \alpha_3 \ket{F_3}
\end{align}
As before, the net Wigner distribution doesn't have cross terms and takes the form
\begin{align}
\underbar u(x_2,p_2) = |\alpha_1|^2 \underbar u^{(1)}(x_2,p_2) + |\alpha_2|^2 \underbar u^{(2)}(x_2,p_2) + |\alpha_3|^2 \underbar u^{(3)}(x_2,p_2)
\end{align}
where $\underbar u^{(1)}$ corresponds to state $\ket{F_1}$ and so on.
\begin{figure}[H]
  \centering
  \includegraphics[scale=0.34]{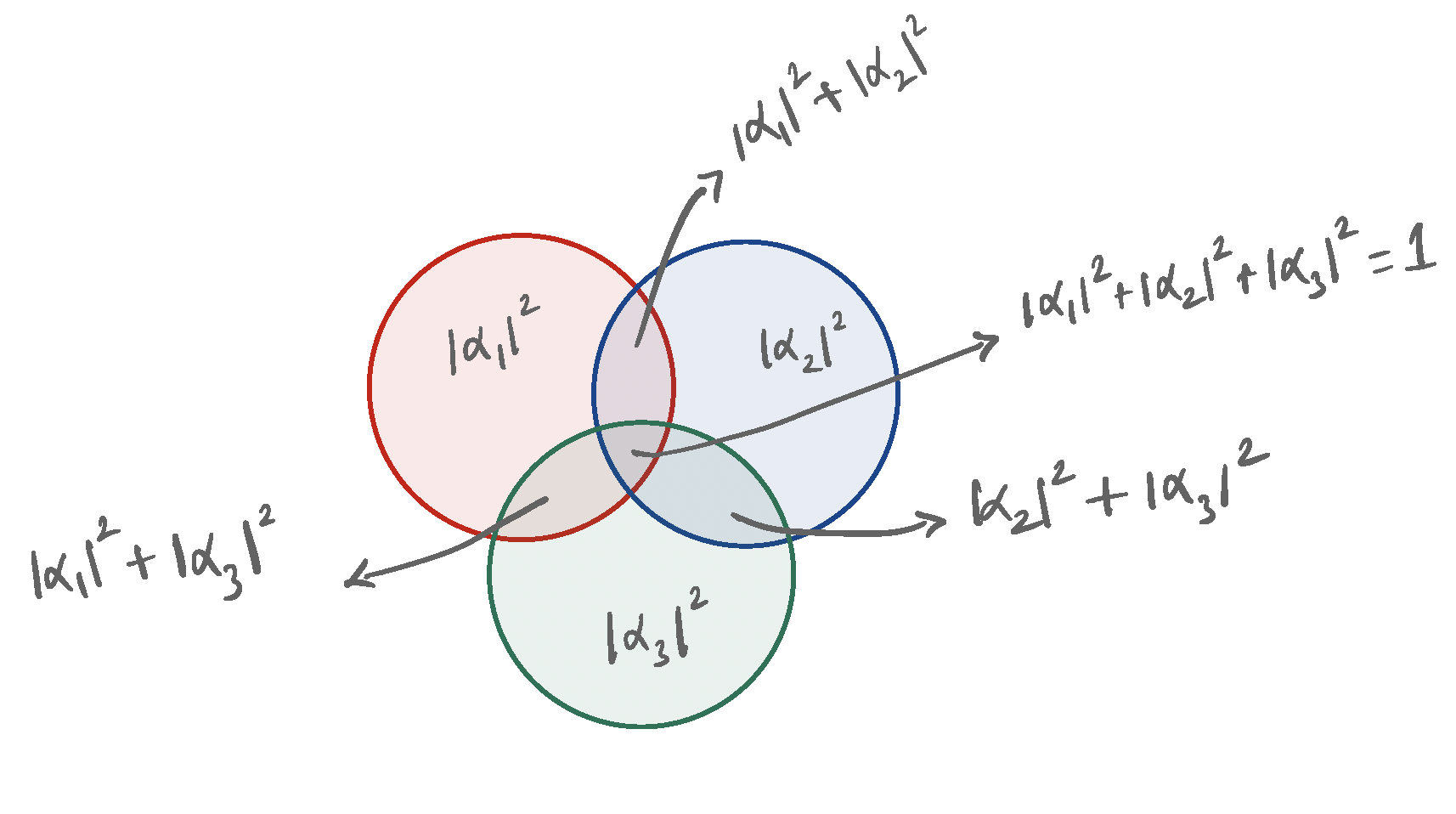}
  \caption{\footnotesize{The red, blue and green discs represent the Wigner distributions of three distinct Slater states. The net Wigner distribution is $1$ in the region where the three circles overlap and it is less than $1$ everywhere else.}}
  \label{fig:3-slater}
\end{figure}

\subsection{Thermal states }\label{sec:rho=u-thermal}

For a thermal mixed state at inverse temperature $\beta$,
\begin{equation}
\hat{\rho}= \frac{e^{-\beta \hat{H}}}{Z}
\end{equation} 
the real space density of electrons is given by 
\begin{align}
\rho_T(x,y) = \langle \psi^{\dagger}(x,y) \psi(x,y) \rangle = \text{Tr}\left( \frac{e^{-\beta \hat{H}}}{Z} \psi^{\dagger}(x) \psi(x) \right)
\end{align}
where $\hat{H} = \sum_{n=0}^{\infty} \epsilon_n \hat{N_n}$. Performing the trace over the occupation number basis $\ket{\{N_n\}}$ and realizing $\bra{\{N_n\}} c_m^{\dagger}c_n \ket{\{N_n\}} = \delta_{nm}  N_m $, we get
\begin{align}
\rho_T(x,y) &= \frac{1}{Z} \sum_{m=0}^{\infty} \sum_{N_0 = 0,1} \sum_{N_1 = 0,1} \cdots  e^{-\beta \sum_n \epsilon_n N_n} N_m \chi_m^*(x,y) \chi_m(x,y) \nonumber \\
&= \sum_{m=0}^{\infty} \langle N_m \rangle |\chi_m(x,y)|^2
\end{align}
where $\langle N_m \rangle$ are given by the Fermi Dirac distribution 
\begin{align}
\langle N_m \rangle = \frac{1}{1 + \exp^{\beta(\epsilon_m - \epsilon_F)}}
\end{align}
For our problem, in the low temperature limit, all the excitations are restricted only to the lowest Landau level. Hence we get
\begin{align}
\rho_T(x,y) = \sum_{l=0}^{\infty} \langle N_{0,l} \rangle | \chi_{0,l}(x,y)|^2 = \sum_{l=0}^{\infty} \langle N_{0,l} \rangle \rho_{0,l}(x,y)
\end{align}
Similarly, we find the Wigner function in this state,
\begin{align}
u_T(x,y,p_x,p_y) &= \int_{\eta_x,\eta_y} \langle \psi^{\dagger} (x + \frac{\eta_x}{2}, y + \frac{\eta_y}{2}) \psi( x - \frac{\eta_x}{2}, y - \frac{\eta_y}{2}) \rangle e^{i p_x \eta_x + i p_y \eta_y}  \nonumber  \\
&= \int_{\eta_x,\eta_y} \sum_{l=0}^{\infty} \langle N_{0,l} \rangle \chi^{*}_{0,l} (x + \frac{\eta_x}{2}, y + \frac{\eta_y}{2}) \chi_{0,l}( x - \frac{\eta_x}{2}, y - \frac{\eta_y}{2}) \rangle e^{i p_x \eta_x + i p_y \eta_y}  \nonumber  \\
&= \sum_{l=0}^{\infty} \langle N_{0,l} \rangle u_{0,l}(x,y,p_x,p_y)
\end{align}
Given that the real space density and the Wigner density can be related for each energy level \eqref{1d-2d-single}
\begin{align}
\rho_{0,l}(x,y) = \frac{1}{ 2\pi \hbar} \int_{x_2,p_2} \frac{1}{\pi \hbar} \exp\left( -\frac{( \sqrt{2} m\omega y + p_2)^2}{\hbar m \omega} \right) \exp\left( - \frac{(\sqrt{2}x - x_2)^2 m \omega}{\hbar} \right)  u_{0,l}(x_2,p_2)
\end{align}
we see that the corresponding relation holds true for the thermal state as well
\begin{align}
\fbox{%
  $\displaystyle
\rho_{T}(x,y) = \frac{1}{ 2\pi \hbar} \int_{x_2,p_2} \frac{1}{\pi \hbar} \exp\left( -\frac{( \sqrt{2} m\omega y + p_2)^2}{\hbar m \omega} \right) \exp\left( - \frac{(\sqrt{2}x - x_2)^2 m \omega}{\hbar} \right)  u_{T}(x_2,p_2)
\label{matching2}$}\end{align}
\begin{figure}[H]
  \centering
  \includegraphics[scale=0.9]{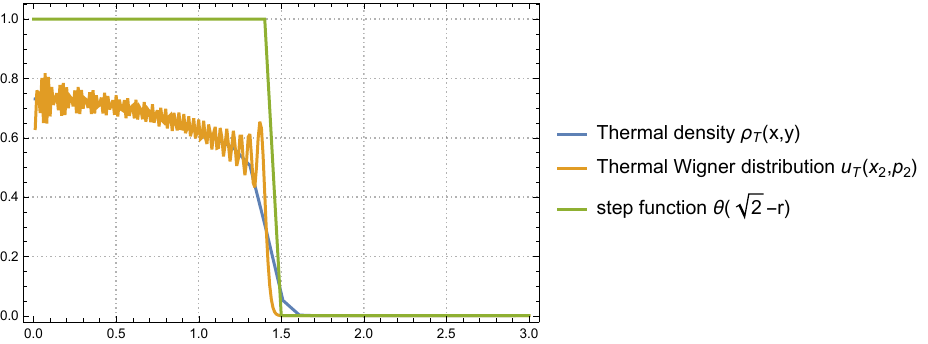}
  \caption{\footnotesize{Classical Limit of thermal states at $\beta=0.1,N=100,\hbar=1/N$.}}
  \label{thermal}
\end{figure}
\subsubsection*{``Diagonal" mixed states}
The above analysis quite easily extends to the class of mixed states we call ``diagonal" mixed states:
\begin{align} \label{eqn:diagonal-mixed}
\rho =  \sum_{\{N_n\}} \lambda_{\{N_n\}} \ket{\{N_n\}} \bra{\{N_n\}}
\end{align}
where $\ket{\{N_n\}}$ are the occupation number basis states. The thermal state is just a special case of \eq{eqn:diagonal-mixed} with 
\begin{align}
\lambda_{\{N_n\}} = \frac{1}{Z} \exp(-\beta \sum_{n=0}^{\infty} \epsilon_n N_n)
\end{align}

\subsection{Dynamics in the classical limit}\label{sec:dynamics}
We construct a time-dependent state by applying a $W_\infty$ transformation \eq{K-transform}, whose classical, differential, version is \eq{K-evolution}. We take $K(\tilde x_2,\tilde p_2)= \fr12(\tilde p^2 - \tilde x^2)$, for which \eq{K-evolution} can be solved:
\begin{align}
  \underbar u(\tilde x_2,\tilde p_2,s)= \underbar u(\tilde x_2 \cosh(s) +\tilde p_2 \sinh(s),\tilde x_2 \sinh(s)+ \tilde p_2 \cosh(s))
  \label{K-finite}
\end{align}
In the plots below we take $s=1$. We start with the above configuration and dynamically evolve it according to \eq{dynamics}. Since the Hamiltonian is that of a harmonic oscillator, the finite-time evolution can be solved:
\begin{align}
  \underbar u(\tilde x_2,\tilde p_2,t)= \underbar u(\tilde x_2 \cos(t) -\tilde p_2 \sin(t),\tilde x_2 \sin(t)+ \tilde p_2 \cos(t),0)
  \label{dynamics-finite}
\end{align}
which is simply a rotation in the phase plane.

The procedure described above can be regarded as a quantum quench, with a pre-quench hamiltonian $K$.

The plots below show the time snapshots of $u$ at times $t=0, \pi/2$. By \eq{rho-vs-u-classical}, the time evolution of the fermion density also behaves in a similar fashion (namely, a rotation in the $x$-$y$ plane).

\begin{figure}[H]
    \centering
    \begin{minipage}{.5\textwidth}
        \centering
        \includegraphics[width=0.48\textwidth, height=0.15\textheight]{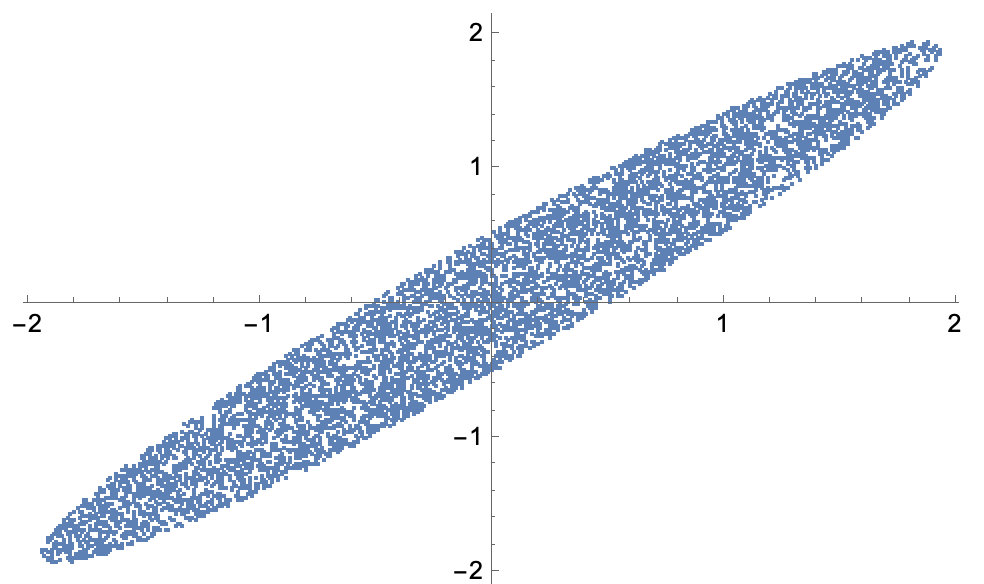}\\
        \vspace{1ex}\vbox{\begin{quote}\baselineskip=9pt{\footnotesize ~~~~~~~~~~~~~~~~~~~~~~~~(a) $t=0$}\end{quote}}
    \end{minipage}%
    \begin{minipage}{0.5\textwidth}
        \centering
        \includegraphics[width=0.48\textwidth, height=0.15\textheight]{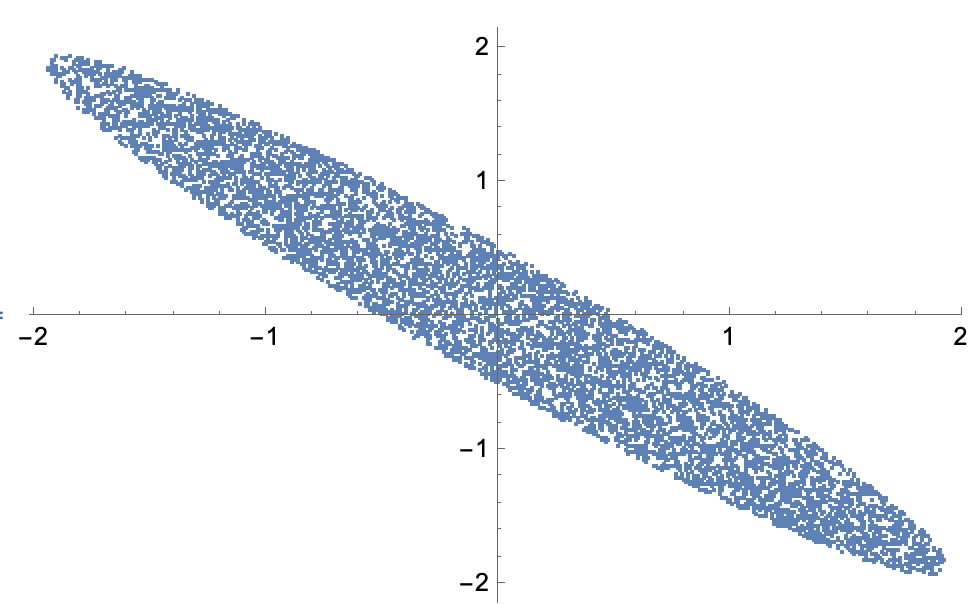}\\
        \kern5pt\vbox{\begin{quote}\baselineskip=9pt {\footnotesize ~~~~~~~~~~~~~~~~~~~~~~~~~(b) $t=\pi/2$.}\end{quote}}
    \end{minipage}
    \caption{\footnotesize{Droplet dynamics in the semiclassical limit (see \eq{dynamics-finite}).}}
    \label{fig:dynamics}
\end{figure}
Since the Hamiltonian dynamics is simply a rotation, starting from any initial configuration, the dynamics is periodic and it cannot show thermalization \cite{Kulkarni:2018ahv}. In \cite{gamma-paper}, we will consider an additional deformation to the dynamics to explore thermalization.\par
The 1D-2D correspondence exists even in the situation where electrons are allowed to fill  higher Landau levels ($n_1=1$).The details of the filling and the explicit calculation of the relation between Wigner distribution and the fermion density is shown in Appendix (\ref{sec:wedding}).

\section{Entanglement Entropy for the generalized lowest landau level}\label{sec:EE}

The ground state entanglement entropy for $N$ fermions in the generalized lowest Landau level (LLL) has already been found in \cite{Lacroix-A-Chez-Toine:2018rwb} using random matrix theory techniques. For a disk in the $(x,y)$ plane of radius $l$, it is shown that the entanglement entropy scales linearly with $r$ as long as $l$ is less than the radius of the Fermi droplet (represented by \eq{ground-u-classical}). Similar results have been obtained in \cite{Tsuchiya_2022} using target space entanglement entropy calculations. The EE has also been calculated in \cite{Das:2022mtb} using the variance method for free fermions. We list below the results obtained in these references.
\begin{itemize}
\item The result from \cite{Lacroix-A-Chez-Toine:2018rwb} is $S = \frac{\alpha_1}{\sqrt{\pi}} \frac{l}{l_0} = 1.80639\fr{l}{l_0}$, where $\alpha_1 = 3.20175$.
\item The result from \cite{Tsuchiya_2022} is $S = 1.81 \frac{l}{l_0}$.
\item The result from  \cite{Das:2022mtb} is $S = \frac{\pi \sqrt{\pi}}{3}\fr{l}{l_0} \approx 1.856 \fr{l}{l_0}$. 
\end{itemize}
We present a calculation below for the EE using the variance method, in a manner similar to but different in some important aspects from \cite{Das:2022mtb}\footnote{In \cite{Das:2022mtb} the LLL band is completely filled, which trivially implies a mass gap in the theory; we consider here an incompletely filled LLL band-- however, we will find that the correlations are nevertheless short-ranged because of the noncommutative structure of space.}. Our result is presented below in Section [\ref{sec:EE-LLL}] and Appendix [\ref{entropy}] and is given by (approximately)
\begin{align}
  S = 1.862 \frac{l}{l_0}.
  \label{ee-lll-value}
\end{align}
The result is good agreement with \cite{Das:2022mtb}. Details of our computation shed light on how the noncommutative nature of the $(x,y)$ plane affects the EE as compared to ordinary 2D fermion systems with a Fermi surface. We explain this below.\footnote{We thank Deepak Dhar and Nikita Nekrasov for important discussions on this issue.}

In a $d$-dimensional free fermion theory the entanglement entropy of the subregion $A \in {\bf R}^2$ is well approximated (in large $N$ limit) by the particle number variance \cite{PhysRevLett.102.100502, PhysRevB.83.161408, PhysRevLett.107.020601}:
\begin{align}
  S_A^{leading} = \frac{\pi^2}{3} \left( \bra{\Omega}\hat N_A^2 \ket{\Omega} - (\bra{\Omega} \hat N_A \ket{\Omega})^2 \right) \qquad \hat N_A = \int_A d^d \vec x\; \Psi^{\dagger}(\vec x) \Psi(\vec x), 
  \label{sa-variance}
\end{align}
where $\ket{\Omega}$ is the $N$-fermion ground state. By using simple manipulations using fermion anticommutation relations, this can be re-expressed as \cite{Das:2022mtb}
\begin{align}
  S_A^{leading} &=\frac{\pi^2}{3}(S_1 - S_2), \nonumber\\
  S_1 &= \bra{\Omega}\hat N_A\ket{\Omega}, \quad S_2= \int_A d^d\vec x \int_A d^2 {\vec x}' |C(\vec x, {\vec x}')|^2,
  \label{var-ee}
\end{align}
where we have introduced the 2-point correlator
\begin{align}
  C(\vec x, {\vec x}')& \equiv \bra{\Omega}{\Psi}^{\dagger}(\vec x){\Psi}({\vec x}')\ket{\Omega} = \int \fr{d^d\vec p}{(2\pi \hbar)^d} \exp[-i\vec p.(\vec x - {\vec x}')/\hbar] u(\fr{\vec x + {\vec x}'}{2}, \vec p) \nonumber\\
 u(\vec x, \vec p) &\equiv \bra{\Omega} \hat U(\fr{\vec x + {\vec x}'}{2}, \vec p)\ket{\Omega}
  \label{c-x-y}
\end{align}
In the second step we have used the relation \eq{u-xp-qft-d} between the Wigner distribution and fermion bilinears.

The EE clearly depends in an important way on the two-point correlator $C(\vec x, {\vec x}')$. Below we will point out how the range of the correlator is qualitatively different between ordinary Fermion systems and LLL fermions, and the consequent effect on the EE. 

\subsection{Review of ordinary 2D fermion systems with a Fermi surface}

In an ordinary 2D fermion system with a Fermi surface, the ground state EE for a disk of radius $l$ behaves as (in the $\hbar \to 0$ limit)
\begin{align}
  S_A \propto \fr{p_F l}{\hbar} \log(\fr{p_F l}{\hbar})
  \label{2D-EE}
\end{align}
A similar logarithm is observed in the 1D case as well (see \eq{1d-EE-log}.

\begin{figure}[H]
    \centering
    \begin{minipage}{0.5\textwidth}
        \centering
        \includegraphics[scale=.5]{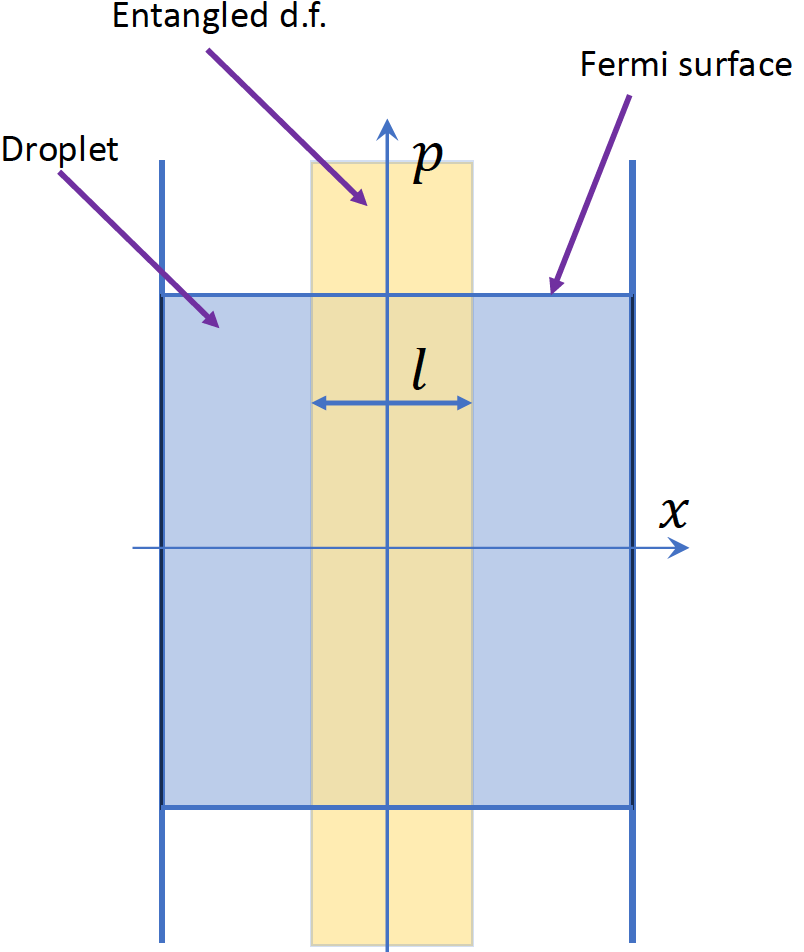}
        
        (a) EE for ordinary fermions
    \end{minipage}%
    \begin{minipage}{.5\textwidth}
      \centering
      \includegraphics[scale=.5]{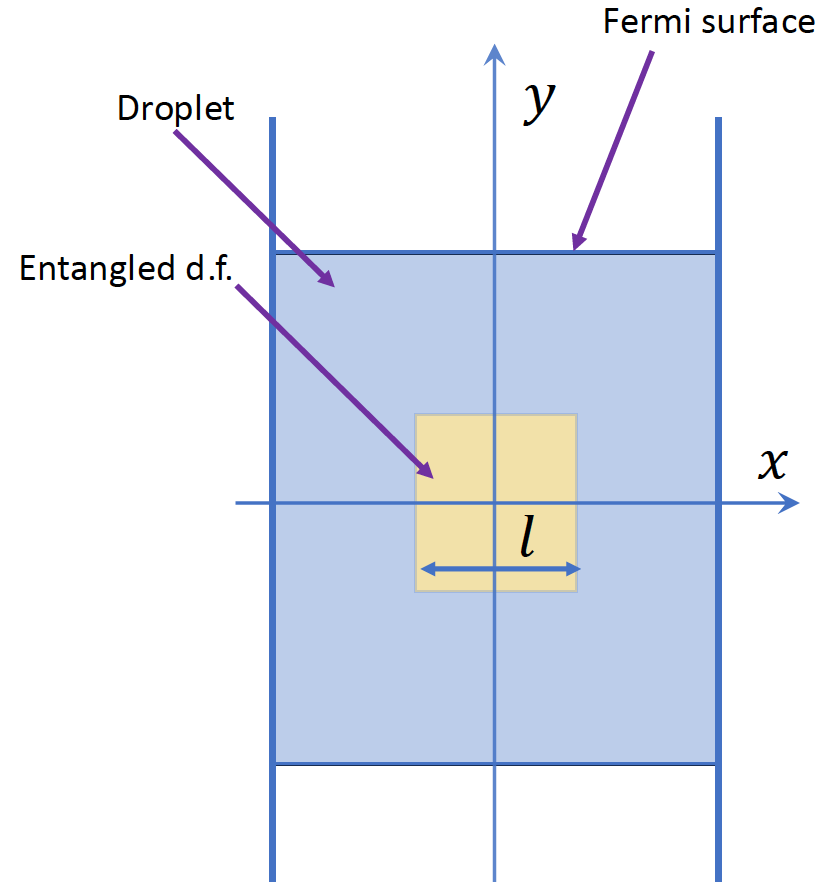}
      
        (b) EE in a noncommutative space
    \end{minipage}    
\caption{\footnotesize{(a) In ordinary fermion systems (taken to be 1D fermions in the figure), the set of entangled {\it phase space} degrees of freedom (d.f.) --- depicted in yellow ---  crosses the Fermi surface at the boundary of the droplet (depicted in blue) where the fermions are. The momentum integrals reach the Fermi surface and lead to a long range two-point correlator $C(x,x')$ whichcan be attributed to massless bosonic fluctuations (sound wave) at the Fermi surface. The long-range correlator leads to a lorarithmic dependence of the EE $\propto \log(p_F l/\hbar)$ which can be identified with EE of a massless boson. For an insight into the calculation, see \eq{logPfL} and remarks around it. In 2D, we get EE $\propto  (p_Fl/\hbar)\ log(p_F l/\hbar)$. (b) In LLL, $y$ is to be identified as the momentum $p$ conjugate to $x$. Hence the specification of an entangling region in {\it space} $(x,y)$ amounts to specifying the entire set of entangled {\it phase space} d.f.; these are marked in yellow and clearly do not see the Fermi surface (in blue). An alternative way to see this is that the noncommutative nature of the $(x,y)$ leads to localized correlators (scale of localization given by the scale of non-commutativity (in spite of of gapless fluctuations near the Fermi surface). The net result is the absence of logarithms much like in a gapped system. For the explicit bevaviour EE $\propto l/l_0$, see the text.}}
    \label{fig:ee}
\end{figure}

The logarithm is contributed by the massless (bosonic) excitations at the Fermi surface. In terms of the specific computations in Appendix \ref{app:fermi-surface} it is straightforward to see that the log comes from the Fermi surface. E.g. for 1D the momentum integration till the Fermi surface leads to the \underbar{long-range correlation $C(x,x')$} \underbar{$ \propto \sin(p_F (x-x')/\hbar)/(x-x')$} in \eq{c-x-y-1d}. This leads to the integral \eq{1D-log-a}, which for large $p_F l/\hbar$, has a logarithmic divergence coming from 
\begin{align}
\int^{p_Fl/\hbar} dt\sin^2(2t)/t = \fr12 \log(p_F l/\hbar), t= p_F (x-x')/\hbar
\label{logPfL}
\end{align}
as explained in \eq{s2-t}). In 2D (see Appendix \ref{app:2d-fermi}), the two-point correlator $C(\vec x, {\vec x}')$, is again long-range, leading to an EE  $\propto t \log(t)$, $t= p_F l/\hbar$.

Scaling analysis: the free fermi systems in the presence of a Fermi momentum $p_F$ has three important dimensionful parameters: $p_F, l$. Hence the EE must be of the form $S_A= f(p_F, l)= f(p_F l)= f(p_F l/\hbar)$.\footnote{In the last step we have reinstated $\hbar$.} Another way to discuss the scaling is to recognize that there are two independent dimensionless parameters in the theory: $N, l/L$.The EE turns out to be of the form $S_A= S_A(N, l/L)= S_A(Nl/L)$. Note that $p_Fl/\hbar= \pi Nl/L$. The specific functional form $f(t) \sim t^{d-1} \log(t)$, $t= p_F l/\hbar$ requires more detailed calculation, but the origin of the logarithm is from the long range correlation as explained above.\\

For non-interacting fermions in $d$-dimensions in an external potential, assumed to be appropriately slowly varying inside the entangling region $A$, it is not difficult to generalize the above argument to arrive at the qualitative form of the EE: $S_A \propto f(t) \sim t^{d-1} \log(t)$, with $t= \overline{p_F}l/\hbar$, where $\overline{p_F}$ represents the average WKB fermi momentum in the entangling region $A$. Once again, the logarithmic behaviour arises from the long-range correlation which can ultimately be traced to the gapless fluctuations at the Fermi surface.

\subsection{Entanglement entropy calculation of the LLL system}\label{sec:EE-LLL}

\paragraph{EE in the noncommutative LLL plane: qualitative remarks}

We emphasized above that for ordinary free fermions, the two-point correlator is long-range, reflecting gapless excitations at the Fermi surface, which eventually gave rise to a logarithmic divergence in the limit of large $p_F l/\hbar$ (see, e.g. \eq{logPfL}). In case of the LLL fermions, the wavefunctions are localized, with \underbar{localization length determined by the} \underbar{scale of non-commutativity $l_0$ (see \eq{x-y-comm})}. The localized wavefunctions eventually lead to a short-range correlator $ \propto \exp[- (\vec x - {\vec x}')^2/(2l_o^2)]$ (see \eq{cxy-lll}). The Fermi surface is not even visible to the entangled degrees of freedom (see remarks below \eq{cxy-lll}). The logarithms which bore the sign of the Fermi surface, are now absent and the EE is simply linear in $l$.

We would like to emphasize the main point again: although we have a partially filled energy band (the LLL band), filled up to a Fermi level, \underbar{and we have essentially gapless fluctuations above the Fermi level}\footnote{The energy gap above the Fermi level is $\hbar \om (1- \nu)$ (see \eq{trap-eigen}) which can be tuned to arbitrarily small values by taking $\nu=\Omega/\om \to 1$).}, we do not have delocalized wavefunctions unlike in ordinary fermion systems with a Fermi surface; the noncommutativity of the $x$-$y$ plane enforces a localization of the wavefunction as explained in the previous paragraph (which is more akin to a gapped system than gapless).

Scaling analysis: The entanglement entropy $S_A$ is a function of $p_F, r, l_0, \hbar$ where we have a new dimensionful parameter $l_0$ which sets the scale of noncommutativity between $x$ and $y$. Thus $S_A= f(p_F r/\hbar, r/l_0)$. As explained above, since the Fermi surface is not seen by the EE, we have $S_A= f(r/l_0)$, where by explicit calculation $f(x) \propto x$.

We now present the explicit calculation of the  EE of the LLL system.

\paragraph{Explicit calculations}

We consider an entangling region $A$ of the shape of a disk of radius $l$ in the $(x,y)$ plane. We will use the equation \eq{var-ee} and \eq{c-x-y}. From \eq{ground-u} and \eq{ground-u-classical}, and noting that $u(x,y,p_x,p_y)= u(x_1,x_2,p_1,p_2)$, we get the following large $N$ formula for the ground state Wigner distribution (for convenience we renamed the radius of the droplet $r_0$ as $R$ in this section):
\begin{align}
  u(x,y,p_x,p_y)
  & =2  \exp{-\left(\frac{\left(p_x-m\omega y\right)^2+\left(p_y + m \omega x\right)^2}{2 m \omega \hbar}\right)} \Theta\left(\bar R^2-\frac{\left(p_x+m\omega y\right)^2+\left(p_y- m \omega x\right)^2}{2 m \omega \hbar}\right) \nonumber\\
  & = 2  \exp{-\left(\frac{\left(\bar p_x- \bar y\right)^2+\left(\bar p_y +\bar x\right)^2}{2}\right)} \Theta\left(\bar R^2-\frac{\left(\bar p_x+ \bar y\right)^2+\left(\bar p_y- \bar x\right)^2}{2}\right), \nonumber\\
  & \bar p_{x,y} = p_{x,y}l_0/\hbar, \bar x= x/l_0, \bar y= y/l_0 
\label{u-x-y-px-py}
\end{align}
Here $\bar R= \sqrt{2N}$ denotes the size of the Fermi fluid droplet (in units of $l_0$) (see \eq{ground-u-classical}).  
Let us now compute $C(\vec x, {\vec x}')$, $\vec x= (x,y), \; {\vec x}'=(x',y')$, using \eq{c-x-y}. 
\begin{align}
 & C(\vec x, {\vec x}')  = \int \fr{dp_x dp_y}{(2\pi\hbar)^2} \, \exp\left(\frac{-i}{\hbar}\left(2 p_x x_- +2 p_y y_-\right)\right) u\left(x_+ ,y_+,p_x,p_y\right), \; x_\pm = \frac{x \pm x'}{2},\ y_\pm = \frac{y\pm y'}{2}
  \nonumber\\
  &= 2\int_{{\bf R}^2} \fr{dp_x dp_y}{(2\pi\hbar)^2} \,\exp\left(\frac{-2i}{\hbar}\left(p_x x_- +p_y y_-\right)\right) \exp{-\left(\frac{\left(p_x-m\omega y_+ \right)^2+\left(p_y + m \omega x_+ \right)^2}{2 m \omega \hbar}\right)} 
  \nonumber\\
  &= 2\int_{{\bf R}^2} \fr{d\tilde p_x d\tilde p_y}{(2\pi\hbar)^2} \, \exp\left(\frac{-2i}{\hbar}\left( \tilde p_x x_- +\tilde p_y y_-\right)\right) \, \exp{-\left(\frac{{\tilde p_x}^2+{\tilde p_y}^2}{2 m \omega \hbar}\right)}, \; \tilde p_x =p_x-m\omega y_+, \tilde p_y =  p_y + m \omega x_+
  \nonumber\\
  &= \fr1{\pi l_o^2} \exp[- (\vec x - {\vec x}')^2/(2l_o^2)],\quad \vec x=(x,y),\; {\vec x}'= (x',y')
  \label{cxy-lll}
\end{align}
In the 2nd step, we have ignored the $\Theta$-function, since the Gaussians ensure that the momenta $p_x, p_y$ are well inside the droplet boundary. In terms of the dimensionless variables introduced above, $\bar p_x$ is centred around $\bar y$ with a half-width of 1 (similarly $\bar p_y$ is centred around $-\bar x$ with a half-width of 1). Since we are interested in $(x,y) \in A$ which is radius $l$, $\bar x, \bar y$ do not exceed $l/l_0$. Hence $\bar p_x, \bar p_y$approximately do not exceed $l/l_0 + 1$, which is far less than the dimensionless size $R= \sqrt{2N}$ of the Fermi fluid droplet. In approximating the $\Theta$ function by 1, we are making only exponentially small errors of order $O(\exp[-\fr{R^2}{l^2}]) \sim O(\exp[-N])$. In the 3rd step we have used the fact that $p_x x_- + p_y y_- = \tilde p_x x_- +\tilde p_y y_-$. The last step follows from trivial Gaussian integration.

\noindent It follows trivially (using $\lan \Omega| \Psi^\dagger (\vec x) \Psi(\vec x)| \Omega \ran$ $= C(\vec x, \vec x)=\fr1{\pi l_o^2}$) that
\begin{align}
  S_1 \equiv \int_A d^2 \vec x \; C(\vec x, \vec x)=\fr{\rm Area~of~A}{\pi l_0^2}= \fr{\pi l^2}{\pi l_0^2}= \fr{l^2}{l_0^2}
  \label{s1-lll}
\end{align}
Using \eq{var-ee} and the expression above for $C(\vec x, {\vec x}')$ in \eq{cxy-lll} we get
\begin{align}
S_2=
& \left(\frac{1}{\pi l_0^2}\right)^2 \int_A dx dy \int_{A} dx' dy' \exp\left[- \fr{(x-x')^2+(y-y')^2}{l_0^2}\right]
 \label{lll-s2-a}
\end{align}
It is tempting to use the delta-function approximation for the Gaussian factor:
\begin{align}
\exp\left[- \fr{(x-x')^2+(y-y')^2}{l_0^2}\right] \to
\pi l_0^2 \delta(x-x') \delta(y-y')
\label{delta-approx}
\end{align}
which would lead to $S_2= \fr{\pi l^2}{\pi l_0^2}= S_1$, implying $S_A=0$! This, of course, proves that the delta-function approximation is too crude. We analyze the issue in a simpler situation below.

\centerline{-----------}

\subsubsection*{Detour: entangling region $A$ of the shape of a square}

Before \eq{lll-s2-a} quantitatively, it is instructive to consider a simpler situation where the entangling region $A$ of the shape of a square of area $l^2$, centred around the origin. In this case, the $x,x'$ and $y,y'$ integrals get separated, and we have 
\begin{align}
  S_2 &= I^2, \quad I= \fr1{\pi l_0^2}\int_{-l/2}^{l/2} dx \int_{-l/2}^{l/2} dx' \exp[-(x-x')^2/l_0^2] \nonumber\\
  &= \fr1{\pi l_0^2}\int_{-l/2}^{l/2}2 dx_-\ (l- 2|x_-|)  \exp[-4x_-^2/l_0^2], \quad x_\pm= \fr{x\pm x'}{2}
  \label{s2-lll-b}.
\end{align}
Here the linear factor $(l- 2|x_-|)$ comes from integrating over $x_+$. Performing the simple Gaussian integral, we get
\begin{align}
  I &= \fr{l}{\sqrt{\pi} l_0} - \fr1{\pi} \nonumber\\
  S_2 &= I^2= \fr{l^2}{\pi l_0^2} - \fr{2}{\pi\sqrt\pi}\fr{l}{l_0} + \fr1{\pi^2}
\label{s2-lll-square}
\end{align}
The first term in $I$ (and hence $S_2$) comes from the $l$ term of $(l- 2|x_-|)$ where the delta-function approximation \eq{delta-approx} is fine. This, expectedly, cancels against $S_1$, which, for the square, becomes $S_1= \fr{l^2}{\pi l_0^2}$ (see \eq{s1-lll}). The second term in $I$ (equivalently $S_2$) comes from the $|x_-|$ in \eq{s2-lll-b}, which is unbounded and invalidates the delta-function approximation.

We get, for the square-shaped entangling region:
\begin{align}
  S_A = \fr{\pi^2}{3}\left(S_1 - S_2\right)= \fr{2\sqrt\pi}{3}\fr{l}{l_0} - \fr13
  \label{ee-square}
\end{align}

\centerline{-----------}

Coming back to the disk problem, we therefore need to compute \eq{lll-s2-a} more accurately. This is done in Appendix \ref{app:x-y}, where we arrive at an analytic expression in terms of an infinite sum \eq{s2-lll-num}. The resulting plot of $S_A$ is presented in Figure \ref{eeboth}. A numerical estimate of $S_A$ is given in \eq{ee-fit}:
\begin{align}
  S_A = 1.86 \fr{l}{l_0}
  \label{ee-fit-txt}
\end{align}

\section{Concluding remarks}\label{sec:conclusion}
The physics of the Landau levels is closely tied to the physics of the integer quantum Hall effect (see, e.g. \cite{tong} for a review). The calculation of the EE and the quench dynamics for the system considered in this paper probably simply translates to the quantum Hall systems. It is worth also exploring the relation to the fractional QHE, where we view the FQHE as the IQHE of composite fermions.\footnote{This possibility was mentioned to us by Jainendra Jain.} 

In this paper, we considered fermions on a plane in a magnetic field and consequences of the non-commutative structure arising from it. Fermions in a magnetic field on other topologies such as the 2-sphere have also been considered in a number of papers (see, e.g. \cite{PhysRevB.80.153303, Murugan:2018hsd, Karabali:2020zap, Nair:2022xxp}).\footnote{Ref. \cite{Karabali:2020zap} also discusses $CP^k$ for $k>1$.} For a disk-shaped entangling region ${\cal D} \in S^2$, it has been reported in \cite{PhysRevB.80.153303, Karabali:2020zap} that the EE is proportional to the perimeter $L$ of the disk. Ref. \cite{PhysRevB.80.153303} shows that the two-point correlation function is short-ranged. As we emphasized in this paper, this is the essential reason for the perimeter law. It is important to mention that Ref. \cite{Nair:2022xxp} describes a fuzzy sphere {\it ab initio}, without prior connection to magnetic fields, and shows behaviour similar to the above references. This indicates that the behaviour of the EE and the fermion density is intrinsically tied to the non-commutative nature of space rather than to specifics of the LLL system.\footnote{The importance of this issue was emphasized to us by Deepak Dhar.} 

Another important area tied to LLL physics is that of half-BPS giant gravitons, studied in great detail by Lin, Lunin and Maldacena \cite{Lin:2004nb}. The LLL conditions arise here as a consequence of the supersymmetry conditions, thus giving rise to the slogan LLM = LLL. A geometric quantization of LLM geometries, both at the level of probe giant gravitons as well as at the level of back-reacted geometries, is carried out in \cite{Mandal:2005wv} in the spirit of our current LLL paper. See also \cite{Mosaffa:2006qk} which proposes a duality between LLM geometries and matrix Chern-Simons theory.

It is interesting to note the relation of the LLL physics to other systems such as the 2D hydrogen atom found by the use of $so(2,3)$ symmetry  \cite{Dereli:2024mxr}. Semiclassical limits of interacting fermions can be obtained by generalizing the phase space techniques of \cite{Das:1991uta, Das:1991ba, Dhar:1992rs, Dhar:1992hr} to include current-current interactions as in \cite{Khveshchenko:1993ug, Dhar:1994ib}.

\paragraph{Generalization and thermalization}

In the present paper, we presented a brief description of dynamics in Section \eqref{sec:dynamics}. The time evolution was given by uniform rigid rotation  of the fermi fluid (see \eq{dynamics-finite}) and was consequently periodic. The uniform rotation is characteristic of a harmonic oscillator hamiltonian. In \cite{gamma-paper}, we will present more general hamiltonians which entail nontrivial dynamics including thermalization.

\subsection*{Acknowledgment}
We deeply appreciate discussions with Subhro Bhattacharjee, Sumit Das, Deepak Dhar,  Emanuel Floratos, Abhijit Gadde, Jainendra Jain, Manas Kulkarni, Shiraz Minwalla, Ganpathy Murthy, Nikita Nekrasov, Onkar Parrikar and David Tong. We also thank Takeshi Morita for collaboration in a related paper \cite{gamma-paper} which has some overlap with the present paper. We thank Manas Kulkarni, Satya Majumdar and Gregory Schehr for inspiring discussions on related work. We thank Dimitra Karabali, Shahin Sheikh-Jabbari, Todor Popov and D. V. Khveshchenko for comments on the first version of this paper and drawing our attention to related literature. One of the authors (G.~M.) would like to acknowledge the Raja Ramanna Chair Professorship.  G.~M. would like to thank the organizers of the 14th regional meeting in String theory (Crete, June 2025), and the Indian Association for Cultivation of Sciences (September 2025) for an opportunity to present part of this work. 
\appendix
\section*{Appendix}
\section{Landau levels}\label{app:ll}
Landau fermions refer to 2D fermions subject to a uniform transverse magnetic field $B$. In the symmetric gauge \cite{PhysRevB.27.3383, Ciftja_2020}
\begin{equation}
\vec{A}(\vec{r})=\frac{B}{2}(-y, x, 0),
\end{equation}
the Hamiltonian becomes
\begin{align}
  H_0 &=\frac{1}{2 m}[{\vec{p}}-q \vec{A}(\vec{r})]^{2}  \nonumber \\
  & =\frac{1}{2m}\left( \left({p_{x}}-\frac{eB {y}}{2}\right)^{2}+\left({p_{y}}+\frac{eB{x}}{2}\right)^{2}\right)
\label{h-ll}\end{align}
As is well-known,  in the Landau problem, one can introduce two special sets of commuting ladder operators:
\begin{align}
    &\hat{a}=\frac{1}{2\sqrt{\hbar m \omega}}\left(\left(m \omega\hat{x}+\hat{p_{y}}\right) +i\left(\hat{p_{x}}-m \omega \hat{y}\right)\right) \nonumber\\
   &\hat{a}^\dagger=\frac{1}{2\sqrt{\hbar m \omega}}\left(\left(m \omega\hat{x}+\hat{p_{y}}\right) -i\left(\hat{p_{x}}-m \omega \hat{y}\right)\right), \label{b-ll}
\end{align}
 \begin{align}
   &\hat{b}=\frac{1}{2\sqrt{\hbar m \omega}}\left(\left(m \omega\hat{x}-\hat{p_{y}}\right) +i\left(\hat{p_{x}}+m \omega \hat{y}\right)\right) \nonumber\\
   &\hat{b}^\dagger=\frac{1}{2\sqrt{\hbar m \omega}}\left(\left(m \omega\hat{x}-\hat{p_{y}}\right) -i\left(\hat{p_{x}}+m \omega \hat{y}\right)\right), \label{a-ll}
\end{align}

where $\omega= eB/(2m)$.
A complete basis of the single-particle Hilbert space can be obtained as simultaneous eigenstates of the number operators $\hat n_1= a^\dagger a$ and $\hat n_2=b^\dagger b$:
\begin{align}
  |n_1, n_2\ran =\frac{1}{\sqrt{n_{1}!}\sqrt{n_{2}!}} (a^\dagger)^{n_1} (b^\dagger)^{n_2} |0\ran, \quad
  a|0\ran = b|0\ran=0
  \label{n1-n2}
\end{align}
The Hamiltonian, however, depends on only the first set of ladder operators:
\begin{align}
H_0 =2\hbar\omega(\hat{a}^\dagger \hat{a} +1/2) \label{spectrum}
\end{align}
Thus the spectrum of the Hamiltonian is given by
\begin{equation}
    E_{n_{1},n_{2}}=2\hbar\om(n_{1} +1/2)\label{spect}
\end{equation}
The wavefunctions corresponding to the states \eq{n1-n2} are (see, e.g., Appendix A of \cite{PhysRevA.103.033321})
\begin{align}
    \psi_{n_1,n_2}(x,y) = \sqrt{\frac{m \omega}{\pi \hbar}} \sqrt{\frac{(n_1+n_2-|n_2-n_1|)!}{\left(\frac{ n_1+n_2+|n_2-n_1|}{2}\right)!}} L_{\frac{n_1+n_2-|n_2-n_1|}{2}}^{|n_2-n_1|}\left((x^2+y^2)\frac{m\omega}{\hbar}\right) e^{-\frac{m \omega}{2 \hbar} (x^2+y^2)} \left((x-i y) \sqrt{\frac{m \omega}{\hbar}}\right)^{n_2-n_1} \label{n1-n2-xy}
\end{align}

\subsection{The $(x_1,x_2)$ coordinates}\label{app:x1-x2}

It is useful to introduce a new set of phase space coordinates $(x_1,p_1,x_2,p_2)$ corresponding to the two sets of ladder operators:
\begin{align}
  \hat{a} = \sqrt{\frac{m\omega}{2\hbar}}\left(\hat{x}_1 + \frac{i}{m\omega}\hat{p}_1\right),
  \quad \hat{a}^\dagger &= \sqrt{\frac{m\omega}{2\hbar}}\left(\hat{x}_1 - \frac{i}{m\omega}\hat{p}_1\right), \nonumber
  \\
\hat{b} = \sqrt{\frac{m\omega}{2\hbar}}\left(\hat{x}_2 +\frac{i}{m\omega}\hat{p}_2\right), \quad 
\hat{b}^\dagger & = \sqrt{\frac{m\omega}{2\hbar}}\left(\hat{x}_2 - \frac{i}{m\omega}\hat{p}_2\right), \label{a-b-x1-x2}
\end{align}
These are related to the original phase space coordinates $(x,y,p_x,p_y)$ as follows:
\begin{align}
 & x_1 = \frac1{\sqrt 2 m\omega}(m\omega x + p_y),
  p_1 = \frac1{\sqrt 2}(p_x - m\omega y)
\nonumber \\
 & x_2 = \frac1{\sqrt 2 m\omega}(m\omega x - p_y),
  p_2 = \frac1{\sqrt 2}(p_x + m\omega y)  \label{x1-x2-x-y}
\end{align}
Note that while the above equation is an operator relation between the two sets of phase space coordinates, it also represents a canonical transformation at the classical level. The wavefunctions for the states \eq{n1-n2} in the $(x_1,x_2)$ basis are
\begin{align}
  \chi_{n_1,n_2}(x_1,x_2) = \sqrt{\frac{m \omega}{\pi \hbar}} \frac{\exp{-m \omega (x_1^2 + x_2^2)/  2 \hbar}}{\sqrt {  2^{n_1+n_2} n_1! n_2! }} H_{n_1}\left(\sqrt{\frac { m  \omega } {  \hbar }} x_1 \right) H_{n_2}\left(\sqrt{\frac { m  \omega } {  \hbar }} x_2 \right)
  \label{n1-n2-x1-x2}
\end{align}
where $H_n(x)$ denotes the Hermite polynomial of degree  $n$.

It is useful to note the inverse transformations to \eq{x1-x2-x-y}:
\begin{align}
  & x = \frac1{\sqrt 2}(x_1 + x_2),
  y = \frac1{\sqrt 2 m\omega}(p_2 - p_1) \nonumber \\
  & p_x = \frac1{\sqrt 2}(p_1 + p_2),
  p_y = \frac{m\omega}{\sqrt 2}(x_1-x_2)
\label{x-y-x1-x2}
\end{align}

\subsubsection{Transformation between $|x_1, x_2 \rangle$ and $|x,y \rangle$}\label{app:transform} 

The ket $\ket{x_1,x_2}$ obeys
\begin{align}
  &\hat{x}_1\ket{x_1,x_2}=x_1\ket{x_1,x_2}\nonumber \\
  &\hat{x}_2\ket{x_1,x_2}=x_2\ket{x_1,x_2} \nonumber
\end{align}
In the $(x,y)$ basis, the above eigenvalue equations become the following. Defining $\psi_{x_1,x_2}(x,y) \equiv \braket{x,y|x_1,x_2}$, and using \eq{x1-x2-x-y}, we get
\begin{align}
&  {x}\psi_{x_1,x_2}(x,y)-\frac{i {\hbar}} {{ m \omega}}\partiald{\psi_{x_1,x_2}(x,y)}{y}=\sqrt{2}x_1 \psi_{x_1,x_2}(x,y) \nonumber\\
&  x\psi_{x_1,x_2}(x,y)+\frac{i {\hbar}} {{ m \omega}}\partiald{\psi_{x_1,x_2}(x,y)}{y}=\sqrt{2}x_2 \psi_{x_1,x_2}(x,y)
\label{x1-x2-eigen}\end{align}
If we add and subtract the above equations, we get 
\begin{align}
 & {x}\psi_{x_1,x_2}(x,y)={(x_1 +x_2) \over \sqrt{2}}\psi_{x_1,x_2}(x,y) \quad \text{and}\nonumber\\
 &\frac{i {\hbar}} {{ m \omega}}\partiald{\psi_{x_1,x_2}(x,y)}{y}=\frac{x_2-x_1}{\sqrt{2}}\psi_{x_1,x_2}(x,y)
\end{align}
It is straightforward to solve the above equations. We get
\begin{equation}
  \psi_{x_1,x_2}(x,y) =  \lan   x,y| x_1, x_2\ran= \frac{1}{\sqrt{2 \pi}} \delta\left(x-{(x_2 +x_1)\over \sqrt{2}}\right)\exp{-iy{\frac{m \omega}{\hbar}} \left(\frac{x_2-x_1}{\sqrt{2} }\right)}
  \label{transform}
\end{equation}
It is straightforward to verify that the wavefunctions $\chi_{n_1,n_2}(x_1,x_2)$ and $\psi_{n_1,n_2}(x,y)$ transform into each other under the above transformation \eq{transform} between the $|x_1, x_2 \rangle$ and $|x,y\rangle$ bases.

\subsection{Lowest Landau Level}\label{app:lll}

Eq. \eq{spect} represents energy bands labelled by $n_1$ (see Fig. \ref{fig:ll-band}) which are called Landau levels.

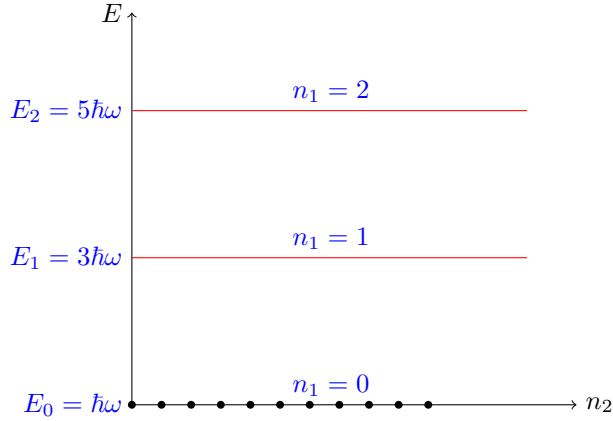
\begin{figure}[H]
  \begin{center}
    \begin{tikzpicture}[scale=0.65]

\draw[->] (0,0) -- (0,8) node[left] {$E$};
\draw[->] (0,0) -- (9,0) node[right] {$n_2$};

\node[left,blue] at (5,0.4) {$n_1=0$};
\node[left,blue] at (0,0) {$E_0=\hbar \omega$};
\node[left,blue] at (5,3.4) {$n_1=1$};
\node[left,blue] at (0,3) {$E_1=3 \hbar \omega$};
\node[left,blue] at (5,6.4) {$n_1=2$};
\node[left,blue] at (0,6) {$E_2=5 \hbar \omega$};
\foreach \i in {0,0.6,...,6.2} {
    \pgfmathsetmacro{\y}{0*\i} 
    \fill[black] (\i,\y) circle (0.08);
}

\draw[red] (0,3) -- (8,3);
\draw[red] (0,6) -- (8,6);

\end{tikzpicture}
  \end{center}
  \caption{\footnotesize{Landau levels of the original Landau problem}}
  \label{fig:ll-band}
\end{figure}

The lowest band, called the lowest Landau level (LLL) consists of states $|n_1, n_2\rangle$ = $|0,n_2\rangle$, which are degenerate, with energy $E_{0,n_2}= \hbar \omega$. 
Wavefunctions for LLL state $|0,n\rangle$ are 
\begin{align}
& \psi_{n}(x,y)= \langle x,y|0,n \rangle=  \frac{e^{\frac{-1}{2l_{0}^
        {2}} \left(x^2+y^2\right)} ((x-i y)/l_{0})^{{n}}}{l_{0}\sqrt{\pi {n}!}} \nonumber\\
  &  \tilde\chi_{n}(x_{1},x_{2}) := \langle x_1, x_2 |0,n \rangle= \sqrt{\frac { m  \omega } {  \pi  \hbar }}  \frac { e^{- ( x_{1}^{  2 } + x_{  2 }^{  2 }) /  2   l_0^2 } } {   \sqrt {  2^n n! } } H_{n}(   x_{  2 }/l_0 )
  \label{lll-wavefunctions}
\end{align}
In the above $l_0$ is the characteristic length defined in \eq{def-l0}.

\section{Constrained quantum mechanics: Dirac bracket}\label{app:dirac}


We would like to find the symplectic form on the constrained phase space defined by \eq{c1-c2}. We will find this by following Dirac's method. To begin,
we compute the matrix $M_{ij} = \{C_i, C_j\}$ of Poisson brackets between the constraints, yielding $M_{12}= 1= -M_{21}$, $M_{11}=0= M_{22}$. Note that this is a non-singular matrix (in other words, the constraints $(C_1, C_2)$ define a pair of second class constraints); the inverse matrix $M^{-1}_{ij}$ is given by $M^{-1}_{12}=-1= - M^{-1}_{21}$. 

The symplectic form on the constrained phase space is given by the Dirac bracket
\[
\{F, G\}_{DB} = \{F, G\} - \{F, C_i\} (M^{-1})_{i,j} \{C_j, G\},
\]
It is straightfoward to see the emergence of a non-trivial Dirac braket between the coordinates of the $x$-$y$ plane:
\begin{align}
    \{x,y\}_{DB}&=\{x,y\}-\sum_{i,j}\{x,C_i\}(M^{-1})_{ij}\{C_j,y\}\nonumber\\ 
    &=-\left(\{x,C_1\}(M^{-1})_{12}\{C_2,y\}+\{x,C_2\}(M^{-1})_{21}\{C_1,y\}\right) \nonumber\\
    &=\frac{1}{2m\omega}= \frac{1}{eB}
    \label{x-y-db}
\end{align}

An alternative way of computing the Dirac bracket is to note that in terms of the phase space coordinates $(x_1, x_2, p_1, p_2)$, the constraints \eq{x1-p1=0} simply tell us to drop the $x_1$ and $p_1$ directions, so that the constrained phase space ${\cal M}$ is given by the two-dimensional space coordinatized by $(x_2,p_2)$; thus the induced symplectic form on ${\cal M}$ is just
\begin{align}
\Omega|_{\cal M} = dx_2 \wedge dp_2
  \label{omega-x2-p2}
\end{align}
Hence, by using \eq{x-y-x1-x2}, we get
\[
  \{x,y\}|_{\cal M} = \fr{\del x}{\del x_2} \fr{\del y}{\del p_2} - \fr{\del x}{\del p_2}\fr{\del y}{\del x_2} = \frac{1}{2m\omega}
\]
which is the same as the Dirac bracket we obtained above. Stated in another way, the symplectic form \eq{omega-x2-p2} leads to
\begin{align}
\Omega_{xy}= \Omega_{x_2 p_2} \fr{\del x_2}{\del x} \fr{\del p_2}{\del y} - (x_2 \leftrightarrow p_2)= -2 m \omega
\label{omega-x-y}
\end{align}
where we have used the relations $x_2 = \sqrt 2 x, p_2 = - \sqrt 2 m\om y$, which are obtained by imposing $x_1=0=p_1$ on \eq{x1-x2-x-y}. 
Eq. \eq{omega-x-y} of course has the same content as \eq{x-y-db}.

\section{Wigner distribution and the Weyl correspondence}\label{app:weyl-wigner}
\subsection{The Weyl correspondence}
Classically a particle moving in $R^1$ is described by a 2D phase space $R^2$. Let us make a choice of canonical phase space coordinates $(x,p)$ such that $\{x,p\}=1$. In the quantum mechanical description, we promote $x,p$ to operators $\hat x, \hat p$, such that $[\hat x, \hat p]=i\hbar$. How does one quantize a general function $f(x,p)$? E.g. for $f(x,p)= xp$, classically the functions $xp, px$ are the same, but the operators $\hat x \hat p, \hat p \hat x$ are different because they are ordered differently. The Weyl correspondence \cite{Weyl1931Theory} assigns to the function $f(x,p)$ a unique operator $\hat f$ and {\it vice versa} as follows:
\begin{align}
  & \hat f= \int \fr{dx\,dp}{2\pi\hbar} f(x,p) \hat g(x,p) \label{weyl-f-hat}\\
  & f(x,p)=  \Tr (\hat f \, \hat g(x,p))
  \label{weyl-f}
\end{align}
where
\begin{align}
  \hat g(x,p) \equiv \int \fr{d\a \,d\b }{2\pi\hbar}\exp\left[i/\hbar\left( \a(\hat x-x)+i\b(\hat p -p)\right)\right]
  \label{g-x-p}
\end{align}
Note that \eq{g-x-p} is essentially like a delta function which indicates replacing $x,p$ by $\hat x, \hat p$, but with a well-defined ordering prescription. The consistency between \eq{weyl-f-hat} and \eq{weyl-f} follows from the fact that
\begin{align}
  \Tr(\hat g(x,p) \hat g(x',p'))= 2\pi \delta(x-x') \delta (p-p')
  \label{ortho-g}
\end{align}
For the function $f(x,p)=xp$, the ``Weyl-ordered'' operator, as defined in \eq{weyl-f-hat}, is the democratic choice $\hat f= \fr12 (\hat x \hat p + \hat p \hat x)$, as is straightforward to verify. 

\subsubsection{Invariance of the Weyl ordering prescription}

\def\hx{{\hat x}}
\def\hp{{\hat p}}
\def\txi{{\tilde \xi}}
\def\tx{{\tilde x}}
\def\tp{{\tilde p}}
\def\htx{{\hat {\tilde x}}}
\def\htp{{\hat {\tilde p}}}

Suppose we choose a different pair of phase space coordinates $\txi_i \equiv (\tx, \tp)$ related by a linear canonical transformation to the original $\xi_i=(x,p)$:
\begin{align}
  \txi_i= M_{ij} \xi_j
  \label{lin-can}
\end{align}
where det $M$=1 (in other words, $M$ is an SL(2,R) matrix).

{\it A priori}, it is not clear that for a given phase space function, the Weyl-ordered operator will be independent of the choice of the phase space coordinates. Let us choose $M_{ij}= \{\{c,-s\},\{s,c\}\}$, $c=\cos\theta, s=\sin\theta$, which is a linear rotation in the phase plane. Under this coordinate transformation, the phase space function $f(x,p)=xp$ becomes the following
\begin{align}
f(x,p)= xp= cs (\tp^2-\tx^2) +(c^2-s^2) \tx\tp \equiv \tilde f(\tx,\tp)
\label{f-ftilde}
\end{align}
The Weyl ordered operators would be
\[
f(x,p) \to \hat f= \fr12(\hx\hp+ \hp\hx), \qquad
\tilde f(\tx,\tp) \to \hat {\tilde f}= cs(\htp^2-\htx^2)+ \fr{c^2-s^2}{2}(\htx\htp+ \htp\htx)
\]
Are the two Weyl-ordered operators $\hat f, \hat {\tilde f}$ the same? The important thing is to note that for a linear canonical transformation \eq{lin-can} between the classical phase space coordinates, the corresponding quantum operators also satisfy the same relations:
\begin{align}
  \hat\txi_i= M_{ij} \hat\xi_j
  \label{lin-can-op}
\end{align}
Applying this fact, we find, by a straightforward calculation, that indeed
\begin{align}
\hat f= \hat{\tilde f}
\label{f-inv}
\end{align}
A general proof of \eq{f-inv} (using the notation $a_i=(\a,\b)$) is based on the following result\footnote{In the rest of this subsection, we will put $\hbar=1$, unless explicitly reinstated.}
\begin{align}
  \hat {\tilde g}(\tilde\xi) & \equiv \int \fr{d^2 a}{2\pi} \exp[i a_i (\hat\txi_i - \txi_i)]
  \nonumber\\
  & = \int \fr{d^2 \tilde a}{2\pi} \exp[i \tilde a_i (\hat\txi_i - \txi_i)]
  \nonumber\\
  & = \int \fr{d^2 a}{2\pi} \exp[i a_i (\hat\xi_i - \xi_i)] \equiv \hat g(\xi)
  \label{g-inv}
\end{align}
Here we have assumed \eq{lin-can} and \eq{lin-can-op}. In the second step, we used a change of dummy variables of integration. In the third step we have chosen $\tilde a_i= M^{-1}_{ij} a_j$ so that $\tilde a_i (\hat\txi_i - \txi_i)$ $= a_i M^{-1}_{ij} M_{jk} (\hat\xi_k - \xi_k)$ $= a_i (\hat\xi_i - \xi_i)$, and the invariance of the measure $d^2 a$ under the SL(2,R) transformation $M^{-1}$.

We thus get \eq{f-inv}, as follows:
\begin{align}
  \hat {\tilde f} & = \int \fr{d^2\tilde \xi}{2\pi} \tilde f(\txi) \hat {\tilde g}(\tilde \xi)\nonumber\\
  & = \int \fr{d^2 \xi}{2\pi} f(\xi) \hat {\tilde g}(\tilde \xi)
  \equiv  \hat f
  \label{f-inv-pf}
  \end{align}
Here we have used $\tilde f(\txi) = f(\xi)$ by definition (see \eq{f-ftilde} as an example), the equation \eq{g-inv} and the invariance of the measure $d^2\tilde \xi= d^2 \xi$ under an SL(2,R) transformation.

\paragraph{The converse statement to \eq{f-inv}}

Suppose we have a given operator $\hat f$. By following steps similar to those in \eq{f-inv-pf}, we find that the corresponding classical phase space functions in different canonical coordinates related by an SL(2,R), are equal: $\tilde f(\tilde \xi)= f(\xi)$:
\begin{align}
  \tilde f(\tilde x, \tilde p) \equiv \tilde f(\tilde \xi)= \Tr( \hat f \hat{\tilde g}(\tilde \xi)
  = \Tr( \hat f \hat{ g}(\xi) = f(\xi) \equiv f(x,p)
  \label{fn-inv}
\end{align}
where $(\tilde x, \tilde p)$ are related $(x,p)$ by a linear canonical transformation as in \eq{lin-can}.

\subsubsection{Some properties of the Weyl correspondence}

In the following we use the definition of the star product (also called Moyal product) and the Moyal bracket \cite{Moyal:1949sk}
\begin{align}
  f*g (x,p) &\equiv  \exp\left[i\fr\hbar{2}\left(\del_x\del_{p'} - \del_{x'}\del_p\right)\right] f(x,p) g(x',p')|_{_{x'=x,p'=p}} 
  \label{star-product} \\
  \{f,g\}_{MB}(x,p) & \equiv \fr1{i\hbar}(f*g - g*f)(x,p)
\label{Moyal-bracket}  
\end{align}
In the following, operators $\hat f$ and the corresponding phase space functions $f(x,p)$ are related as in \eq{weyl-f-hat} and \eq{weyl-f}.

\begin{table}[H]
\begin{align}
  \hbox{Operator}   & \kern20pt & \hbox{phase space function} \nonumber\\
  \hat f_1  &\kern20pt &  f_1(x,p) \nonumber \\
  \hat f_2 &\kern20pt &  f_2(x,p) \nonumber\\
  \hat f_1 \hat f_2  &\kern20pt &  f_1*f_2(x,p)\nonumber \\
       [\hat f_1, \hat f_2] &\kern20pt & i\hbar \{f_1, f_2\}_{MB}(x,p) \nonumber\\
  &\kern20pt & \nonumber\\     
  \hbox{Density matrix} & \kern20pt & \hbox{Wigner distribution} \nonumber\\
  \hat \rho  & \kern20pt & u(x,p)= \Tr(\hat \rho\, \hat g(x,p))\nonumber\\
  |\psi \rangle \langle \psi |  & \kern20pt & u(x,p)= \langle \psi| \hat g(x,p))| \psi \rangle\nonumber \\
  &\kern20pt & \nonumber\\     
  \hbox{Traces} & \kern20pt & \hbox{Integrals} \nonumber\\
  \Tr \hat f_1 =&\kern20pt &  \int \fr{dx\,dp}{2\pi\hbar} f_1(x,p)\nonumber\\
  \Tr (\hat f_1 \hat f_2)= &\kern20pt & \int \fr{dx\,dp}{2\pi\hbar} f_1(x,p) f_2(x,p)
  \nonumber
\end{align}
\caption{\footnotesize{Weyl correspondence.}}
\label{tab:weyl}
\end{table}
In the last line we have used \eq{ortho-g}. We have included some statement about the Wigner distribution which will be explained below.

\subsection{The Wigner distribution}\label{app:wigner}

The Wigner distribution for a general mixed state $\hat\rho$ is defined as the classical phase space function, denoted $u(x,p)$, related by the Weyl correspondence to the state $\hat\rho$:
\begin{align}
  u(x,p)= \Tr(\hat \rho\, \hat g(x,p))
  \label{def-mixed-wigner}
\end{align}
A pure state corresponds to $\hat \rho= \hat P_\psi \equiv |\psi \rangle \langle \psi|$, for which the Wigner distribution becomes
\begin{align}
  u(x,p)= \Tr(\hat P_\psi \hat g(x,p))=\int dx_1\,dx_2\, \psi^*(x_1) \psi(x_2) \delta(x- \fr{x_1+x_2}2) \exp[ip(x_1-x_2)/\hbar]
  \label{def-wigner}
\end{align}
The last expression comes by writing the Trace as
\[
\int dx_1\,dx_2\, \langle x_2 | \psi \rangle \langle \psi |x_1 \rangle \langle x_1 | \hat g(x,p) |x_2 \rangle
\]
and noting that
\begin{align}
  \langle x_1 | \hat g(x,p) |x_2 \rangle = \delta(x- \fr{x_1+x_2}2) \exp[ip(x_1-x_2)/\hbar]
  \label{g-xp-matrix}
\end{align}
which is straightforward to derive by applying the Baker-Campbell-Hausdorff formula to the exponential in $\hat g(x,p)$.

A more conventional expression for the Wigner distribution, equivalent to \eq{def-wigner} is
\begin{align}
  u(x,p)=\int_{-\infty}^\infty d\eta, \psi^*(x+\eta/2) \psi(x- \eta/2) \exp[ip\eta/\hbar]
  \label{def2-wigner}
\end{align}

\subsubsection{Properties of the Wigner distribution}\label{app:wigner-prop}

It is easy to derive the main property of the Wigner distribution:
\begin{align}
  \langle \psi |\hat f | \psi \rangle= \int \fr{dx\,dp}{2\pi\hbar} u(x,p) f(x,p)
  \label{wigner-average}
\end{align}
To see this, note that the LHS= $\Tr P_\psi \hat f$. By using the phase space representations the trace of operator products given in Table \ref{tab:weyl}, we obtain the RHS.

The formula \eq{wigner-average} establishes the Wigner distribution as some kind of a phase space distribution. It is not {\it strictly} a phase space density since it is not positive definite; in fact the negativity of the Wigner distribution carries important information (related to the extent of non-classicality of the wavefunction). The marginal distributions are positive definite, however:
\begin{align}
  & \int \fr{dp}{2\pi\hbar} u(x,p)= \rho(x) \equiv |\psi(x)|^2
  \label{real-density}\\
  & \int dx\, u(x,p)= \tilde\rho(p) \equiv |\tilde\psi(p)|^2
  \label{mom-density}
\end{align}
where $\tilde\psi(p)=\int dx\, \psi(x) \exp[-ipx/\hbar]$ is the momentum-space wavefunction.

\paragraph{Invariance property of the Wigner distribution:}

If $u(x,p)$ and $\tilde u(\tilde x,\tilde p)$ are the Wigner distributions for the same state $|\psi \rangle$ in two different phase space coordinates which are related to each other by a linear canonical transformation \eq{lin-can}, then, following the logic of \eq{fn-inv}, we must have
\begin{align}
  u(x,p)= \tilde u(\tilde x, \tilde p)
  \label{u-inv}
\end{align}

\paragraph{Three important properties:}\label{app:3prop}

Let us recall the following properties of a density matrix $\hat \rho$:
\begin{align}
  \Tr (\hat \rho)=1 \label{norm}\\
  i\hbar \del_t \hat \rho =  [\hat H, \hat \rho] \label{eom}
\end{align}
and for a pure state $\hat \rho= |\psi \rangle \langle \psi |$
\begin{align}
  (\hat \rho)^2 = \hat \rho \label{pure}
\end{align} 
By using the definition \eq{def-mixed-wigner} of the Wigner distribution, and the properties of the Weyl correspondence listed in Table \ref{tab:weyl}, we get, from the above properties
\begin{align}
  \int \fr{dx\,dp}{2\pi\hbar} u(x,p)= 1 \label{u-norm}\\
  \del_t u(x,p)= \{H,u \}_{MB} \label{u-eom}\\
  u \star u (x,p)= u(x,p) \label{u-pure}
\end{align}  

\subsubsection{Many-body Wigner distribution}\label{app:many-body-wigner}

Let us consider a single Slater state, of the form
\begin{align}
  |F \rangle= c^\dagger_{f_N}...c^\dagger_{f_2} c^\dagger_{f_1} |0\rangle
  \label{slater}
\end{align}
where $c^\dagger_{f_k}|0\ran$ creates the single particle state $|f_k\ran$, with wavefunction $\chi_{f_k}$. Using the definition of the second quantized Wigner distribution
\begin{align}
  U(x,p)&=\int d\eta_x  {\hat\Psi}^\dagger (x + \eta/2){\hat\Psi} (x -\eta/2)\exp[i\eta p /\hbar ],
  \label{u-xp-qft}
\end{align}
it is easy to see that the many-body Wigner function is a sum over Wigner functions of the individual filled states
\begin{align}
  u(x,p) \equiv \lan F| U(x,p)|F \ran = \sum_k u_{f_k}(x,p)
  \label{qft-uxp}
\end{align}
By using Table \ref{tab:weyl} and the orthogonality of the states $|f_k\ran$, we find that the many-body Wigner distribution for single Slater states satisfies properties similar to the single-particle Wigner distribution:
\begin{align}
  \int \fr{dx\,dp}{2\pi\hbar} u(x,p)= N \label{U-norm}\\
  \del_t u(x,p)= \{H,u \}_{MB} \label{U-eom}\\
  u \star u (x,p)= u(x,p) \label{U-slater}
\end{align}

\paragraph{Large $N$ limit:}

In the large $N$ limit, defined by $N\to \infty, \hbar\to 0$ with $N\hbar =1$, the 3 properties above become
\begin{align}
  \int \fr{dx\,dp}{2\pi} u(x,p)= 1 \label{ucl-norm}\\
  \del_t u(x,p)= \{H,u \}_{PB} \label{ucl-eom}\\
  u^2 (x,p)= u(x,p) \label{ucl-slater}
\end{align}
where we have used the fact that in this limit the star product becomes ordinary product and the Moyal bracket becomes regular Poisson bracket.\footnote{The $\hbar$-dependent higher order derivatives drop out in the $\hbar\to 0$ limit, since, as shown in the text, the many-body Wigner distribution $u(x,p)$ has a smooth  $\hbar\to 0$ limit; this is something that does not happen for single-particle Wigner distribution which does not have a smooth  $\hbar\to 0$ limit.}
The property \eq{ucl-slater} implies the \underbar{droplet property} of $u(x,p)$ i.e. since $u(x,p)$ can only assume values 0 or 1, it is equally well specified by a region $D$ of the phase space (the ``droplet'') where $u=1$ (elsewhere $u=0$). Thus, the semiclassical Wigner distribution is of the form of a characteristic function of the droplet region \cite{Iso:1992ca, Iso:1992aa}
\begin{align}
  u(x,p)= {\cal C}_D(x,p),~~  \hbox{where} \nonumber\\
  {\cal C}_D(x,p) = 1 ~~\hbox{if}~~ (x,p) \in D, \ =0 ~~ \hbox{Otherwise}.
  \label{u-droplet}
\end{align}
Now, what is the droplet region $D$ for a specific single Slater state \eq{slater}?

\paragraph{Ground state:}
For the ground state of $N$ noninteracting fermions in a confining potential (including the case of a box), the region $D$ turns out to be a region in phase space bounded by the Fermi surface. As an example, let us consider fermions in a periodic box of length $L$. Let $N=2M+1$; the ground state consists of states with momenta $p= \hbar 2\pi n/L$, $n=0, \pm 1,..., \pm M$. A simple calculation for the state $|n\ran$ using \eq{def-wigner}\footnote{The integral over $\eta$ is to be suitably modified for a periodic box; the modification is straightforward and is not detailed here.} gives $u_n(x,p)= \delta_{n_p,n}$, where $n_p \equiv L p/(2\pi \hbar)$. Summing over these over the filled states following \eq{qft-uxp}, and taking the large $N$ limit, we get
\begin{align}
  u(x,p)= \theta(p_F - |p|), \quad p_F \equiv  2\pi \hbar M/L \to \pi/L 
  \label{uxp-box}
\end{align}
For a more general confining potential $V(x,p)$, we get
\begin{align}
  u(x,p)= \theta(p_F(x) - |p|), 
  \label{uxp-V}
\end{align}
where the local Fermi momentum $p_F(x)$ is given by the large $N$ limit of the WKB momentum $\sqrt{2(E_F - V(x))}/\hbar$; $E_F$ is the Fermi energy, {\it viz.} the $N$-th energy eigenvalue. The expression \eq{uxp-V} can be arrived at by using the WKB form of the wave-functions $\sim \exp[\pm i \sqrt{2(E_F - V(x))}/\hbar]$.

\paragraph{Band state}
Suppose the state \eq{slater} corresponds to a filled band ({\it i.e.}, with filling $f_i = f_1 +(i-1)$,  $i=1,2,...,N$. By following similar logic as above, we can show that in the large $N$ limit
\begin{align}
  u(x,p)= \theta(p_{f_N}(x) - |p|)\theta(p- p_{f_1}(x)), 
  \label{uxp-V-band}
\end{align}
which corresponds to a droplet $D$ with the topology of an annulus. 

\paragraph{Linear combination of Slater states}
As discussed in the text, in this case, typically, (see \eq{u-lt-1})
\begin{align}
  u(x,p)< 1
  \label{uxp-lt-1}
\end{align}

\subsection{General dimensions}

The preceding discussion in this section can be simply generalized to higher dimensional phase spaces and the corresponding Hilbert spaces. The $d$-dimensional counterpart of \eq{u-xp-qft} is
\begin{align}
  U(\vec x,\vec p)&=\int d^d \vec\eta  {\hat\Psi}^\dagger (\vec x + \vec\eta/2){\hat\Psi} (\vec x -\vec \eta/2)\exp[i\vec \eta. \vec  p /\hbar ],
  \label{u-xp-qft-d}
\end{align}
The generalization of \eq{u-inv} to higher dimensions is simply
\begin{align}
  u(\vec x,\vec p)= \tilde u(\tilde{\vec x}, \tilde{\vec p})
  \label{u-inv-d}
\end{align}
where $(\vec x, \vec p)$ denote a 2$d$-dimensional phase space. $(\tilde{\vec x}, \tilde{\vec p})$ is another set of phase space coordinates which are related to the former ones by a linear canonical transformation $M$:
\begin{align}
\{\tilde{\vec x}, \tilde{\vec p}\}^T= M\cdot \{\vec x, \vec p\}^T
\end{align}

\section{Representations of delta and theta functions}

The delta and theta functions can be represented in a distributional sense as follows:
\begin{align}
  \frac{e^{-\frac{(x-x_1)^2}{h}}}{\sqrt{\pi } \sqrt{h}} & \stackrel{\hbar\to 0}{\to} \delta(x-x_1) \label{delta-rep}\\
  \frac{1}{2} \left(\text{erf}\left(\frac{x-x_1}{\sqrt{h}}\right)+1\right) & \stackrel{\hbar\to 0}{\to} \theta(x-x_1) \label{theta-rep}
\end{align}
Here erf(.) means the Error function. The above limits are to be interpreted in terms of integration with test functions:
\begin{align}
\int_{-\infty}^\infty dx_1\, \frac{e^{-\frac{(x-x_1)^2}{h}}}{\sqrt{\pi } \sqrt{h}} f(x_1)
\stackrel{\hbar\to 0}{\to} f(x)
\label{delta-meaning}\\
\int_{-\infty}^\infty dx_1\,  \frac{1}{2} \left(\text{erf}\left(\frac{x-x_1}{\sqrt{h}}\right)+1\right) f(x_1) \stackrel{\hbar\to 0}{\to} \int_{-\infty}^x dx_1\, f(x_1)
\label{theta-meaninng}
\end{align}
where $f(.)$ is a test function (a continuous function, with compact support or an appropriate fall-off condition). It is important that the test function does not depend on $\hbar$.
    
\section{Ground state density and Wigner distribution and large $N$}\label{rhoproof}

\subsection{Fermion density}

In the large \( N \) limit, we claim that the ground state fermion density is given by:
\begin{align}
\rho(x, y) = \frac{m \omega}{\pi \hbar} \theta(x^2 + y^2 - 1),
\label{rho=u}
\end{align}
where \( \theta \) is the Heaviside step function. The proof goes as follows:

The fermion density can be expressed as (using the form \eq{lll-gen-psi} of the wavefunctions $\psi_n(x,y)$) (using Mathematica)
\[
\rho(x, y) = \sum_{n=0}^{N-1} \psi^\dagger_n(x, y) \psi_n(x, y) =\frac{m \omega}{\pi \hbar} \frac{\Gamma(N, \frac{m \omega r^2}{\hbar} )}{\Gamma(N)},
\]
where \( r^2 = x^2 + y^2 \). Here the incomplete gamma function is defined as:
\[
\Gamma(N, x) = \int_{x}^{\infty} dt \, e^{-t} t^{N-1}, \quad x= \frac{m \omega r^2}{\hbar}
\]
To analyze the large $N$ limit, it is useful to define $t= N s, x= N y$, and convert the above integral as 
\[
\Gamma(N, x) = N^N \int_{y}^{\infty} ds \, e^{-N \Tilde f(s)}, \quad
\tilde f(s)= f(s)+ \frac{1}{N}\log s \quad f(s)= s- \log s \quad  y=  \frac{m \omega r^2}{N \hbar}= r^2/r_0^2, \; r_0^2 = \fr1{m \om}
\]
We will solve this integral using saddle point analysis. We Taylor expand  $f(s)$ as $f(s)=f(s_0)+\frac{f''(s_0)}{2!}(s-s_0)^2+\frac{f'''(s_0)}{3!}(s-s_0)^3...$ where $f'(s)=0|_{s=s_0}$. For the case above, $s_0=1$. To evaluate this at large $N$, we need to consider the following two cases.\\
\textbf{Case 1:} The saddle point $(s_0=1)$ lies  in between the integration limits i.e $s_0\in (y,\infty) \implies    \frac{m \omega r^2}{N \hbar} \leq 1 $ . We use the scaling $s-1=\frac{u}{\sqrt{N}}$ and express
\begin{align}
 N^N \int_{y}^{\infty} ds \, e^{-N \Tilde f(s)}
 &=  N^N \int_{y}^{\infty} ds \, e^{-N f(s)} \frac{1}{s} \nonumber \\
& =\frac{N^N}{\sqrt{N}}  \int_{-\infty}^{\infty} du\;  e^{-N \left(1+\frac{u^2}{2N}+\frac{u^3}{6 N \sqrt{N}}\right)}\left(1-\frac{u}{\sqrt{N}} + \mathcal{O}(1/N)\right) \nonumber \\
 &= \frac{N^N}{\sqrt{N}} e^{-N}\int_{-\infty}^{\infty} \sqrt{2 \pi}\left(1+ O(1/N)\right) \nonumber \\
 &=\sqrt{\frac{2 \pi}{N}}N^{N} e^{-N}\left(1-O(1/N)\right)
\end{align}
Therefore, for $x < N$ 
\[
\Gamma(N, x) = \sqrt{\frac{2 \pi}{N}}N^{N} e^{-N}\left(1-O(1/N)\right)
\]
\textbf{Case 2:} $s_0$ does not lies between integration limits $ \implies   \frac{m \omega r^2}{N \hbar} \geq 1 $ 

In the range  $y<s<\infty, f(s)$ is monotonically increasing. Hence the integral will be dominated by $f(s)|_{s=y}$. Therefore, for $x>N$
\begin{align}
&\Gamma(N, x)  \approx N^N e^{-Nf(y)} y^{-1} 
\end{align}
The {\it Stirling's approximation} of gamma function gives
\begin{align}
\Gamma(N) \approx \sqrt {\frac{2 \pi}{N}} N^N e^{-N} 
\end{align}
We obtain:
\[
\frac{\Gamma(N, x)}{\Gamma(N)} =
\begin{cases} 
1+ O(\frac{1}{N}) & \text{if } x < N, \\
\sqrt{N} \exp\left(-N \left( \frac{x}{N} - \log(\frac{x}{N}) - 1 \right)\right) x^{-1} \approx O(e^{-N}) & \text{if } x \geq N.
\end{cases}
\]
We have used the fact that $f(y)>f(s_0)=1$ for the case of $x \geq N$. Thus, the fermion density is:
\[
\rho(x, y) = \frac{m \omega}{\pi \hbar} \times
\begin{cases} 
1 -O(\frac{1}{N})& \text{if } \frac{ m\omega r^2}{\hbar} < N, \\
\exp\left(-N \left( r^2 - \log(r^2) - 1 \right)\right) r \approx O(e^{-N}) & \text{if }  \frac{ m\omega r^2}{\hbar}  \geq N.
\end{cases}
\]
which gives not only \eq{rho=u} but also the subleading corrections at large $N$.
\subsection{Wigner Distribution}\label{wign-xy}
\textbf{Proof of equation(\ref{eq:wign-xy}})\par

The wave function for LLL state is given by equation (\ref{lll-gen-psi}).We will use the expression (\ref{def2-wigner}) to calculate Wigner distribution for LLL state.It is useful to visit some mathematical identities
\begin{align}
\int_{-\infty}^{\infty} dx e^{-\left(\frac{x}{2}-a\right)^2}\left(\frac{x}{2}-b\right)^{2m}=\frac{\sqrt{\pi}}{2^{(2m-1)}}(-1)^m H_{2m}\left(i (a-b)\right)
\label{id1}\end{align}
\begin{align}
\sum_{k=0}^{n}{}^{n}C_{k}H_{2k}(x)H_{2(n-k)}(y)=2^{2n}n!(-1)^n L_{n}(x^2+y^2)
\label{id2}\end{align}
\begin{align}
\int_{-\infty}^{\infty} dx e^{-(x/a-b)^2} H_{2k}\left(\frac{x}{a}-c\right)=\sqrt{\pi}a \left(2 (b-c)\right)^{2k}
\label{id3}\end{align}
Where $H_{n}(x),L_{n}(x)$ are $n^{th}$ order Hermite and Lagurre polynomial respectively.
\begin{align}
&\psi^{*}_{n}\left(\vec{x}-\frac{\vec{\eta}}{2}\right)\psi_{n}\left(\vec{x}+\frac{\vec{\eta}}{2}\right)
=\frac{1}{l_0^2 \pi n!}\exp
 \left \{-\frac{1}{2l_0^2}\left[(x-\frac{\eta_1}{2})^2
+(y-\frac{\eta_2}{2})^2
+(x+\frac{\eta_2}{2})^2
+(y+\frac{\eta_2}{2})^2\right]
\right\}\nonumber \\
& \times \left(\frac{1}{l_0}\left((x-\frac{\eta_1}{2})+i(y-\frac{\eta_2}{2})\right)\right)^n
\left(\frac{1}{l_0}\left((x+\frac{\eta_1}{2})-i(y+\frac{\eta_2}{2})\right)\right)^n\nonumber \\
&=\frac{1}{l_0^2 \pi n!}\exp \left\{-\frac{1}{l_0^2}\left(x^2+y^2+\frac{\eta_1^2}{2}+\frac{\eta_2^2}{2}\right)\right\}\left(\frac{1}{l_0^2}\left(x^2+y^2-\frac{\eta_1^2}{4}-\frac{\eta_2^2}{4}+i \left(\eta_y-\eta_2x\right)\right)\right)^n \nonumber \\ 
&=\frac{(-1)^n}{l_0^2 \pi n!}\exp \left\{-\frac{1}{l_0^2}\left(x^2+y^2+\frac{\eta_1^2}{2}+\frac{\eta_2^2}{2}\right)\right\}
\left(
\left(\frac{\eta_1}{2l_0}-\frac{iy}{l_0}\right)^2
+\left(\frac{\eta_2}{2l_0}+\frac{ix}{l_0}\right)^2
\right)^n
\end{align}
\begin{align}
&\psi^{*}_{n}\left(\vec{x}-\frac{\vec{\eta}}{2}\right)\psi_{n}\left(\vec{x}+\frac{\vec{\eta}}{2}\right)\exp \left \{\frac{-i}{\hbar}\left(\eta_1 p_x+\eta_2 p_y\right)\right\}
=\frac{(-1)^n}{l_0^2 \pi n!}\exp \left\{-\frac{1}{l_0^2}\left(x^2+y^2+\frac{\eta_1^2}{2}+\frac{\eta_2^2}{2}\right)\right\}
\exp \left \{\frac{-i}{\hbar}\left(\eta_1 p_x+\eta_2 p_y\right)\right\}\nonumber \\
& \times \left(
\left(\frac{\eta_1}{2l_0}-\frac{iy}{l_0}\right)^2
+\left(\frac{\eta_2}{2l_0}+\frac{ix}{l_0}\right)^2
\right)^n\nonumber \\
&=\frac{(-1)^n}{l_0^2 \pi n!} \exp \left \{-\left(\frac{x^2+y^2}{l_0^2}+\frac{l_0^2}{\hbar^2}(p_x^2+p_y^2)\right)\right\}
\exp \left \{-\left(\frac{\eta_1}{2l_0}+i \frac{l_0 p_x}{\hbar}\right)^2-
\left(\frac{\eta_2}{2l_0}+i \frac{l_0 p_y}{\hbar}\right)^2\right \} \nonumber \\
&\times \left(
\left(\frac{\eta_1}{2l_0}-\frac{iy}{l_0}\right)^2
+\left(\frac{\eta_2}{2l_0}+\frac{ix}{l_0}\right)^2
\right)^n
\end{align}
We introduce (using identity (\ref{id1}))
\begin{align}
&I_1=\int_{-\infty}^{\infty} d\eta_1 \exp\left \{-\left(\frac{\eta_1}{2l_0}+i \frac{l_0 p_x}{\hbar}\right)^2\right\}
\left(\frac{\eta_1}{2l_0}-\frac{iy}{l_0}\right)^{2k}=\frac{\sqrt{\pi}l_0}{2^{(2k-1)}}(-1)^k H_{2k}\left(\frac{l_0 p_x}{\hbar}-\frac{y}{l_0}\right) \nonumber \\
&I_2=\int_{-\infty}^{\infty} d\eta_2 \exp\left \{-\left(\frac{\eta_2}{2l_0}+i \frac{l_0 p_y}{\hbar}\right)^2\right\}
\left(\frac{\eta_2}{2l_0}+\frac{ix}{l_0}\right)^{2(n-k)}=\frac{\sqrt{\pi}l_0}{2^{(2(n-k)-1)}}(-1)^{(n-k)} H_{2(n-k)}\left(\frac{l_0 p_y}{\hbar}+\frac{x}{l_0}\right) \nonumber \\
& \text{and} \nonumber \\
& I_1 \times I_2=\frac{\pi l_0^2(-1)^n}{2^{2n-2}}H_{2k}\left(\frac{l_0 p_x}{\hbar}-\frac{y}{l_0}\right) H_{2(n-k)}\left(\frac{l_0 p_y}{\hbar}+\frac{x}{l_0}\right)
\end{align}
Therefore
\begin{align}
u_n(\vec{x},\vec{p})&=\sum_{k=0}^{n} {}^{n}C_{k}
\frac{(-1)^n}{l_0^2 \pi n!} \exp \left \{-\left(\frac{x^2+y^2}{l_0^2}+\frac{l_0^2}{\hbar^2}(p_x^2+p_y^2)\right)\right\}I_1 \times I_2\nonumber \\
&=\exp \left \{-\left(\frac{x^2+y^2}{l_0^2}+\frac{l_0^2}{\hbar^2}(p_x^2+p_y^2)\right)\right\}
\frac{1}{n! 2^{2n-2}}\sum_{k=0}^{n} {}^{n}C_{k}H_{2k}\left(\frac{l_0 p_x}{\hbar}-\frac{y}{l_0}\right) H_{2(n-k)}\left(\frac{l_0 p_y}{\hbar}+\frac{x}{l_0}\right)
\label{u-herm}\end{align}
Now using the identity (\ref{id2}),we reach the final result,
\begin{align}
  \label{u-n-app}
u_n(\vec{x},\vec{p})=4(-1)^n \exp \left \{-\left(\frac{x^2+y^2}{l_0^2}+\frac{l_0^2}{\hbar^2}(p_x^2+p_y^2)\right)\right\}L_n\left(\left(\frac{l_0 p_x}{\hbar}-\frac{y}{l_0}\right)^2+\left(\frac{l_0 p_y}{\hbar}+\frac{x}{l_0}\right)^2\right)
\end{align}
\subsubsection{Large N behaviour}
The $N$-particle analog of the above result is given by summing $u_n$ over the filled states $n=0,1,...,N-1$. The result is derived in \cite{PhysRevA.97.063614, PhysRevA.104.013314}, which we quote below:
\[
U(\vec{x},\vec{ p}) =\sum_{n=0}^{N-1}u_n(\vec{x},\vec{p})=
\begin{cases} 
1 & \text{if } \frac{ m\omega r^2}{\hbar} < 2N, \\
\frac{e^{-\frac{4 a^{3/2}}{3}}}{a^{3/4}}& \text{if }  \frac{ m\omega r^2}{\hbar}  \geq 2 N.
\end{cases}
\]
where $a=\sqrt{2}N^{1/6}\left(\sqrt{r^2}-\sqrt{\frac{2N \hbar}{m \omega}}\right)$
\section{Linear combination of Slater states} \label{app:lin-comb-slater}
Here we analyze simpler versions of Eq.~(\ref{gen-slate}) and examine the contribution of the off-diagonal terms by choosing the following states to construct $|F_{\text{gen}}\rangle$:
\begin{align}
&|F_{1}\rangle=c_{N+1}^\dagger c_{N-1}^\dagger \dots c_1^\dagger| 0\rangle \nonumber \\
&|F_2\rangle=c_{N}^\dagger c_{N-1}^\dagger \dots c_1^\dagger|0\rangle \nonumber\\
&|F_3\rangle=c_{N+2}^\dagger c_{N}^\dagger    \dots c_{2}^\dagger| 0\rangle
\label{F123}\end{align}
\begin{figure}[H]
  \begin{center}
  \begin{tikzpicture}[scale=0.65]

\draw[->] (0,0) -- (0,14) node[left] {$E$};
\draw[->] (0,0) -- (14,0) node[right] {$n_2$};

\draw[blue] (0,10) -- (12,10);
\node[left,blue] at (0,0)  {$E_0=\frac{1}{2}\hbar \omega$};
\node[left,blue] at (0,10) {$E_1=(2+\nu)\hbar \omega$};

\foreach \x in {2,6,10} {
  \foreach \y in {0,1,2,5,6,7,8} {
    \draw[red] (\x,\y) -- (\x+2,\y);
  }
}

\foreach \x in {3,7,11} {
  \foreach \y in {3,3.5,4} {
    \fill[black] (\x,\y) circle (0.08);
  }
}

\foreach \x/\y in {
  3/0, 3/1, 3/2, 3/5, 3/7,   
  7/0, 7/1, 7/2, 7/5, 7/6,   
  11/1, 11/2, 11/5, 11/6, 11/8 
}{
  \fill[black] (\x,\y) circle (0.08);
}

\foreach \y/\name in {
  0/{c_1^{\dagger}},
  1/{c_2^{\dagger}},
  2/{c_3^{\dagger}},
  5/{c_{N-1}^{\dagger}},
  6/{c_{N}^{\dagger}},
  7/{c_{N+1}^{\dagger}},
  8/{c_{N+2}^{\dagger}}
}{
  \node[left,blue] at (1.8,\y) {$\name$};
}

\foreach \x/\name in {3/{|F_1\rangle}, 7/{|F_2\rangle}, 11/{|F_3\rangle}} {
  \node[below] at (\x,0) {$\name$};
}

\end{tikzpicture}

 \end{center}
  \caption{\footnotesize{Figure illustrates the fillings defining the states $|F_1\rangle,\ |F_2\rangle,\ |F_3\rangle$ in \eq{F123}.}}
  \label{fig:multi-slater}
\end{figure}
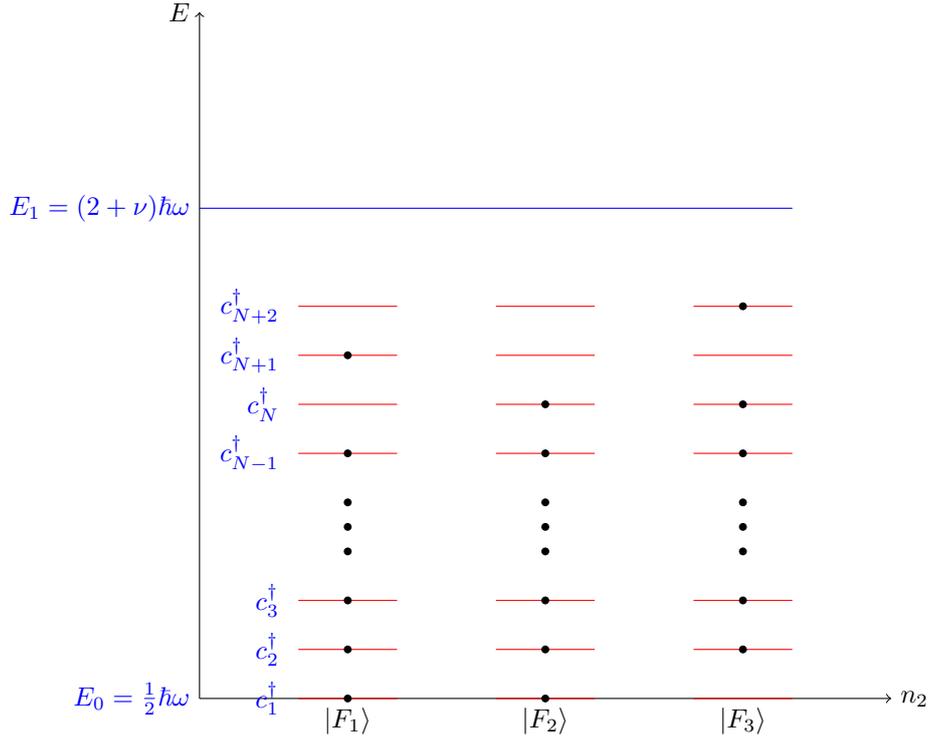

\begin{itemize}
  
    \item \textbf{Case 1: Linear combination of two Slater states}
  \begin{enumerate}[label=(\roman*)]
   \item If we construct $|F\rangle_{gen}=\alpha_1|F_1\rangle+\alpha_2 |F_2\rangle$, we can easily see that such a linear combination can be rewritten as follows 
   \begin{align}
|F'\rangle=\left(\alpha_1 c_{N+1}^\dagger+\alpha_2 c_{N}^\dagger \right)c_{N-1}^\dagger c_{N-2}^\dagger...c_{2}^\dagger c_{1}^\dagger|0\rangle \label{f-prime}\end{align}    
This is a single Slater determinant constructed from a new set of single-particle states. The Wigner distribution function and the fermion density corresponding to the state $|F\rangle_{gen}$ are given by Eqs.~(\ref{lll-u-slater}) and (\ref{slater-rho}). Hence, the equivalence relation given in Eq.~(\ref{1d-2d-N}) holds.
 \item We now choose  $|F\rangle_{gen}=\alpha_2|F_2\rangle+\alpha_3 |F_3\rangle$. In this case, we do not have the freedom to find a basis in which $|F\rangle_{gen}$ is a Slater state. Hence we evaluate,
 \begin{align}
 &\rho_{gen}(x,y)=|\alpha_2|^2 \rho^{(2)}(x,y)+|\alpha_3|^2 \rho^{(3)}(x,y) \label{rho-23}\\
 &\tilde U_{gen}(\vec{x},\vec{p})=|\alpha_2|^2U_{22}+|\alpha_3|^2U_{33}
 =|\alpha_2|^2\underbar u^{(2)}(x_2,p_2)+|\alpha_3|^2\underbar u^{(3)}(x_2,p_2)
 \label{u-23}\end{align}
 From this, it follows that:
\begin{align}
        \tilde U_{gen} * \tilde U_{gen}= |\alpha_2|^4 \tilde U_{22}+|\alpha_3|^4\tilde U_{33} \implies \tilde U_{gen} * \tilde U_{gen}-\tilde U_{gen} \leq 0 \implies  0 \leq \tilde U_{gen} \leq 1
\label{u-gen2}\end{align}
 \end{enumerate}
Hence, we see that when we take a linear combination of just two Slater states, the correspondence \eq{1d-2d-N} holds and $\rho \leq 1$ can be shown. 
    \item \textbf{Case 2: Linear combination of more than two Slater states}\label{3_slate}

 For a state like $|F\rangle_{gen}=\alpha_1|F_1\rangle+\alpha_2 |F_2\rangle+\alpha_3 |F_3\rangle$, 
 \begin{align}
\rho_{gen}(x,y)=&|\alpha_1|^2\rho^{(1)}(x,y)+|\alpha_2|^2\rho^{(2)}(x,y) +|\alpha_3|^2\rho^{(3)}(x,y) \nonumber \\
&+\frac{2}{l_0^3 \pi N! \sqrt{(N+1)}}\left(\text{Re}(x\alpha_1^* \alpha_2)+ \text{Im}(y\alpha_2^* \alpha_1)\right)e^{-\frac{x^2+y^2}{l_0}}\left(\frac{x^2+y^2}{l_0^2}\right)^N
\label{rho-gen3} \end{align}
 The total Wigner distribution is:
    \begin{align}
    \tilde U_{gen} & =|\alpha_1|^2\sum_{i=1}^N u_{m_i}(x_2,p_2) + |\alpha_2|^2\sum_{i=1}^N u_{n_i}(x_2,p_2)+|\alpha_3|^2\sum_{i=1}^N u_{k_i}(x_2,p_2)\nonumber \\
    & \quad +\left(\frac{1}{l_0}\text{Re}(x_2\alpha_1^* \alpha_2)+\frac{l_0}{\hbar}\text{Im}(p_2\alpha_2^* \alpha_1)\right)\frac{8}{\sqrt{2  (N+1)}}(-1)^N e^{-\left(x_2^2/l_0^2+\frac{p_2^2 l_0^2}{\hbar^2}\right)}L_{N}^{1}\left(2\left(x_2^2/l_0^2+\frac{p_2^2 l_0^2}{\hbar^2}\right)\right)
    \label{u-gen3}
    \end{align}
 The contributions from the off-diagonal terms are:
    \begin{align}
        & \alpha_1^*\alpha_2  U_{12}+\alpha_2^* \alpha_1U_{21}\nonumber \\
        & \quad =\frac{\alpha_1^*\alpha_2}{2^N N!\sqrt{2 \pi \hbar(N+1) }}\int d\eta e^{-i \eta p/\hbar}H_N(x-\eta/2)H_{N+1}(x+\eta/2) e^{-\frac{\left(x-\eta/2\right)^2+\left(x+\eta/2\right)^2}{2\hbar}}\nonumber \\
        & \quad +\frac{\alpha_2^* \alpha_1}{2^N N!\sqrt{2 \pi \hbar(N+1) }}\int d\eta' e^{-i \eta' p/ \hbar}H_{N+1}(x+\eta'/2)H_{N}(x+\eta'/2)e^{-\frac{\left(x-\eta'/2\right)^2+\left(x+\eta'/2\right)^2}{2\hbar}}\nonumber \\ & \quad =+\left(\frac{1}{l_0}\text{Re}(x_2\alpha_1^* \alpha_2)+\frac{l_0}{\hbar}\text{Im}(p_2\alpha_2^* \alpha_1)\right)\frac{8}{\sqrt{2  (N+1)}}(-1)^N e^{-\left(x_2^2/l_0^2+\frac{p_2^2 l_0^2}{\hbar^2}\right)}L_{N}^{1}\left(2\left(x_2^2/l_0^2+\frac{p_2^2 l_0^2}{\hbar^2}\right)\right)
 \label{cross}
    \end{align}
    The other off-diagonal terms ($U_{13}$, $U_{23}$, and their complex conjugates) are zero.
 \end{itemize}
In section [\ref{large_n_slater}],we will show that the contribution of the off diagonal term in the above expressions ((\ref{rho-gen3}) and (\ref{u-gen3})) is negligible compared to the  contribution from the diagonal terms.
The similar strategy will be useful for any linear combination of the Slater states.
\subsection{Large $N$ behavior for linear combination of Slater states}\label{large_n_slater}
\begin{itemize}
\item \textbf{Case 1: Linear combination of two Slater states}

 \begin{enumerate}[label=\textbf{(\Roman*)}]
   \item Since equation (\ref{f-prime}) is a Slater state in new basis, equality (\ref{rho-vs-u-classical2}) holds.
   \item For the state$|F_{gen}\rangle=\alpha_2|F_2\rangle+\alpha_3|F_3\rangle$,In the large $N$ limit, again by (\ref{rho-vs-u-classical2}) we get,
\begin{align}
   \rho_{gen}(x,y)=\frac{m \omega}{\pi \hbar}\tilde U_{gen}(\vec{x},\vec{p})
\end{align}
\begin{figure}[H]
    \centering
    \begin{minipage}{.5\textwidth}
        \centering
        \includegraphics[width=0.68\textwidth, height=0.15\textheight]{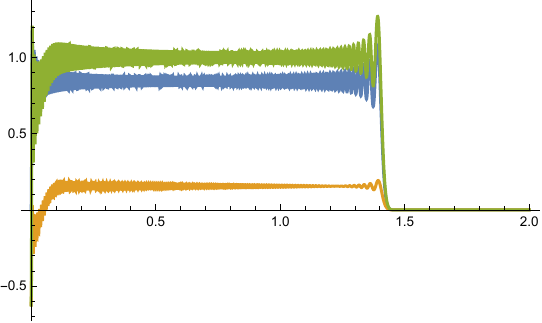}\\
        \vspace{1ex}\vbox{\begin{quote}\baselineskip=9pt{\footnotesize (a) The green curve shows the $\tilde U_{gen}$(\ref{u-23}) in large $N$ limit of the Wigner distribution.  The blue and orange curve shows $|\alpha_2|^2\underbar u^{(2)}(x_2,p_2)$and $|\alpha_3|^2\underbar u^{(3)}(x_2,p_2)$ respectively.  On the $x$-axis is plotted $\tilde r$.}\end{quote}}
    \end{minipage}%
    \begin{minipage}{0.5\textwidth}
        \centering
        \includegraphics[width=0.68\textwidth, height=0.15\textheight]{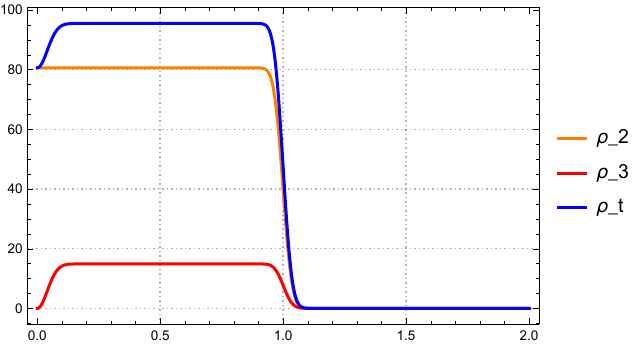}\\
        \kern5pt\vbox{\begin{quote}\baselineskip=9pt {\footnotesize (b) The solid blue curve shows the large $N$ limit of the Fermion density $\rho_{gen}$. On the $x$-axis is plotted $r$.}\end{quote}}
    \end{minipage}
     \caption{\footnotesize{Classical limit of linear combination of slater states properties in the presence of off diagonal term. $N=300, \hbar=1/N, m\om=1$.}}
\end{figure}
 \end{enumerate}
 
\item \textbf{Case 2: Linear combination of more than two Slater staes}\newline
As can be seen from Figs.~(\ref{fig:multiple-slater}) and (\ref{fig:offd}), the $\tilde U_{gen}$ given by equation (\ref{u-gen3}) and $\rho_{gen}(x,y)$ given by equation (\ref{rho-gen3}) show negligible contributions from the off-diagonal terms compared to the diagonal ones.\begin{figure}[H]
    \centering
    \begin{minipage}{.5\textwidth}
        \centering
        \includegraphics[width=0.68\textwidth, height=0.15\textheight]{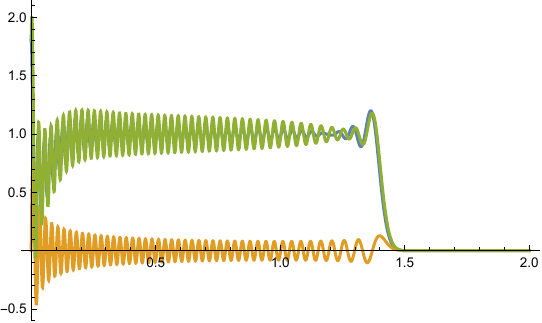}\\
        \vspace{1ex}\vbox{\begin{quote}\baselineskip=9pt{\footnotesize (a) The green curve shows the large $N$ limit of the Wigner distribution.  The orange curve shows the contribution from off diagonal term.  On the $x$-axis is plotted $\tilde r$.}\end{quote}}
    \end{minipage}%
    \begin{minipage}{0.5\textwidth}
        \centering
        \includegraphics[width=0.68\textwidth, height=0.15\textheight]{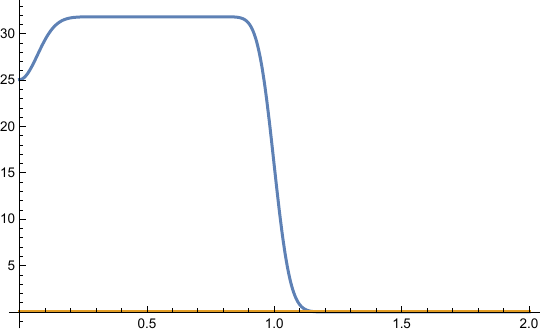}\\
        \kern5pt\vbox{\begin{quote}\baselineskip=9pt {\footnotesize (b) The solid blue curve shows the large $N$ limit of the Fermion density. The orange curve shows the contribution from off diagonal term. On the $x$-axis is plotted $r$.}\end{quote}}
    \end{minipage}
    \caption{\footnotesize{Classical limit of linear combination of slater states in the presence of off diagonal terms (see Section \ref{large_n_slater}). $N=300, \hbar=1/N, m\om=1$.}}
    \label{fig:multiple-slater}
\end{figure}
\begin{figure}[H]
  \centering
  \includegraphics[scale=1]{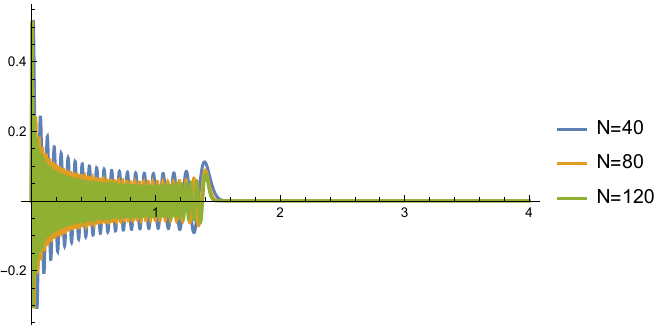}
  \caption{\footnotesize{Scaling of off diagonal terms with $N$.The $y$ axis shows equation(\ref{cross}) and $x $ axis is $\tilde r$.}}
  \label{fig:offd}
\end{figure}

\end{itemize}
Hence,
 \begin{align}
  \tilde U_{gen} \approx \sum_{i=1,2,3} |\alpha_i|^2 u^{(i)}(x,p),\quad \rho_{gen}(x,y) \approx   \sum_{i=1,2,3} |\alpha_i|^2 \rho^{(i)}(x,y),    \label{largen-u-gen3}
    \end{align}
Therefore the equation (\ref{rho-vs-u-classical}) still continues to hold for equation(\ref{largen-u-gen3}).

Note that generically, in the large $N$ limit,
\begin{align}
  \tilde U_{gen} < 1
  \label{u-lt-1}
\end{align}
To see this from \eq{largen-u-gen3}, suppose that each of the states $|f_1\ran, |f_2\ran, |f_3\ran$, by itself, corresponds to a filled band, and that the corresponding Wigner distributions $u_{(i)}$, in the large $N$ limit, are represented by droplets $R_1, R_2, R_3$, respectively (see \eq{uxp-V-band}). In case the droplets do not overlap, \eq{u-lt-1} is clearly true; {\it e.g.}, in the region of droplet $R_1$, $\tilde U_{gen} = |\alpha_1|^2 < 1$.

\section{Entanglement Entropy calculations}
\label{entropy}
\subsection{Entanglement entropy of the LLL system}\label{app:EE-LLL}
\subsubsection{Behavior of entanglement entropy in  $(x,y)$ plane}\label{app:x-y}
We consider the entangling region to be a disk $A$ of radius $l$ in the $(x,y)$-plane. We will use the relation \eq{var-ee}
\begin{align}
S_A= \fr{\pi^2}{3} \left(S_1 - S_2\right)
\label{var-ee-app}
\end{align}
We have already seen in the text (see \eq{s1-lll}) that
\begin{align}
  S_1= \fr{l^2}{l_0^2}
  \label{s1-lll-app}
\end{align}
In this appendix, we will describe the computation of the integral \eq{lll-s2-a} for $S_2$:
\begin{align}
  S_2=
& \left(\frac{1}{\pi l_0^2}\right)^2 \int_A dx dy \int_{A} dx' dy' \exp\left[- \fr{(x-x')^2+(y-y')^2}{l_0^2}\right]
 \label{lll-s2-app}
\end{align}
Using polar coordinates in the $x,y$ plane, we get
\begin{align}
&S_2=\frac1{\pi^2 l_0^4} \int_0^l dr\ \int_0^l dr' r r' \exp{-\fr{1}{l_0^2}(r^2+r'^2)}\int_0^{2 \pi} d\theta\ \int_0^{2 \pi} d\theta'\ e^{\fr{ 2}{l_0^2}r r' \cos(\theta-\theta')} \end{align}
The angular integral is easily computed 
\begin{align}
\int_0^{2 \pi} d\theta d\theta' e^{\fr{ 2}{l_0^2}r r' \cos(\theta-\theta')}=(2 \pi)^2I_{0}\left(\frac{2 r r'}{l_0^2}\right)\end{align}
To compute the resultant radial integration, we expand the $I_0$ function in a power series, so that the double integral becomes a sum of factorized integrals. The final expression for $S_2$ becomes
\begin{align}
  S_2 = \fr4{l_0^4} \int dr\ dr'\ r r' \exp{-\fr{1}{l_0^2}(r^2+r'^2)}I_{0}\left(\frac{2 r r'}{l_0^2}\right)= \sum_{k=0}^{k=\infty} \left(\frac{\gamma\left(k+1,\frac{l^2}{l_0^2}\right)}{\Gamma\left(k+1\right)}\right)^2
  \label{s2-lll-num}
\end{align}
We are not able to compute the sum exactly. A numerical plot of the EE \eq{var-ee-app} looks like
\begin{figure}[H]
  \centering
  \includegraphics[scale=0.6]{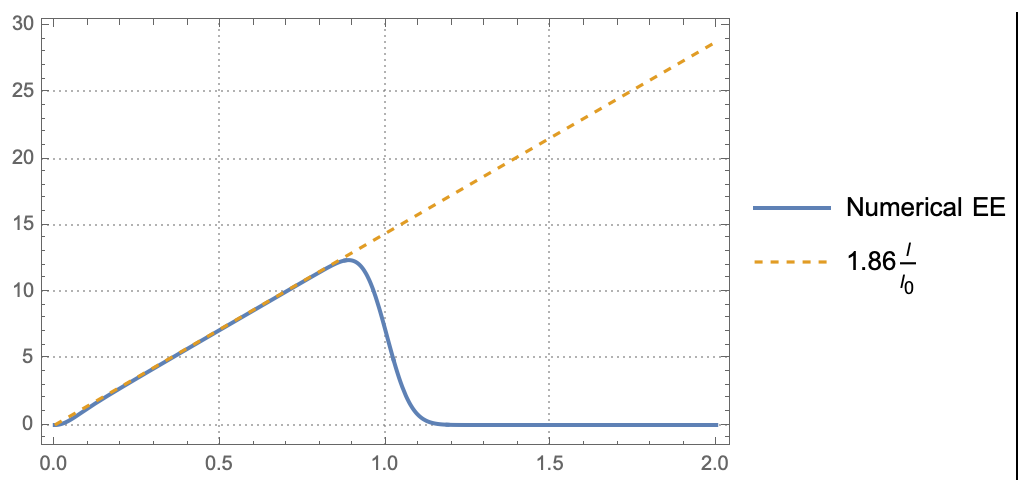}
  \caption{\footnotesize{Entanglement entropy for $N=60$.The orange dashed line shows the linear behavior of entanglement entropy in the case where subregion is smaller than the Fermi droplet.}}
  \label{eeboth}
\end{figure}
The fit to the linear regime of the plot is
\begin{align}
  S= 1.86 \fr{l}{l_0}
  \label{ee-fit}
\end{align}

\subsubsection{Behavior of entanglement entropy in  $(x_1,x_2)$ plane}
We will now encounter a rather different (logarithmic) behaviour of the EE when we consider a strip geometry in the $(x_1,x_2)$ plane (see Figure \ref{fig:strip}).
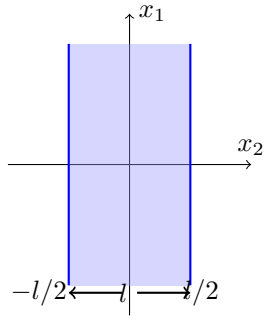
\begin{figure}[H]
  \begin{center}
    \begin{tikzpicture}[scale=0.2]
\draw[->] (-8,0) -- (8,0) node[above] {$x_2$};
\draw[->] (0,-10) -- (0,10) node[right] {$x_1$};
  \fill[blue!40, opacity=0.4] (-4,-8) rectangle (4,8);
  \draw[blue, thick] (-4,-8) -- (-4,8);
  \draw[blue, thick] (4,-8) -- (4,8);
  \draw[->,thick] (-0.5,-8.5) -- (-4,-8.5);
  \draw[->,thick] (0.5,-8.5) -- (4,-8.5);
\node[left,black] at (0.5,-8.5)  {$l$};
\node[left,black] at (-3.5,-8.5)  {$-l/2$};
\node[left,black] at (6.5,-8.5)  {$l/2$};
 \end{tikzpicture}
\end{center}
  \caption{\footnotesize{Strip in the $x_1$-$x_2$ plane.}}
  \label{fig:strip}
\end{figure}
As we have seen, in terms of the $(x_1,x_2)$ variables, the LLL system essentially reduces to 1D QM in the $x_2$. The $x_1$ direction is localized, and it is essentially factored out from the 4D Wigner distribution. The $x_1$ remains irrelevant because of the strip geometry, as we will see below. Thus we expect that the EE of the strip should reflect the EE of an ordinary 1D fermi fluid with a Fermi surface, and should show the standard logarithmic behaviour. This is what we indeed find below.

The Wigner function is given by
\begin{align}
u(x_1,x_2,p_1,p_2)=e^{-\left(\frac{x_1^2 m \omega}{\hbar}+\frac{p_1^2}{m \omega \hbar}\right)}\Theta\left(2N-\frac{(m \omega x_2)^2+p_2^2}{ m \omega \hbar}\right) 
\end{align} 
We write  $S_1$ and $S_2$ as 
\begin{align}
   S_1=\frac{1}{(2 \pi \hbar)^2}\int_A dx_1dx_2 \int dp_1 dp_2 \, u(x_1,x_2,p_1,p_2)
\end{align}
and 
\begin{align}
    S_2=&\frac{1}{(2 \pi \hbar)^4} \int_A dx_1 dx_2 \int_{A'} dx_1' dx_2' \int dp_1 dp_2 dp_1' dp_2' \, \exp\left(\frac{-i}{\hbar}\left((p_1-p_1')(x_1-x_1')+(p_2-p_2')(x_2-x_2')\right)\right) \nonumber \\ 
    & \quad \times u\left(\frac{x_1+x_1'}{2},\frac{x_2+x_2'}{2},p_1,p_2\right)u\left(\frac{x_1+x_1'}{2},\frac{x_2+x_2'}{2},p_1',p_2'\right)
\end{align}
Again we will do $S_1$ calculation first:
\begin{align}
  S_1  = \frac{2}{(2 \pi \hbar)^2}\int_A dx_1 dx_2 \int dp_1 dp_2 \exp{-\left(\frac{(m \omega x_1)^2+p_1^2}{ m \omega \hbar}\right)} \Theta\left(2N-\frac{(m \omega x_2)^2+p_2^2}{ m \omega \hbar}\right) 
\end{align}
Now for $S_2$ term 
\begin{align}
&\int_{-\infty}^{\infty}d p_1 e^{-\frac{p_1^2}{m\omega \hbar}-\frac{i p_1(x_1-x_1')}{\hbar}}
=e^{-\frac{m \omega (x_1-x_1')^2}{4 \hbar}} \sqrt{\pi m \omega \hbar}\quad ,
\int_{-\infty}^{\infty}d p_1' e^{-\frac{p_1'^2}{m\omega \hbar}-\frac{i p_1'(x_1-x_1')}{\hbar}}
= e^{-\frac{m \omega (x_1-x_1')^2}{4 \hbar}} \sqrt{\pi m \omega \hbar} \nonumber \\
& \int_{-\infty}^{\infty} dp_2 e^{\frac{-i p_2(x_2-x_2')}{\hbar}}\Theta\left(2N-\frac{(m \omega \frac{x_2+x_2'}{2})^2+p_2^2}{ m \omega \hbar}\right)
=\int_{-m \omega\sqrt{\frac{2N \hbar}{m \omega}-\left(\frac{x_2+x_2'}{2}\right)^2\frac{m \omega}{\hbar}}}^{m \omega\sqrt{\frac{2N \hbar}{m \omega}-\left(\frac{x_2+x_2'}{2}\right)^2\frac{m \omega}{\hbar}}}
 e^{\frac{-i p_2(x_2-x_2')}{\hbar}} dp_2
 \nonumber \\
& =2 \frac{\sin \left(\left(\frac{x_2-x_2'}{\hbar}\right) \left(m \omega\sqrt{\frac{2N \hbar}{m \omega}-\left(\frac{x_2+x_2'}{2}\right)^2\frac{m \omega}{\hbar}}\right)\right)}{\frac{x_2-x_2'}{\hbar}}
\end{align}
Therefore
\begin{align}
S_2=\frac{4 \pi m \omega \hbar}{(2 \pi \hbar)^4}\int dx_1dx_1' e^{\frac{-m \omega}{2 \hbar}\left((x_1-x_1')^2+(x_1+x_1')^2\right)}
\int dx_2 dx_2' \frac{\sin^2 \left(\left(\frac{x_2-x_2'}{\hbar}\right) \left(m \omega\sqrt{\frac{2N \hbar}{m \omega}-\left(\frac{x_2+x_2'}{2}\right)^2\frac{m \omega}{\hbar}}\right)\right)}{\left(\frac{x_2-x_2'}{\hbar}\right)^2}
\end{align}
For slowly varying potential ($V(x)=\frac{1}{2}m \omega^2 x^2 $),and under the condition that both $( x_2 )$ and $( x_2' )$ are away from the classical turning points, the following approximation holds (see \cite{Das:2022mtb} for further details):
\begin{align}
m \omega\sqrt{\frac{2N \hbar}{m \omega}-\left(\frac{x_2+x_2'}{2}\right)^2\frac{m \omega}{\hbar}} \approx \sqrt{2N \hbar m \omega}
\end{align}
Now consider the subregion as a strip in $(x_1-x_2)$ plane, such as $x_1 \in (-\infty,\infty)$ and $x_2 \in (0,l)$ see figure (\ref{fig:strip})
\begin{align}
S_1=\int_{-l}^l dx_2 dp_2 \Theta\left(2N-\frac{(m \omega x_2)^2+p_2^2}{ m \omega \hbar}\right) \end{align}
\begin{align}
S_2 \approx \frac{1}{(2 \pi \hbar)^2}\int_{-l}^l dx_2\int_{-l}^l dx_2' \frac{\sin^2 \left(\frac{x_2-x_2'}{\hbar}\right)}{\left(\frac{x_2-x_2'}{\hbar}\right)^2}
\end{align}
Now as we can observe, above problem has reduced to calculation of entanglement entropy for 1D harmonic oscillator, this problem is already been solved in \cite{Das:2022mtb, PhysRevE.103.L030105}, we will just rewrite the leading order result here,
\begin{align}
S \approx \frac{1}{3} \log\left[\frac{\sqrt{2N}l}{l_0}\right]
\end{align}

\subsection{Review of ordinary fermions with Fermi surfaces}\label{app:fermi-surface}

\subsubsection{1D Fermi fluid}\label{app:1d-fermi}

We consider 1D fermions in a periodic box of size $L$: $\{ x\in (-L/2,L/2)\}$; the fermi surface is given by $p_F = \fr{N\pi\hbar}{L}$. We consider an entangling region $A$ which is an interval of size $l\ll L$:  $\{ x\in (-l/2,l/2)\}$. 
\begin{figure}[H]
  \begin{center}
\begin{tikzpicture}[scale=0.45]
\draw[->] (-8,0) -- (8,0) node[above] {$x$};
\draw[->] (0,-10) -- (0,10) node[right] {$P_x$};
\fill[pattern=north east lines, pattern color=blue!40] (-4,-4) rectangle (4,4);
  \draw[red, thick] (-4,-4) rectangle (4,4);
\node[left,black] at (6.5,0.5)  {$x=\frac{L}{2}$};
\node[left,black] at (-4.5,0.5)  {$x=-\frac{L}{2}$};
\node[left,black] at (-4.5,-4)  {$-p_F$};
\node[left,black] at (-4.5,4)  {$p_F$};
\node[left,black] at (-4.5,4)  {$p_F$};
 \draw[black, very thick] (-2,0) -- (2,0);
  \node[above] at (0,0) {Region A};
  \node[left,black] at (-2,0)  {$\frac{-l}{2}$};
\node[right,black] at (2,0)  {$\frac{l}{2}$};
 \end{tikzpicture}
\end{center}
  \caption{\footnotesize{Spatial subregion $A$ for a 1D fermi fluid.}}
  \label{fig:subregion_1D}
\end{figure}
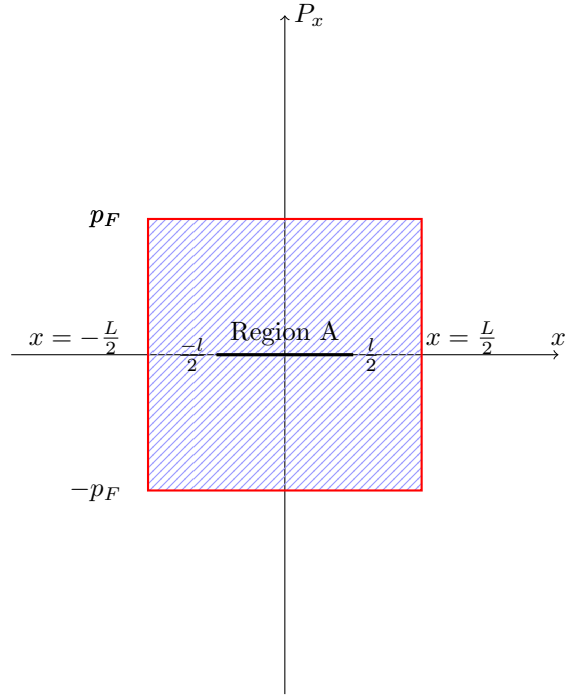

The EE is given by
\begin{align}
  S_A= \fr{\pi^2}{3}\log(\fr{p_F l}{\hbar})
  \label{1d-EE-log}
\end{align}
A brief derivation is given below.

\noindent The ground state Wigner distribution is given by (see Figure \ref{fig:subregion_1D})
\begin{align}
  u(x,p)= \theta(p_F - |p|), \; x\in [-L/2, L/2], \; p_F= N\pi\hbar/L
  \label{1d-u}
\end{align}
which leads to the following two-point function (using \eq{c-x-y})
\begin{align}
  C(x,x')= \int_{-p_F}^{p_F} \fr{dp}{2\pi\hbar} \exp[-i p(x-x')/\hbar]
  = \fr1{\pi} \fr{\sin(p_F(x-x')/\hbar)}{(x-x')}
  \label{c-x-y-1d}
\end{align}
Using these and \eq{var-ee}, we can easily compute
\begin{align}
S_1=\frac{1}{2 \pi \hbar} \int_{-l/2}^{l/2} dx \int dp\  u(x,p) =\frac{l p_F}{\pi \hbar}= \fr{l}{L}N, 
\end{align}
\begin{align}
  S_2=\frac{1}{\pi^2}\int_{-l/2}^{l/2}dx\int_{-l/2}^{l/2}dx' \, \left(\fr{\sin(p_F(x-x')/\hbar)}{(x-x')}\right)^2.
  \label{1D-log-a}
\end{align}
Using a change of variable $x_{ \pm} =\fr12(x \pm x')$, as we did earlier in section (\ref{sec:EE-LLL}), 
\begin{align}
S_2 &=\frac{1}{\pi^2}\int_{0}^{l/2}dx_-\int_{-l/2+x_-}^{l/2-x_-}dx_+ \left(\fr{\sin(2p_Fx_-/\hbar)}{x_-}\right)^2 \nonumber\\
&=\frac{1}{\pi^2} \int_{-l/2}^{l/2}dx_-(l-2|x_-|)\frac{ \sin^2 (\frac{2p_Fx_-}{\hbar})}{x_-^2}
\label{1D-log-b}
\end{align}
Putting  $t=\frac{p_Fx_-}{\hbar}$, we get
\begin{align}
S_2&= \frac{1}{\pi^2}\int_0^{\frac{\tilde p_F}{2}}dt\;\left(\frac{\tilde p_F}{t^2}-\frac{2}{t}\right)\sin^2(2 t)
\label{s2-t}\end{align}
where $\frac{p_F l}{\hbar}=\tilde p_F$
We will now evaluate equation (\ref{s2-t} for large $\tilde p_F$.
\begin{itemize}
\item The first term in the R.H.S of equation (\ref{s2-t}) can be evaluated in the limit $p_F \to \infty$
\begin{align*}
\int_0^{\infty} \frac{\sin^2(2t)}{t^2} =\pi
\end{align*}
\item We cannot use such a limit in the second term of equation (\ref{s2-t}) because of the logarithmic divergence at $t \to \infty$. Under the change of variable $y=2t$ we evaluate the second term as
\begin{align*}
\int_0^{p_F}dy \frac{1-\cos(2y)}{2y} =\frac{1}{2} \log \tilde p_F
\end{align*}
\end{itemize}
Therefore, to leading order in large $\tilde p_F$, we get
\begin{align}
  S_2 =  \frac{1}{\pi^2}\left(\pi (l p_F/\hbar)- \log (l p_F/\hbar) \right)
\end{align}
Thus,
\begin{align}
S= \fr{\pi^2}{3} (S_1 - S_2)= \fr13 \log(p_F l/\hbar)
\label{1d-ee-app}
\end{align}
which reproduces equation (\ref{1d-EE-log}). 


\subsubsection{2D Fermi fluid}\label{app:2d-fermi}

We consider 2D fermions in a periodic box of size $L^2$:
$\{ x\in (-L/2,L/2)\}$, $\{ y\in (-L/2,L/2)\}$; the fermi surface is given by $p_F \propto \fr{\sqrt N\pi\hbar}{L}$. We consider an entangling region $A$ which is a square of size $l^2, \; \ll L$:  $\{ x\in (-l/2,l/2)\}$, $\{ y\in (-l/2,l/2)\}$.
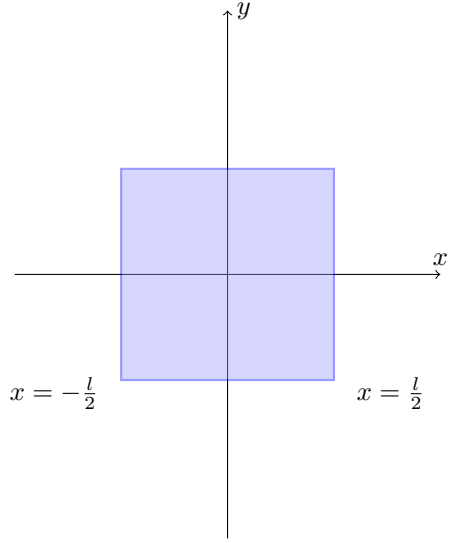
\begin{figure}[H]
  \begin{center}
\begin{tikzpicture}[scale=0.35]
\draw[->] (-8,0) -- (8,0) node[above] {$x$};
\draw[->] (0,-10) -- (0,10) node[right] {$y$};
\filldraw[blue, fill=blue!40, opacity=0.4, thick] (-4,-4) rectangle (4,4);
\node[right,black] at (4.5,-4.5)  {$x=\frac{l}{2}$};
\node[left,black] at (-4.5,-4.5)  {$x=-\frac{l}{2}$};
 \end{tikzpicture}
\end{center}
  \caption{\footnotesize{Entangling region in the shape of a square for a 2D fermi fluid.}}
  \label{fig:subregion_new}
\end{figure} 

The EE is given by
\begin{align}
  S_A \propto \fr{p_F l}{\hbar} \log(\fr{p_F l}{\hbar})
  \label{2d-EE-log}
\end{align}

A brief derivation is given below. In this derivation we could proceed as in the 1D case, doing the momentum integrals first and arriving at \eq{var-ee}. For the square droplet in Figure \ref{fig:fermi-circle}, the $x$ and $y$ directions factorize. Both $S_1$ and $S_2$ become squares of the respective 1D formulae, leading to the above expression.

In the following we will present the derivation a bit differently, doing the $x,y$ integrals first and then addressing the momentum integrals. An advantage of this approach is that it explicitly shows that the logarithms appear when the momentum integrals reach the Fermi surface.

\paragraph{Derivation}
\begin{align}
S_1=\frac{1}{(2 \pi \hbar)^2}\int_{-l/2}^{l/2} dx \int_{-l/2}^{l/2} dy \int_{-\infty}^{\infty}dp_x \int_{-\infty}^{\infty}dp_y u(x,y,p_x,p_y)
\end{align}
 where $u(x,y,p_x,p_y)=\Theta \left(\frac{p_F^2}{2 m}-\left(\frac{p_x^2+p_y^2}{2m}\right) \right)$. In the radial coordinate $(p_x,p_y)=p(\cos \theta,\sin 
 \theta)$. Therefore
 \begin{align}
 & S_1=\frac{2\pi }{(2 \pi \hbar)^2}l^2 \int_{0}^{p_F}pdp =\frac{l^2 p_F^2}{4 \pi \hbar^2},\quad p_F^2=\frac{4 \pi N}{L^2}\hbar^2 \nonumber \\
 & \text{hence,}\nonumber\\
  &S_1= \frac{N l^2}{L^2} \quad \text{as expected.}
 \end{align}
 
 \begin{align}
 S_2=&\frac{1}{(2 \pi \hbar)^4}\int_A dx dy \int_A dx' dy' \int_{\mathbb{R}^2} dp_x dp_y\int_{\mathbb{R}^2} dp_x' dp_y'\Theta (p_F^2-(p_x^2+p_y^2))\Theta (p_F^2-(p_x'^2+p_y'^2)) \nonumber \\
&\quad \times  \exp{-\frac{i}{\hbar}(x-x')(p_x-p_x')-\frac{i}{\hbar}(y-y')(p_y-p_y')} \nonumber \\
& =\frac{1}{(2 \pi \hbar)^4} \int_{0}^{p_F}dp p \int _{0}^{2 \pi}d\theta\int_{0}^{p_F}dp' p' \int_{0}^{2\pi}d\theta' \left(\frac{2 \sin (
\frac{(p \cos \theta -p'\cos \theta')l}{2 \hbar})}{\frac{p \cos \theta-p'\cos \theta'}{\hbar}}\right)^2
\left(\frac{2 \sin (
  \frac{(p \sin \theta -p'\sin \theta')l}{2 \hbar})}{\frac{p \sin \theta-p'\sin \theta'}{\hbar}}\right)^2
\label{2D-log-a}
 \end{align}
 Again we used here radial coordinate. We now do the following scaling
 \begin{align*}
 \tilde p=\frac{p l}{\hbar}, \quad \tilde p'=\frac{p' l}{\hbar} ,\quad \tilde p_F=\frac{p_Fl}{\hbar}
 \end{align*}
 \begin{align}
S_2=\frac{1}{\pi^4}\int_{0}^{\tilde p_F} d\tilde p \tilde p \int_{0}^{\tilde p_F} d\tilde p' \tilde p' \int_{0}^{2 \pi}d \theta \int_{0}^{2 \pi}d \theta'
\left(\frac{ \sin (
\frac{\tilde p \cos \theta -\tilde p'\cos \theta'}{2 })}{\tilde p \cos \theta-\tilde p'\cos \theta'}\right)^2
\left(\frac{ \sin (
\frac{\tilde p \sin \theta -\tilde p'\sin \theta'}{2})}{\tilde p \sin \theta-\tilde p'\sin \theta'}\right)^2
 \end{align}
 We are interested  in $\hbar \to 0$ limit i.e., the $\tilde p_F \to \infty $ limit. In this limit, the Fermi surface goes to $\infty$ and as a result, we can replace the Fermi surface by box.
\begin{figure}[H]
  \begin{center}
\begin{tikzpicture}[scale=0.2]
\draw[->] (-8,0) -- (8,0) node[above] {$p_x$};
\draw[->] (0,-10) -- (0,10) node[right] {$p_y$};
\draw[black, very thick] (0,0) circle (5);
\draw[red, very thick, ->] (0,0) -- (4,3);
  \node[above] at (4.2,3) {$P_F$};  
 \draw[->] (10,0) -- (26,0) node[above] {$p_x$};
\draw[->] (18,-10) -- (18,10) node[right] {$p_y$};
\draw[black,very thick] (14,-4) rectangle (22,4);
\node[below,black] at (17,-4)  {$-p_F$};
\node[left,black] at (14,0.5)  {$-p_F$};
\node[right,black] at (22,0.5)  {$p_F$};
\node[above,black] at (17,4 )  {$p_F$};
 \end{tikzpicture}
\end{center}
  \caption{\footnotesize{On the left we consider the original droplet which has the shape of a disk of radius $p_F$, which represents the ground state of a free fermion problem. On the right we consider a square droplet of side $p_F$; In the limit of $p_F \to \infty$ both shapes will fill out the plane. For large enough sizes of both figures, we expect the EE for both shapes to show similar qualitative behaviour.}}
  \label{fig:fermi-circle}
 \end{figure}
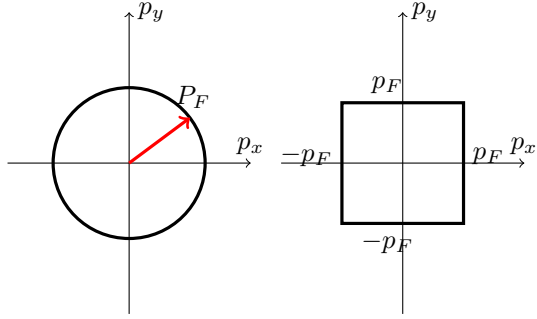
 In such a case $u(x,y,p_x,p_y)=\Theta (p_F-|p_x|) \Theta(p_F-|p_y|) \implies S_1=\frac{l^2 N}{L^2}$.
 \begin{align}
 S_2 &=\frac{1}{(2 \pi \hbar)^4}\int_{-p_F}^{p_F}dp_x \int_{-p_F}^{p_F}dp_x '\int_{-p_F}^{p_F}dp_y\int_{-p_F}^{p_F}dp_y' \left(\frac{2 \sin\left( \frac{(p_x-p_x')l}{2\hbar}\right)}{(p_x-p_x')/\hbar}\right)^2
 \left(\frac{2 \sin\left( \frac{(p_y-p_y')l}{2\hbar}\right)}{(p_y-p_y')/\hbar}\right)^2 \nonumber \\
 & =\frac{p_F^2l^2}{\pi^2 \hbar^2}-\frac{2}{\pi^3}\frac{p_F l}{\hbar}\log(\frac{p_Fl}{\hbar})+\frac{1}{\pi^4}\left(\log (\frac{p_Fl}{\hbar})\right)^2
 \label{2D-log}
 \end{align}
 \begin{align}
   S=\frac{2}{3\pi}\frac{p_F l}{\hbar}\log\frac{p_F l}{\hbar}
   \label{2D-log-b}
 \end{align}
 For the original spherical Fermi surface, we expect the numerical coefficient in above equation to be different.
 
\section{$1$D-$2$D correspondence for the  first landau level: Wedding cake structure}\label{sec:wedding}
In this section we will allow the electrons to access the first landau level. Consider a state of $N$ electrons, where first $N_1$ number of electrons are in lowest landau level and remaining $N_2$ electrons are in the first Landau level such that $N=N_1+N_2$. The wedding cake structure of the Wigner distribution similar to what we will discuss below has appeared in \cite{PhysRevA.103.033321}.
 \par
Of the $N_1$ electrons in the LLL, $N_{max}$ electrons fill all states with energies below the bottom of the first Landau level. The remaining $m=N_1-N_{max}$ LLL electrons have energies greater than the bottom of the first Landau level. We then fill the first Landau level with $N_2=m$ electrons, such that the total number of electrons are $N = N_1+N_2 = N_{max}+2m$ (see Figure (\ref{cake-spect}))
\begin{figure}[H]
  \centering
  \begin{tikzpicture}[>=stealth, thick]

\draw[->] (0,0) -- (0,6) node[left] {$E$};
\draw[->] (0,0) -- (8,0) node[right] {$n_2$};

\draw[blue] (0,3) -- (7,3);
\node[left,blue] at (0,0) {$n_1=0$};
\node[left,blue] at (0,3) {$n_1=1$};

\draw[red] (0,0) -- (7,5);
\foreach \i in {0,0.428,...,6} {
    \pgfmathsetmacro{\y}{5/7*\i} 
    \fill[black] (\i,\y) circle (0.08);
}

\draw[red] (0,3) -- (7,8);
\foreach \i in {0,0.428,...,2} {
    \pgfmathsetmacro{\y}{3+5/7*\i} 
    \fill[black] (\i,\y) circle (0.08);
}

\draw[decorate,decoration={brace,amplitude=5pt}] (-0.1,3.1) -- (2,10/7+3+0.19)
    node[midway,above,yshift=5pt,blue] {$m$};

\draw[decorate,decoration={brace,amplitude=6pt}] (-0.1,0.1) -- (6.2,4.5)
    node[pos=0.25,above,yshift=6pt,black] {$N_1$};
    
\draw[decorate,decoration={brace,amplitude=6pt,mirror}] (4.2,2.9) -- (6.1,4.25)
    node[midway,below,yshift=-5pt,blue] {$m$};
    
\draw[dashed] (4,0) -- (4,6);
\node[below] at (4,0) {$N_{\max}$};

\end{tikzpicture}

    \caption{\footnotesize{Spectrum for multiple filled Landau levels.}}
  \label{cake-spect}
\end{figure}
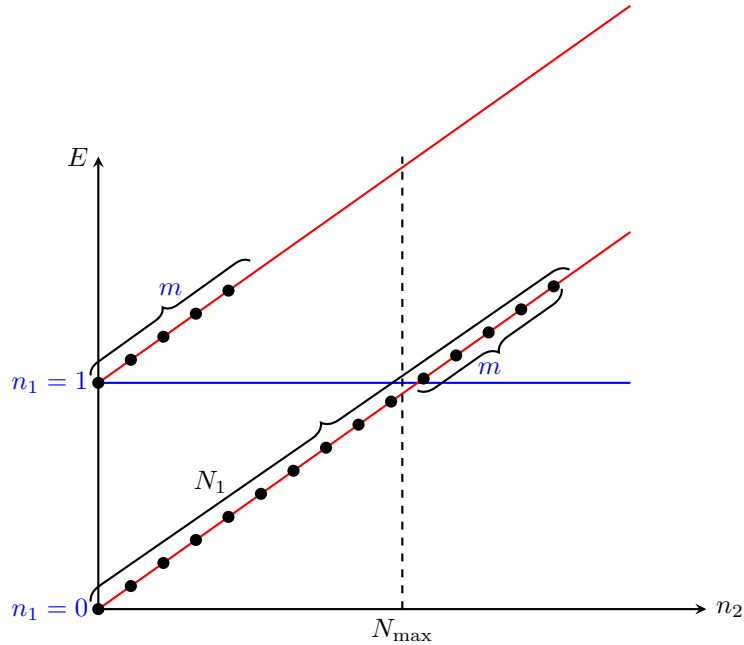
The Wigner distribution function representing above state is 
\begin{align}
U(x_2,p_2)=\sum_{n_1,n_2}u_{n_1,n_2}(x_2,p_2)=\sum_{n_2=0}^{N_{max}+m-1}u_{0,n_2}(x_2,p_2)+\sum_{n_2=0}^{m-1}u_{1,n_2}(x_2,p_2)
\label{fll}\end{align}
In the large $N$ limit, the first term of the RHS of the above equation is  \begin{align}\sum_{n_2=0}^{N_{max}+m-1}u_{0,n_2}(x_2,p_2)=\theta(\tilde r_{N_1}^2-\tilde r^2), \tilde r_{N_1}=\sqrt{N_1\hbar}\end{align}
The second term in the RHS of the above equation is simplified as
\begin{align}
\sum_{n_2=0}^{m-1}u_{1,n_2}(x_2,p_2) &=u_1(x_1,p_1)\sum_{n_2=0}^{m-1}u_{n_2}(x_2,p_2) \nonumber \\ 
&= (-2) \exp\left(-\frac{p_1^2 }{m \omega \hbar}- \frac{x_1^2 m \omega}{\hbar}\right) \left(1-2\left(\frac{p_1^2 }{m \omega \hbar}+ \frac{x_1^2 m \omega}{\hbar}\right)\right)\sum_{n_2=0}^{m-1}u_{n_2}(x_2,p_2) \nonumber \\
&= \left(\hbar \left(\frac{\partial^2 }{m \omega \partial x_1^2}+\frac{m \omega \partial^2 }{\partial p_1^2}\right)+2\right)
\exp\left(-\frac{p_1^2 }{m \omega \hbar}- \frac{x_1^2 m \omega}{\hbar}\right)\sum_{n_2=0}^{m-1}u_{n_2}(x_2,p_2)
\end{align}
The partial derivative operator vanishes in $\hbar \rightarrow 0$ limit.Hence in the large N limit
\begin{align}
U(x_2,p_2)=\theta(\tilde r_{N_1}^2-\tilde r^2)+\exp\left(-\frac{p_1^2 }{m \omega \hbar}- \frac{x_1^2 m \omega}{\hbar}\right)\theta(\tilde r_{N_2}^2-\tilde r^2)
\end{align}
In the large N limit fermion density is
\begin{align}
\rho(x,y)=\frac{m \omega}{\pi \hbar}\left(\theta(\tilde r_{N_1}^2-\tilde r^2)+\theta(\tilde r_{N_2}^2-\tilde r^2)\right)
\end{align}
\begin{figure}[H]
    \centering
    \begin{minipage}{.5\textwidth}
        \centering
        \includegraphics[width=1.28\textwidth, height=0.15\textheight]{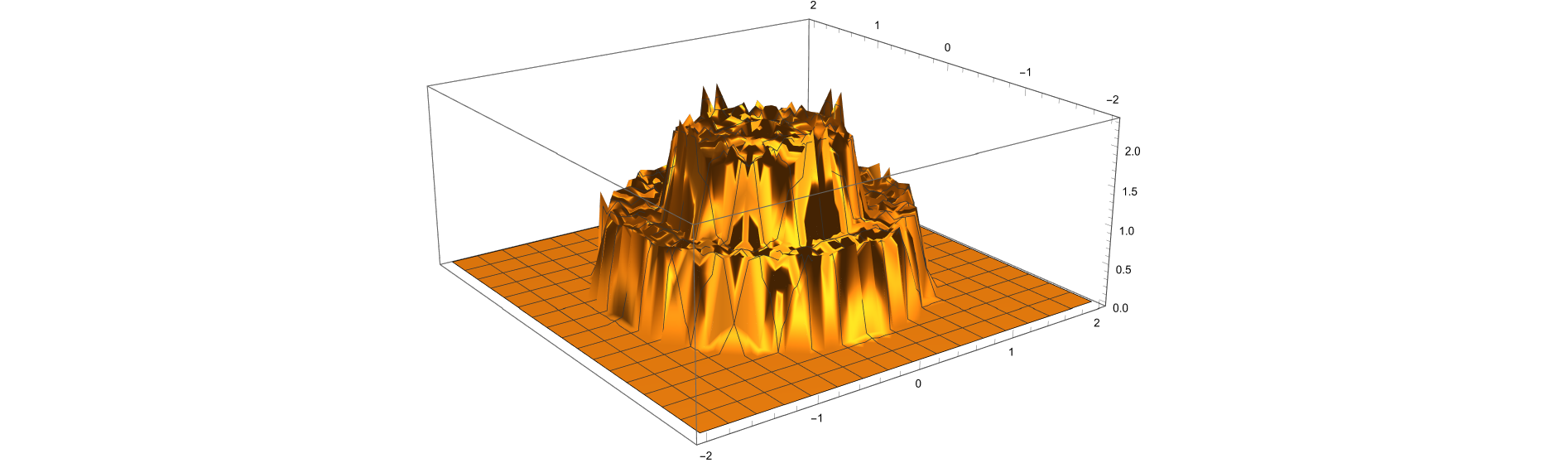}\\
        \vspace{1ex}\vbox{\begin{quote}\baselineskip=9pt{\footnotesize (a) 3D plot of the Wigner distribution. The horizontal axes represent $\tilde x_2, \tilde p_2$.}\end{quote}}
    \end{minipage}%
    \begin{minipage}{0.5\textwidth}
        \centering
        \includegraphics[width=0.48\textwidth, height=0.15\textheight]{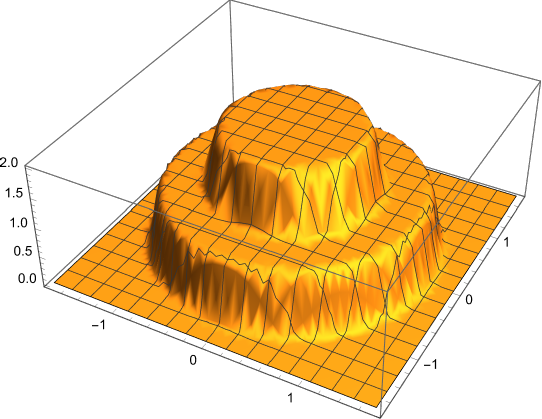}\\
        \kern5pt\vbox{\begin{quote}\baselineskip=9pt {\footnotesize (b) 3D plot of the fermion density. The horizontal axes represent $x,y$.}\end{quote}}
    \end{minipage}
    \caption{\footnotesize{Classical limit of ground state properties (3D plot). $N=100, \hbar=1/N, m\om=5$.}}
    \label{fig:cake}
\end{figure}
\newpage
\bibliography{references.bib}
\bibliographystyle{JHEP} 

\end{document}